\documentclass[11pt]{article}

\usepackage[usenames,dvipsnames,table]{xcolor}
\usepackage[utf8]{inputenc}

\pdfoutput=1 
\usepackage{jheppub} 

\usepackage{graphicx}
\usepackage{amsmath, amssymb}

\newcommand{\be}{\begin{eqnarray}}
\newcommand{\ee}{\end{eqnarray}}
\newcommand{\bea}{\begin{eqnarray}}
\newcommand{\eea}{\end{eqnarray}}

\newcommand{\bn}{\begin{enumerate}}
\newcommand{\en}{\end{enumerate}}

\def\SO{\mathrm{SO}}

\def\Z{\mathbb{Z}}

\def\e{\epsilon}

\def\half{\frac{1}{2}}

\title{Sequences of $6d$ SCFTs on generic Riemann surfaces}

\preprint{}

\author{Shlomo S. Razamat}
\author{and Evyatar Sabag}

\affiliation{Department of Physics, Technion, Haifa, 32000, Israel}

\emailAdd{razamat@physics.technion.ac.il}
\emailAdd{sevyatar@campus.technion.ac.il}

\abstract{We consider compactifications of $6d$ minimal $(D_{N+3},D_{N+3})$ type conformal matter SCFTs on a generic Riemann surface. We derive the theories corresponding to three punctured spheres (trinions) with three maximal punctures, from which one can construct models corresponding to generic surfaces. 
The trinion models are simple quiver theories with $\mathcal{N}=1$ $SU(2)$ gauge nodes. One of the three puncture non abelian symmetries is emergent in the IR. The derivation of the trinions proceeds by analyzing RG flows between conformal matter SCFTs with different values of $N$ and relations between their subsequent reductions to $4d$. In particular, using the flows we first derive trinions with two maximal and one minimal punctures, and then we argue that collections of $N$ minimal punctures can be interpreted as a maximal one.
This suggestion is checked by matching the properties of the $4d$ models such as 't Hooft anomalies, symmetries, and the structure of the conformal manifold to the expectations from $6d$. We then use the understanding that collections of minimal punctures might be equivalent to maximal ones to construct trinions with three maximal punctures, and then $4d$ theories corresponding to arbitrary surfaces, for $6d$ models described by two $M5$ branes probing a $\mathbb{Z}_k$ singularity. This entails the introduction of a novel type of maximal puncture. Again, the suggestion is checked by matching anomalies, symmetries and the conformal manifold to expectations from six dimensions. These constructions thus give us a detailed understanding of compactifications of two sequences of six dimensional SCFTs to four dimensions.
}

\begin{document} 

\maketitle
\flushbottom

\section{Introduction}

Following the seminal work of Gaiotto \cite{Gaiotto:2009we} in recent years several instances of dictionaries between  $4d$ SCFTs and compactifications on a Riemann surface of $6d$ SCFTs  have been worked out. For general Riemann surfaces examples include $A_1$ $(2,0)$ \cite{Gaiotto:2009we}, $A_2$ $(2,0)$ \cite{Gadde:2015xta}, $A_1$ $(2,0)$ probing $\Z_2$ singularity \cite{Gaiotto:2015usa,Razamat:2016dpl}, $SU(3)$ and $SO(8)$ minimal SCFTs   \cite{Razamat:2018gro}, and the rank one E-string \cite{Kim:2017toz,Razamat:2019vfd}. Much more is known for special surfaces. For example $(2,0)$ theories on surfaces with irregular punctures in some cases give Argyres-Douglas theories, for which Lagrangians have been worked out in \cite{Maruyoshi:2016tqk,Maruyoshi:2016aim,Agarwal:2016pjo}. Other examples involve  $(2,0)$ SCFTs probing ADE singularities and higher rank E-string on surfaces of genus zero with at most two punctures or tori \cite{ Kim:2018bpg,Kim:2018lfo,Bah:2017gph,Chen:2019njf,Pasquetti:2019hxf}, some more esoteric $6d$ SCFTs on tori \cite{Razamat:2018gbu,Zafrir:2018hkr}, as well as $(2,0)$ on  spheres with special collections of punctures \cite{Razamat:2019vfd}. Such dictionaries between four and six dimensions lead to many insights into duality properties of four dimensional theories as well as to some systematic understanding of emergent IR symmetries. 

In some cases a six dimensional SCFT compactified on a circle, possibly with holonomies and twists, leads to an effective gauge theory description in five dimensions. The four dimensional theories resulting from a further compactification on a segment or a circle can be rather systematically derived once duality domain walls in five dimensions \cite{Gaiotto:2014ina} are understood \cite{Kim:2017toz,Kim:2018bpg,Kim:2018lfo}.  The duality walls in question interpolate between two different effective five dimensional theories obtained by compactifications of same $6d$ SCFT on a circle albeit with different values of holonomies. Upon further compactification to four dimensions these non trivial profiles of holonomies can be interpreted as flux for the global symmetry of the $6d$ SCFT supported on the two dimensional Riemann surfaces \cite{Chan:2000qc,Kim:2017toz}. 

Such direct analysis of compactification of six dimensional SCFTs to four dimensions is applicable for surfaces (with flux supported on them) of genus zero and less than three punctures, or a torus with no punctures. A systematic way to construct surfaces with more punctures was suggested in \cite{Razamat:2019mdt}. To do so one considers flux compactifications of a $6d$ SCFT together with a vacuum expectation value deformation triggering a flow to a different SCFT in six dimensions. The basic observation is that first reducing to four dimensions on a certain surface with flux and then flowing is equivalent to first flowing to the new SCFT in six dimensions and then reducing on a {\it different} Riemann surface. This new surface has the same genus as the first one but possibly more punctures. The number of additional punctures is related to the details of the flow and on the value of the flux. This observation was used in \cite{Razamat:2019mdt} to study relations between type $A_{N-1}$ $(2,0)$ theories probing $\Z_k$ singularity with different values of $k$. In this case theories corresponding to tori with arbitrary number of minimal punctures are known \cite{Gaiotto:2015usa,Bah:2017gph} and the procedure of generating punctures from flux can be put to the test.

In the current paper we {\it first } apply the same procedure to compactifications of $(1,0)$ six dimensional SCFT residing on a single M5 brane probing $D_{N+3}$ singularity. This model is known as the $(D_{N+3}, D_{N+3})$ minimal conformal matter \cite{DelZotto:2014hpa} and the case of $N=1$ is the (rank 1) E-string. We will refer to the set of $4d$ theories obtained by compactifying the minimal $(D_{N+3},D_{N+3})$ conformal matter on punctured Riemann surfaces as theories in {\it minimal class ${\cal S}_{D_{N+3}}$}. For brevity we will also often refer to  $(D_{N+3}, D_{N+3})$ minimal conformal matter as $D_{N+3}$ type conformal matter.
 Here using the duality domain wall picture the theories corresponding to sphere with two maximal punctures are known for general $N$ \cite{Kim:2018bpg,Kim:2018lfo} and for $N=1$ compactifications for arbitrary surfaces were derived in \cite{Kim:2017toz,Razamat:2019vfd}. However no theories corresponding to surfaces with more than two punctures for $N>1$ are known.  By studying flows between $D_{N+3}$ minimal conformal matter theories with different values of $N$ we will derive theories corresponding to compactifications on spheres with two maximal punctures, with $SU(2)^N$ symmetry, and arbitrary number of minimal punctures. The minimal punctures we will obtain have $SU(2)$ global symmetry.
The puncture symmetry appears explicitly in the UV Lagrangian as the Cartan $U(1)$ which enhances to $SU(2)$ in the IR.  These models are simple quiver theories with $SU(2)$ $N_f=4$ gauge nodes.

In the case of $N=1$ the minimal puncture coincides with the maximal one and the construction will give us a sphere with three maximal punctures from which any surface can be constructed. In this case we will be able to test our results against the known constructions of \cite{Kim:2017toz,Razamat:2019vfd}. To obtain theories corresponding to arbitrary surfaces we will need to gauge the emergent $SU(2)$ symmetries. In particular the Lagrangian constructions of this paper  are dual to the ones of \cite{Kim:2017toz,Razamat:2019vfd}.
 
\begin{figure}[t]
	\centering
  	\includegraphics[scale=0.32]{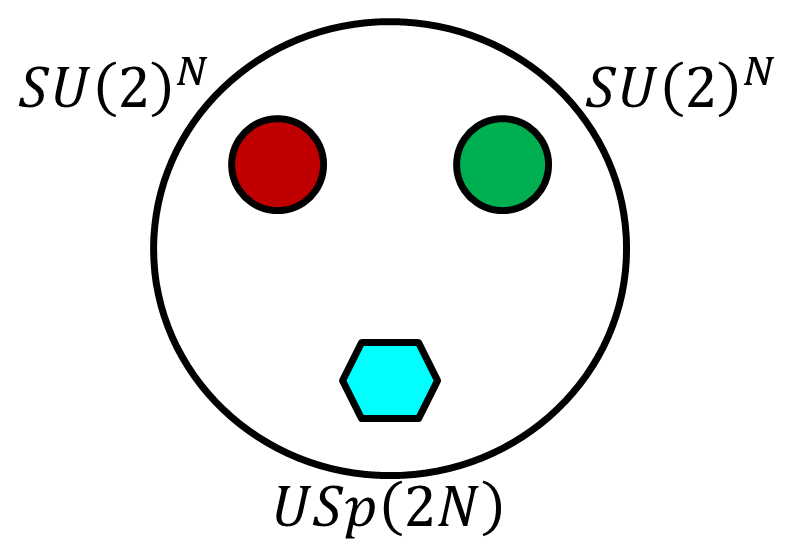}
    \caption{The  trinion for the $D_{N+3}$ minimal conformal matter obtained in this paper. The trinion has a quiver description with ${\cal N}=1$ $SU(2)$ gauge nodes. The two factors of $SU(2)^N$ global symmetry are manifest in the UV Lagrangian. The $USp(2N)$ symmetry appears only in the IR as enhancement of a UV $U(1)^N$ global symmetry.}
    \label{F:DinteractingTrinion}
\end{figure}

For $N>1$ we will argue that a theory with $N$ minimal punctures has a locus on the conformal manifold where the $N$ copies of $SU(2)$ symmetries enhance to $USp(2N)$. The $D_{N+3}$ type minimal conformal matter has actually, at least, three different types of maximal punctures with different symmetries $SU(2)^N$, $SU(N+1)$, and $USp(2N)$ \cite{Hayashi:2015fsa,Hayashi:2015zka}. We show that the $USp(2N)$ symmetry we obtain corresponds to a maximal puncture. Thus, in particular we obtain theories corresponding to spheres with three maximal punctures, two $SU(2)^N$ and an additional $USp(2N)$ one, see Figure \ref{F:DinteractingTrinion}. 
We then use these to construct theories corresponding to arbitrary Riemann surfaces. We verify that the resulting theories have the correct anomalies and the expected structure of the superconformal index. In particular the conformal manifold is given by the space of flat connections for the $SO(4N+12)$ global symmetry and the complex structure moduli. By using the Lagrangians for spheres with two maximal punctures and some value of flux \cite{Kim:2018bpg,Kim:2018lfo} we then consider theories with non zero flux.

Let us  stress that although the trinion theories we construct are usual Lagrangian models, to construct theories corresponding to general Riemann surfaces we need to gauge symmetries which only appear in the IR. This is a recurring theme in many constructions of dictionaries between six and four dimensions. There are numerous examples by now \cite{Gadde:2015xta,Razamat:2016dpl,Agarwal:2018ejn,Kim:2017toz} with some of the more recent ones being the sphere with two puncture compactifications of rank $Q>1$ E-string \cite{Pasquetti:2019hxf}. There one of the two $USp(2Q)$ puncture symmetries is an enhanced symmetry in the IR from the $SU(2)^Q$ UV Lagrangian symmetry.

\begin{figure}[t]
	\centering
  	\includegraphics[scale=0.32]{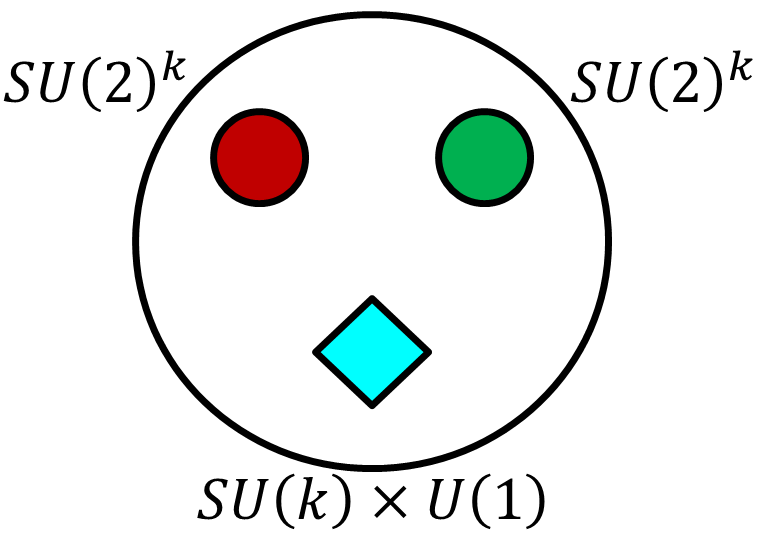}
    \caption{The formal depiction of the trinion we suggest. The trinion has two maximal punctures with $SU(2)^k$ symmetry and a novel maximal puncture with $SU(k)\times U(1)$ symmetry. The novel puncture is maximal in the sense that the symmetry associated to it can be gauged to glue together different surfaces.}
    \label{F:AtypeTrinion}
\end{figure}

In the {\it second} part we then use the observation that collections of minimal punctures can combine into a maximal puncture\footnote{See also \cite{Razamat:2016dpl,Razamat:2018gro}  and \cite{Chacaltana:2012ch,Chacaltana:2013oka,Chacaltana:2016shw}  for earlier appearances of this effect. In the latter set of references such smooth transitions between collectionsns of smaller punctures into bigger ones in the framework of class $\mathcal{S}$  were called atypical degenerations.} and apply it to the case of compactifications of two $M5$ branes probing a $\Z_k$ singularity.
Here the theories corresponding to spheres with at most two maximal and many minimal punctures are known \cite{Gaiotto:2015usa}. We  argue that a collection of $k$ minimal punctures can be  interpreted as a novel type of maximal puncture which has $SU(k)\times U(1)$ symmetry, see Figure \ref{F:AtypeTrinion}. The minimal punctures have a $U(1)$ symmetry associated to them and the claim is that somewhere on the conformal manifold of these models $U(1)^k$ symmetry enhances to $SU(k)\times U(1)$. We claim that theories corresponding to surfaces with these novel punctures can be glued to each other by first gauging the $SU(k)$ symmetry, and then the remaining $U(1)$ symmetry is conjectured to enhance to $SU(2)$ which is also gauged. We check this sequence of conjectures by computing anomalies and comparing them to six dimensions and by verifying the expected symmetry properties of the superconformal indices. 

We want to stress in this paper three different important points. {\it First}, as was discussed in \cite{Razamat:2019mdt}, understanding compactifications of sequences of theories with flux but maybe with small number of punctures can lead to understanding of surfaces with any number of certain types of punctures by studying RG flows. {\it Second}, collection of smaller punctures can be equivalent to bigger punctures. Here by size of a puncture we refer to the dimension of the symmetry group associated with it. {\it Third}, there might be different types of maximal punctures along which surfaces can be glued together. All three of these points appeared before in various works and our results suggest that  these should be crucial in understanding general types of compactifications of six dimensional SCFTs to four dimensions.

The organization of the paper is as follows. In section \ref{sec:6d} we collect the general robust expectations about the $4d$ SCFTs obtained in compactifications of the $D_{N+3}$ minimal conformal matter that can be obtained from properties of the $6d$ SCFT.  As the derivation of the four dimensional models from six dimensional flows is rather technical and builds heavily on previous results, we next discuss our main four dimensional results in sections \ref{sec:estring} and \ref{sec:dnmatter}  and postpone the derivation to section five.
In section \ref{sec:estring} we present our main results in the case of $D_{4}$ minimal conformal matter. In particular we discuss the $4d$ trinion models, the dualities,  symmetry enhancements, the matching of anomalies with $6d$, and expectations from indices. In section \ref{sec:dnmatter} we discuss the results for $D_{N+3}$ minimal conformal matter with $N>1$. Here some new features appear. In particular the basic trinion in addition to two maximal punctures has a minimal puncture. We argue that $N$ minimal punctures correspond to an $USp(2N)$ maximal puncture and repeat the checks detailed above for these cases. In section \ref{sec:derivation} we discuss the derivation of the results in the previous two sections studying various RG flows in $6d$ and $4d$.  In section \ref{sec:atrinions} we discuss the construction of four dimensional theories obtained by compactifying on a general Riemann surface the $(1,0)$ SCFT residing on two $M5$ branes probing $\Z_k$ singularity. In section \ref{sec:discussion} we summarise and discuss our results. Several appendices complement the main part of this paper with additional details.

\section{Expectations from six dimensions}\label{sec:6d}

Let us briefly review some robust expectations of the four dimensional models obtained in the compactifications of the $6d$ minimal $D_{N+3}$ SCFT. These expectations involve symmetries, 't Hooft anomalies, and some properties of protected operators. 

The six dimensional models have $SO(4N+12)$ symmetry group for $N>1$ which enhances to $E_8$ for $N=1$.\footnote{We are  not going to be careful about the global structure of the symmetry groups in this paper.} Throughout this paper the $4d$ residual symmetries from the $6d$ symmetry will be often referred to as the {\it internal symmetries}.
Mostly we will consider compactifications without turning on flux supported on the Riemann surface and thus for closed Riemann surfaces the above $6d$ symmetry is the one expected of the four dimensional theories. Punctures typically break this symmetry group to some subgroup depending on the properties of the punctures \cite{Razamat:2016dpl}. Turning on flux for abelian subgroups of $SO(4N+12)$ also breaks the symmetry to the $U(1)$ with the flux times its commutant. The four dimensional models also inherit an R-symmetry from six dimensions. This will be the R-symmetry we will use for the $4d$ models although this is not necessarily the conformal one.

We can compute the 't Hooft anomalies for all the global symmetries by considering the anomaly polynomial of the six dimensional model and integrating it over the Riemann surface \cite{Benini:2009mz}. In case punctures are present one also needs to take into account the anomaly inflow contributions at the punctures, see for example \cite{Kim:2017toz}. The anomaly eight-form polynomial of $D_{N+3}$ minimal conformal matter SCFT is given by \cite{Ohmori:2014kda},

\be
\label{E:minD6dAP}
I_{8} & = & \frac{N(10 N + 3)}{24}C^2_2 (R) - \frac{N(2N+9)}{48} p_1 (T) C_2 (R) - \frac{N}{2} C_2 (R) C_2(SO(4N+12))_{V} \nonumber \\  & + & \frac{(N+2)}{24} p_1 (T) C_2(SO(4N+12))_{V} + \frac{(2N+1)}{24} C^2_2(SO(4N+12))_{V} \nonumber \\  & - & \frac{(N-1)}{6} C_4(SO(4N+12))_{V} + (29+(N-1)(2N+13))\frac{7p_1 (T) - 4p_2 (T)}{5760} 
\ee
Here $C_2 (R)$ denotes the second Chern class in the fundamental representation of the $SU(2)_R$ R-symmetry of the six dimensional model, $C_n(G)_{\bold{R}}$ is the $n$-th Chern class of the global symmetry $G$, evaluated in the representation $\bold{R}$ (here $V$ stands for vector), and $p_1 (T), p_2 (T)$ are the first and second Pontryagin classes, respectively.

We wish to compactify the $6d$ theory on a genus $g$ Riemann surface. In this process we can turn on flux for abelian subgroups of the $6d$ $SO(4N+12)$ global symmetry supported on the compactification surface. For the sake of brevity we will only discuss in detail flux in a single particular $U(1)$. We choose this flux  such that it breaks the $6d$ symmetry down to $U(1) \times SU(r) \times \SO(4N+12-2r)$. In order to compactify the $6d$ anomaly polynomial eight-form to $4d$, we first need to decompose the $SO(4N+12)$ characteristic classes to the commutant of the $U(1)$'s first Chern class, we find that
\be
\label{E:SOdecomp}
C_2(SO(4N+12))_{V} & = & - r C^2_1 (U(1)) + C_2(SO(4N+12-2r))_{V} + 2 C_2(SU(r))_{F} ,\nonumber\\
C_4(SO(4N+12))_{V} & = & \frac{r(r-1)}{2} C^4_1 (U(1)) + 2 C_2(SU(r))_{F} C_2(SO(4N+12-2r))_{V} \nonumber\\
 & &  + C^2_2(SU(r))_{F}- r C^2_1 (U(1)) C_2(SO(4N+12-2r))_{V}\nonumber \\
 & & + 2(3-r) C^2_1 (U(1)) C_2(SU(r))_{F} - 6 C_1 (U(1)) C_3(SU(r))_{F}\nonumber\\ 
 & & + C_4(SO(4N+12-2r))_{V} + 2 C_4(SU(r))_{F}.  
\ee 
Where $C_1(U(1))$ is the first Chern class of the chosen $U(1)$. All our normalizations follow the conventions of \cite{Kim:2018bpg}.

Carrying on with the compactification process, we now wish to compactify the  above anomaly polynomial on a Riemann surface $\Sigma$ of genus $g$ with flux under the chosen $U(1)$. To that end we set $\int_{\Sigma} C_1(U(1))=-z$ where $z$ is an integer. The natural R-symmetry from $6d$ (not necessarily the superconformal R-symmetry in $4d$) under the embedding $U(1)_{R}\subset SU(2)_{R}$ in general does not preserve supersymmetry. We can preserve supersymmetry by twisting the $SO(2)$ acting on the tangent space of the Riemann surface with the Cartan of $SU(2)_{R}$, and find the decomposition $C_{2}\left(R\right)=-C_{1}\left(R'\right)^{2}+2(1-g)tC_{1}\left(R'\right)+\mathcal{O}\left(t^{2}\right)$. Finally we set 
\be
\label{twisting}
C_{1}\left(U(1)\right)=-zt+\epsilon C_{1}\left(U(1)_{R}\right)+C_{1}\left(U(1)_{F}\right).
\ee 
The first term is the flux multiplied by $t$ a unit flux two form on $\Sigma$, meaning $\int_{\Sigma} t=1$. The second term is needed for possible mixing of the $4d$ global $U(1)$ with the superconformal R-symmetry, the mixing parameter $\epsilon$ will be determined by a-maximization. The last term is the $4d$ curvature of the $U(1)$.

Plugging the aforementioned decompositions to \eqref{E:minD6dAP}, we find the $4d$ anomaly polynomial six-form, from which we can extract the 't Hooft anomalies
\be
\label{E:4dAP}
&& Tr(U(1)_R) = -N(2N+9)(g-1)\,, \qquad Tr(U(1)^3_R) = N(10N+3)(g-1), \nonumber\\
&& Tr(U(1)_R U(1)^2_F) = -2Nr(g-1)\,, \qquad Tr(U(1)_F U(1)^2_R) = 2 N r z,\nonumber\\
&& Tr(U(1)_F) = -2 r (N+2) z\,, \qquad Tr(U(1)^3_F) =-(3r + 2N - 2) r z,\nonumber\\
&& Tr(U(1)_R SO(4N+12-2r)^2) = -N(g-1)\,, \qquad Tr(U(1)_R SU(r)^2) = -N(g-1) ,\nonumber\\ 
&& Tr(U(1)_F SO(4N+12-2r)^2) = - \frac{r z}{2} \,, \qquad  Tr(U(1)_F SU(r)^2) = - \frac{(2N+r-2) z}{2}, \nonumber\\
&& Tr(SU(r)^3) = - 2 z (N-1)\,.
\ee
In the above anomalies $U(1)_R$ refers to the $6d$ R-symmetry. From these results we can determine $\epsilon$ in \eqref{twisting} via $a$-maximization \cite{Intriligator:2003jj}. We find,
\be
\epsilon=\frac{2\sqrt{9N^{2}(1-g)^{2}+z^{2}(5N+1)(2N+3r-2)}-3N(g-1)}{3(2N+3r-2)}.
\ee
This gives the $a$ and $c$ anomalies
\be
a & = & \frac{3}{32}\left(2N\left(16N-9r\epsilon^{2}+9\right)(g-1)-rz\epsilon\left(2N\left(3\epsilon^{2}-10\right)+3(3r-2)\epsilon^{2}-4\right)\right)\\
c & = & \frac{1}{32}\left(2N\left(50N-27r\epsilon^{2}+36\right)(g-1)-rz\epsilon\left(2N\left(9\epsilon^{2}-32\right)+9(3r-2)\epsilon^{2}-20\right)\right).
\ee Specifically with no flux the six dimensional R-symmetry is also the superconformal one in $4d$ as there are no abelian symmetries for it to mix with. The conformal anomalies are,
\be
a= \frac{3N}{16}\left(16N+9\right)(g-1)\,,\qquad c=  \frac{N}{16}\left(50N+36\right)(g-1)\,.
\ee

\

We would also like to add the anomalies in case of breaking $SU(r)$ to $U(1)^{r-1}$ by fluxes, as it will be useful for some of the computations we will perform. To this end, we will need to decompose the $SU(r)$ characteristic classes to the $U(1)$'s characteristic classes in the $6d$ eight-form anomaly polynomial after using the decompositions of \eqref{E:SOdecomp}. The $SU(r)$ decompositions are given by,
\be
C_2(SU(r))_F & = & -\half \sum_{i=1}^r C_1(U(1)_{a_i})^2\,,\nonumber\\
C_3(SU(r))_F & = & \frac{1}{3} \sum_{i=1}^r C_1(U(1)_{a_i})^3\,,\nonumber\\
C_4(SU(r))_F & = & -\frac{1}{4} \sum_{i=1}^r C_1(U(1)_{a_i})^4 +\frac{1}{8}\sum_{i,j=1}^r C_1(U(1)_{a_i})^2 C_1(U(1)_{a_j})^2\,,
\ee 
where $C_1(U(1)_{a_r})= - \sum_{i=1}^{r-1} C_1(U(1)_{a_i})$.
From here the compactification process is similar to the one before only we additionally set,
\be
C_1(U(1)_{a_i})=-N_{a_i} t + \epsilon_i C_1(U(1)_R)+C_1(U(1)_{F_i})\,.
\ee
As before, the first term is the flux multiplied by $t$, a unit flux two form on $\Sigma$ ($\int_{\Sigma} t=1$). The second term is needed for possible mixing of the $4d$ global $U(1)_{a_i}$ with the superconformal R-symmetry. The mixing parameter $\epsilon_i$ will be determined by a-maximization. The last term is the $4d$ curvature of each $U(1)_{a_i}$.

After this process we can find the $4d$ anomaly polynomial six-form, and extract from it the 't Hooft anomalies involving $U(1)_{F_i}$,\footnote{In the last line here and in the  last line of \eqref{E:4dAPwPunc}, in the previous version of the paper the anomalies were given implicitly for the Cartan generators of the group. Here these are given for the group as written. This change of notations also propagates to  comparisons of various  anomalies in the following sections.}
\be
\label{E:AnomGenFlux}
&& Tr(U(1)_{F_i}) = Tr(U(1)_{F_i}^3) = -2(N+2)(N_{a_i}-N_{a_r})\,,\nonumber\\
&& Tr(U(1)_{F_i}U(1)_{F_j}^2) = -2(N-1)z+2(N+1)N_{a_r} - 2N_{a_i} - 2N_{a_j}\,, \nonumber\\
&& Tr(U(1)_{F_i}U(1)_{R}^2) = 2N(N_{a_i}-N_{a_r})\,, \qquad Tr(U(1)_{F_i}^2 U(1)_{R}) = -4N(g-1)\,, \nonumber\\
&& Tr(U(1)_{F_i}U(1)_{F}^2) = -(2N+r-2)(N_{a_i}-N_{a_r})\,, \nonumber\\ 
&& Tr(U(1)_{F_i}^2 U(1)_{F}) = -2(2N+r-2)z +2(N-1)(N_{a_i}+N_{a_r})\,, \nonumber\\
&& Tr(U(1)_{F_i} SO(4N+12-2r)^2) = -\half(N_{a_i}-N_{a_r})\,,
\ee
where $U(1)_R$ refers to the $6d$ R-symmetry.

Next we can consider including surfaces with punctures. A way to define punctures is to reduce the six dimensional SCFT to five dimensions on a circle and study boundary conditions in five dimensions. The circle reductions of $D_{N+3}$ conformal matter were studied by various authors, see \cite{Ohmori:2015pua,Ohmori:2015tka,Hayashi:2015fsa,Zafrir:2015rga,Hayashi:2015zka}.
Choosing different holonomies when compactifying leads to different five dimensional effective gauge theory descriptions \cite{Hayashi:2015fsa,Hayashi:2015zka}. One may obtain gauge theory descriptions with gauge groups being  $SU(2)^N$, $USp(2N)$, or $SU(N+1)$. The relevant boundary conditions then freeze the gauge degrees of freedom on the boundary and produce theories with global symmetries $SU(2)^N$, $USp(2N)$, or $SU(N+1)$ associated to punctures. The two former descriptions will play a role in this paper. The punctures and their properties were discussed in detail in \cite{Kim:2018bpg}, here we will only need the 't Hooft anomalies of various symmetries in the presence of punctures.

In order to find the puncture contribution to the anomaly polynomial we can simply replace $g\to g+s_{tot}$ where $s_{tot}$ is the total number of maximal punctures. In addition we need to consider the contribution from the anomaly inflow, this was calculated in \cite{Kim:2018bpg} and \cite{Kim:2018lfo} for the $USp(2N)$ and $SU(2)^N$ maximal punctures, respectively. Some of the 't Hooft anomalies considering all the above are\footnote{The remaining anomalies depend on finer details of the definition of the puncture we did not yet specify. For example the punctures break some of the $6d$ symmetry to a subgroup and the choice of the imbedding of this subgroup affects the anomalies. This choice is often called the color of the puncture, \cite{Razamat:2016dpl}.}
\be
\label{E:4dAPwPunc}
Tr(U(1)_R) & = & -N(2N+9)(g-1+\frac{s_{tot}}{2})-\half N(2N+1)s_{USp(2N)}-\frac{3}{2}N s_{SU(2)^N},\nonumber\\
Tr(U(1)^3_R) & = & N(10N+3)(g-1+\frac{s_{tot}}{2})-\half N(2N+1)s_{USp(2N)}-\frac{3}{2}N s_{SU(2)^N}, \nonumber\\
Tr(U(1)_R U(1)^2_F) & = & -2Nr(g-1+\frac{s_{tot}}{2}),\ \  Tr(U(1)_F U(1)^2_R) = 2 N r z,\nonumber\\
Tr(U(1)_R SU(r)^2) & = & -N(g-1+\frac{s_{tot}}{2}) ,\ \ Tr(U(1)_R SO(4N+12-2r)^2) = -N(g-1+\frac{s_{tot}}{2}),\nonumber\\ 
Tr(U(1)_R SU(2)^2_s) & = & -1, \ \ Tr(U(1)_R USp(2N)^2_s) = -\frac{1}{2}(N+1). 
\ee
In the notation we used $s_G$ is the number of maximal punctures with symmetry $G$, and $G_s$ is a simple factor of symmetry which is part of the symmetry $G$ associated with a maximal puncture.

Finally, let us also mention that the details of the six dimensional theory also give rise to expectations for relevant and marginal operators of the four dimensional models. In particular the dimension of the conformal manifold for a theory corresponding to compactification with no punctures is expected to be,

\be
\biggl[dim \, G_{\{{\bf F}\}} (g-1)+\# Ab._{\{{\bf F}\}}\biggr]+(3g-3 )\,.
\ee Here  $G_{\{{\bf F}\}}$ is the subgroup of the global symmetry group commuting with the flux and $\# Ab._{\{{\bf F}\}}$ is the number of abelian factors in $G_{\{{\bf F}\}}$ \cite{Razamat:2016dpl}. 
The first term above arises from flat connections on the surface and the second from complex structure moduli. Alternatively, one can view the first term as coming from KK reduction of the conserved current in six dimensions and the second term as KK reduction of the stress-energy tensor.

For generic values of flux and genus one can expect \cite{babuip,talknazareth,Beem:2012yn} the index for closed Riemann surfaces to be equal in low orders of expansion in superconformal fugacities to,\footnote{For index definitions see Appendix \ref{A:indexdefinitions}.}
\be
\label{fluxindex}
{\cal I} = 1+\biggl(3g-3+\sum_{i=1}^{dim \, G} \biggl(\sum_{j=1}^{rank \, G} {\frak Q}^i_j \mathcal{F}_j + g-1\biggr) \prod_{\ell=1}^{rank\, G} a_\ell^{{\frak Q}^i_l }\biggr) pq +\cdots\,.
\ee 
where we compute the index using the six dimensional R-symmetry, $\mathcal{F}_j$ is the flux for the $U(1)_j$ and ${\frak Q}^i_j$ is 
defined through, 
\be
\chi_{adj} (\{ a_{\ell} \})=\sum_{i=1}^{dim \, G} \prod_{\ell=1}^{rank\, G} a_\ell^{{\frak Q}^i_\ell }\,.
\ee Here $\chi_{adj} (\{a_\ell\})$ is  the character of the adjoint representation of the flavor symmetry $G$ and
  $a_\ell$ being the fugacities for the Cartan subgroup. In particular for the case with no flux the index has the form,

\be\label{expindex}
{\cal I}= 1+ (\chi_{adj} (\{a_\ell\})\, (g-1)+3g-3) pq+\cdots\,.
\ee
For low values of genus and/or flux there might be additional terms but for generic compactifications this is the expected behavior. 

\section{Compactifications of  the $D_{4}$ SCFT: the E-string}\label{sec:estring}
In this section we propose a $4d$ Lagrangian for the three punctured sphere, or trinion, of the $D_{N+3}$ minimal conformal matter SCFT with $N=1$ also known as the rank $1$ E-string theory. We postpone the derivation of the results from RG flows to section \ref{sec:derivation}. Here we will discuss the properties of the trinion, the combination of trinions to arbitrary Riemann surfaces, duality properties, and comparison of the properties of the four dimensional models with six dimensional expectations discussed in the previous section. 

Let us note that much is known about compactifications of the E-string theory on Riemann surfaces with flux. In particular explicit constructions of Lagrangian $4d$ SCFTs corresponding to compactifications on two punctures spheres with flux are known  \cite{Kim:2017toz} and also Lagrangians for compactifications on closed Riemann surfaces  \cite{Razamat:2019vfd}. Moreover in \cite{Kim:2017toz} also a construction for a trinion with some value of flux was derived. However that trinion had only a rank five symmetry visible in the UV Lagrangian with the remaining three Cartan generators accidental in the IR. The advantage of the construction we will present here is that it is much simpler and all the rank of the global symmetry is visible in the UV with only the nonabelian structure enhancing in the IR. We will use the previous results on E-string compactifications to perform a variety of consistency checks to our results.

\subsection{The trinion}

\begin{figure}[t]
	\centering
  	\includegraphics[scale=0.25]{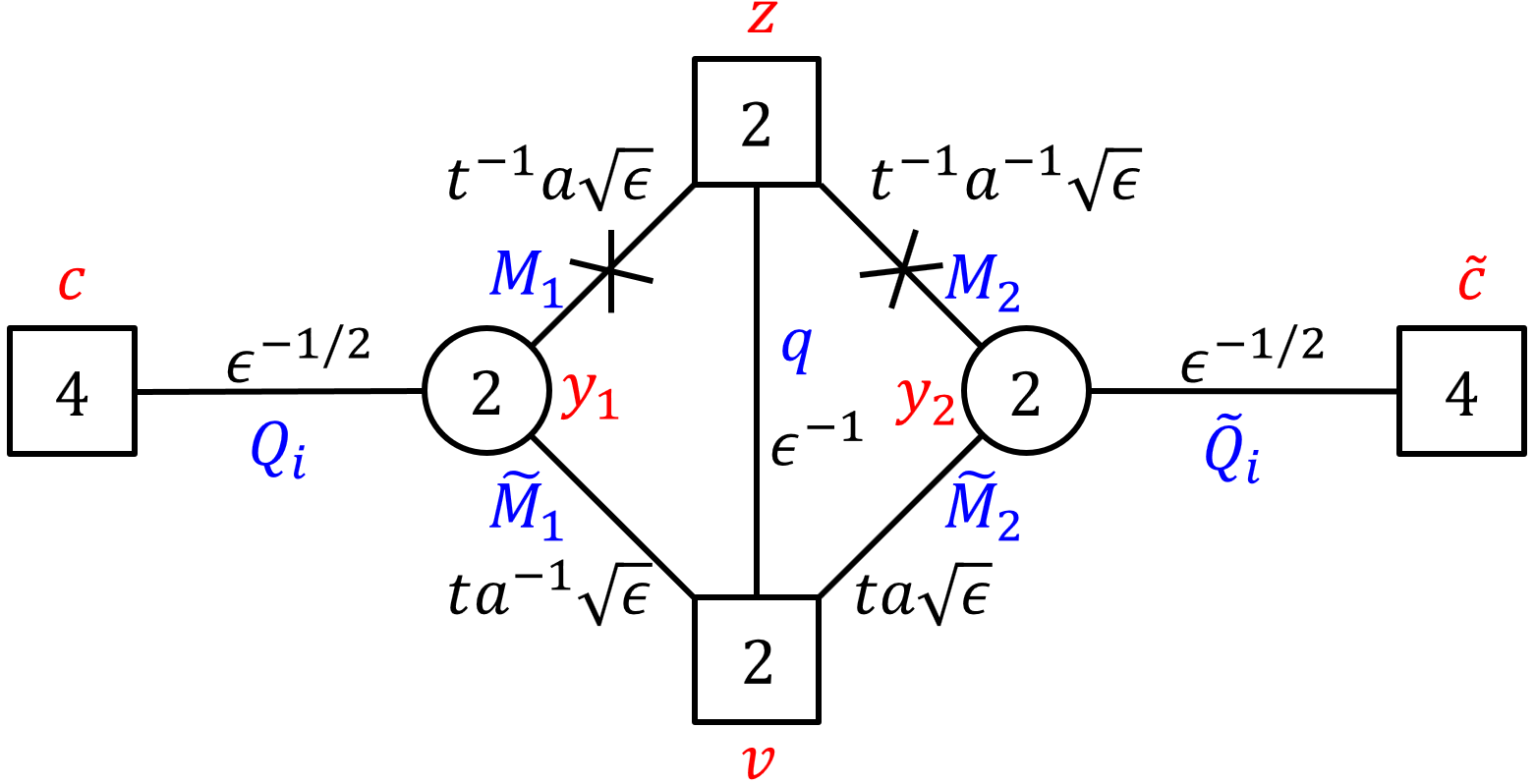}
    \caption{A quiver diagram of a trinion with three maximal punctures for rank one  E-string. The squares and circles denote $SU(n)$ global and gauge symmetries, respectively, where $n$ is the number inside the shapes. All fields have R-charge $1/2$ except the gauge singlet fields that have R-charge $1$. $\epsilon$ is the fugacity associated to the additional maximal puncture, while $t$, $a$ and the two $SU(4)$'s are related to the internal symmetries that arise from $6d$. In blue we write the field names, in red we write the symmetries associated fugacities, and in black we write the charges of each field. The $X$'s on bifundamental fields denote flip fields coupled through the superpotential to baryonic operators built from the bifundamentals. Six additional flip fields need to be included as well, three flipping half of the six operators in the symmetric representation of each of the $SU(4)$'s. These break each of the two $SU(4)$ symmetries to $SU(3)\times U(1)$. In addition, each triangle has a  superpotential term turned on for it.}
    \label{F:EStringTrinion}
\end{figure}

The E-string trinion quiver is shown in Figure  \ref{F:EStringTrinion}. Two of the maximal punctures appear explicitly as $SU(2)$ global symmetries while the third appears as a $U(1)_{\epsilon}$ symmetry that enhances to $SU(2)_{\epsilon}$ as we will soon discuss. The theory has  a superpotential which takes the following form,
\be
W & = & (M_1 \widetilde{M}_1 + M_2 \widetilde{M}_2)q + F_1 M_1^2 + F_2 M_2^2 + F_{12} Q_1 Q_2 + F_{13} Q_1 Q_3 + F_{23} Q_2 Q_3\nonumber\\
 & & + \widetilde{F}_{12} \widetilde{Q}_1 \widetilde{Q}_2 + \widetilde{F}_{13} \widetilde{Q}_1 \widetilde{Q}_3 + \widetilde{F}_{23} \widetilde{Q}_2 \widetilde{Q}_3.
\ee
We suppressed the $SU(2)$ indices for clarity as they contract in a trivial way.  The different fields are defined in Figure \ref{F:EStringTrinion}.
The $F$ fields are flip fields, which are gauge singlets, needed for the enhancement of $U(1)_{\epsilon}$ to $SU(2)_{\epsilon}$. 

The full information about the fields, gauge symmetries, superpotential and charges under global symmetries can be neatly encoded in the expression for the superconformal index which for the trinion is,
\be
\label{E:EstringTrinion}
\mathcal{I}_{v,z,\epsilon}^{trinion} & = & \kappa^{2}\oint\frac{dy_{1}}{4\pi iy_{1}}\oint\frac{dy_{2}}{4\pi iy_{2}}\frac{\prod_{i=1}^{4}\Gamma_{e}\left(\left(pq\right)^{\frac{1}{4}}\epsilon^{-\half}c_{i}y_{1}^{\pm1}\right)\Gamma_{e}\left(\left(pq\right)^{\frac{1}{4}}\epsilon^{-\half}y_{2}^{\pm1}\widetilde{c}_{i}\right)}{\Gamma_{e}\left(y_{1}^{\pm2}\right)\Gamma_{e}\left(y_{2}^{\pm2}\right)}\nonumber\\
 & & \prod_{i=1}^{3}\underline{\Gamma_{e}\left(\sqrt{pq}\epsilon c_{i}c_{4}\right)\Gamma_{e}\left(\sqrt{pq}\epsilon \widetilde{c}_{i}\widetilde{c}_{4}\right)}\,\underline{\Gamma_{e}\left(\sqrt{pq}t^{2}a^{\pm2}\epsilon^{-1}\right)}\nonumber\\
 & & \Gamma_{e}\left(\left(pq\right)^{1/4}t^{-1}a\epsilon^{1/2}y_{1}^{\pm1}z^{\pm1}\right)\Gamma_{e}\left(\left(pq\right)^{1/4}t^{-1}a^{-1}\epsilon^{1/2}z^{\pm1}y_{2}^{\pm1}\right)\Gamma_{e}\left(\sqrt{pq}\epsilon^{-1}z^{\pm1}v^{\pm1}\right)\nonumber\\
 & & \Gamma_{e}\left(\left(pq\right)^{1/4}ta^{-1}\epsilon^{1/2}y_{1}^{\pm1}v^{\pm1}\right)\Gamma_{e}\left(\left(pq\right)^{1/4}ta\epsilon^{1/2}v^{\pm1}y_{2}^{\pm1}\right)\,.
\ee
The contribution of the flip fields appearing in the above superpotential are underlined for clarity. 

The fact that the $U(1)_\epsilon$ symmetry here enhances to $SU(2)$ can be proven using IR dualities. The gauge nodes in the description of the trinion are $SU(2)$ with four flavors. This gauge theory enjoys an action of a large set of IR dualities. In particular there are $72$ duals which have the same gauge sector but different sets of gauge singlet fields and superpotentials \cite{Seiberg:1994pq,Intriligator:1995ne,Csaki:1997cu,Spiridonov:2008zr,Dimofte:2012pd,Razamat:2017hda}. Since the fundamental representation of $SU(2)$ is pseudoreal we have a choice which four of the eight fundamentals we call quarks and which antiquarks. Therefore the model has $\frac12 {8 \choose 4}= 35$ Seiberg duals \cite{Seiberg:1994pq} with the gauge singlets corresponding to the mesons. 
Using these dualities on both gauge nodes it can be shown that, for instance, the $SU(2)_v$ symmetry can be exchanged with the $U(1)_\epsilon$ symmetry. In particular we can have three IR dual frames in which any two of the three $SU(2)_v$, $SU(2)_z$, and $SU(2)_\epsilon$ are explicit in the Lagrangian while the third appears in the IR. We will discuss the sequence of dualities in Appendix \ref{sec:duality}.

Next we wish to combine the trinions to construct theories corresponding to higher genus Riemann surfaces. First, we note that every puncture has an octet of operators in the fundamental representation of the puncture symmetry. For $SU(2)_v$ these are $\widetilde M_1 Q_i$ and  $\widetilde M_2 \widetilde Q_i$, for $SU(2)_z$ these are $M_1Q_j$ and $M_2\widetilde Q_i$. For $SU(2)_\epsilon$ these are all the  flip fields, the baryonic operators $\widetilde M_i^2$, and the operators $Q_4Q_i$, $\widetilde Q_4\widetilde Q_i$. We refer to the collection of these operators as ``moment maps'' by abuse of terminology and denote them as $\widehat M_i^{(X)}$ with $X$ standing for the type of puncture.
 Note that the charges of the operators for the three punctures are different,
\be
\widehat M^{(v)} &:&\qquad \{ t a^{-1} c_1\,,\; t a^{-1} c_2\,,\; t a^{-1} c_3\,,\; t a^{-1} c_4\,,\; t a \widetilde c_1\,,\; t a \widetilde c_2\,,\; t a \widetilde c_3\,,\; t a \widetilde c_4\}\,,\nonumber\\
\widehat M^{(z)} &:&\qquad \{ t^{-1} a c_1\,,\; t^{-1} a c_2\,,\; t^{-1} a c_3\,,\; t^{-1} a c_4\,,\; t^{-1} a^{-1} \widetilde c_1\,,\; t^{-1} a^{-1} \widetilde c_2\,,\; t^{-1} a^{-1} \widetilde c_3\,,\; t^{-1} a^{-1} \widetilde c_4\}\,,\nonumber\\
\widehat M^{(\epsilon)} &:&\qquad \{ c_1^{-1} c_2^{-1}\,,\; c_1^{-1} c_3^{-1}\,,\; c_2^{-1}c_3^{-1}\,,\; \widetilde c_1^{-1} \widetilde c_2^{-1}\,,\; \widetilde c_2^{-1} \widetilde c_3^{-1}\,,\;  \widetilde c_3^{-1} \widetilde c_1^{-1}\,,\;t^2a^{-2}\,,\; t^2 a^2\}\,.
\ee 
Here the charges are encoded in the powers of the fugacities for various symmetries.

The fact that the operators are charged differently under the eight abelian symmetries means that the punctures are different. We refer to the punctures as being of a different type due to this difference.\footnote{Usually the various types are divided to different properties such as {\it colors} and {\it signs} \cite{Kim:2017toz}, but we will avoid this terminology and its meaning here.} Note also that the punctures break the $E_8$ symmetry of the six dimensional theory to $SU(8)\times U(1)$ \cite{Kim:2017toz}.
We glue two trinions together by identifying two punctures of the same type, one on each trinion, and gauging the diagonal $SU(2)$ symmetry of the two puncture symmetries. In addition we introduce an octet of fundamental fields  of $SU(2)$, $\Phi_i$, and turn on the superpotential,
\be\label{supglue}
W=\sum_{i=1}^8\Phi_i\;\biggl({\widehat M}^{(X)}_i-{{\widehat N}^{(X)}}_i\biggr)\,,
\ee where $\widehat M^{(X)}$ and $\widehat N^{(X)}$ are the two moment maps of the two punctures. We stress that we only glue two punctures of same type together. We will refer to this gluing procedure as $\Phi$-gluing. Note that the gluing is  a relevant operation. Each $SU(2)$ puncture symmetry has eight fundamentals with R-charge $\frac12$ and $2+2+8$ fundamentals with R-charge $1$ when gauged using the $\Phi$ gluing. Computing the superconformal R-symmetry of the trinion at the fixed point one finds that only the symmetries $U(1)_t$, $U(1)_{c_4}$ and $U(1)_{\widetilde c_4}$ mix with the R symmetry, see Appendix  \ref{sec:duality}. We then first introduce the fields $\Phi$ and turn on the superpotential \eqref{supglue}. This is a relevant deformation which locks the symmetries of the $\Phi$ fields with the ones of the trinion. Finally we gauge the puncture $SU(2)$ symmetry. Let us denote the mixing coefficient of the $U(1)_t$ with the R-symmetry as $\alpha_t$ and the mixing coefficient of $U(1)_{c_4}$, which is the same as the one for $U(1)_{\widetilde c_4}$, as $\alpha_c$.
The $\beta$ function is proportional to $Tr U(1)_R SU(2)^2$ which is given for $SU(2)_v$ by,
\be
&& \underbrace{2}_{\lambda_a}+\underbrace{\frac12(\frac12+\alpha_t-1)\times 8}_{2\times (\widetilde M_1\text{\,and\,} \widetilde M_2)}+\underbrace{\frac12(1-1)\times 4}_{2\times q}\nonumber\\
&& +\underbrace{\frac12(1-\alpha_t-\alpha_c-1)\times 2}_{4th\text{\,and\,}8th\text{\, component of\,} \Phi}+\underbrace{\frac12(1-\alpha_t+\frac13\alpha_c-1)\times 6}_{\text{other\, components of\,} \Phi} =0\,,
\ee  
where $\lambda_\alpha$ are the gaugino. Thus the gauging is conformal at the fixed point after turning on the superpotential.
 One can verify that no operators fall below the unitarity bound while gluing theories together.

As a further illustration of the gluing procedure, the index of the four punctured sphere using $\Phi$-gluing of two trinions along a $z$ type of puncture is given by,
\be
\label{E:EstringFourPz}
\mathcal{I}_{v,u,\epsilon,\delta} & = & \kappa\oint\frac{dz}{4\pi i z}
\mathcal{I}_{v,z,\epsilon}^{trinion}\mathcal{I}_{u,z,\delta}^{trinion}
\frac{\prod_{i=1}^{4}\Gamma_{e}\left(\sqrt{pq}t a^{-1} c_{i}^{-1}z^{\pm1}\right)\Gamma_{e}\left(\sqrt{pq}t a \widetilde c_{i}^{-1}z^{\pm1}\right)}{\Gamma_{e}\left(z^{\pm2}\right)}\,,
\ee 
and along an $\epsilon$ type puncture we have,
\be
\label{E:EstringFourPeps}
\mathcal{I}_{v,z,u,w} & = & \kappa\oint\frac{d\epsilon}{4\pi i \epsilon}
\mathcal{I}_{v,z,\epsilon}^{trinion}\mathcal{I}_{u,w,\epsilon}^{trinion}\times \nonumber\\
 & & \frac{\Gamma_{e}\left(\sqrt{pq}t^{-2} a^{\pm2} \epsilon^{\pm1}\right)\prod_{i=1}^{3}\Gamma_{e}\left(\sqrt{pq}(c_{i}c_4)^{-1}\epsilon^{\pm1}\right)\Gamma_{e}\left(\sqrt{pq}(\widetilde c_{i} \widetilde c_4)^{-1}\epsilon^{\pm1}\right)}{\Gamma_{e}\left(\epsilon^{\pm2}\right)}\,.
\ee

Another type of gluing one can discuss is the so called $S$-gluing \cite{Benini:2009mz,Gaiotto:2015usa,Hanany:2015pfa,Razamat:2016dpl}. This gluing involves two punctures of different types,\footnote{Specifically different sign and same color.} and we gauge the puncture $SU(2)$ symmetry and add the following superpotential without introducing additional matter fields,
\be
\label{supglues}
W=\sum_{i=1}^8{\widehat M}^{(X)}_i{{\widehat N}^{(X)}}_i\,,
\ee 
where $\widehat M^{(X)}$ and $\widehat N^{(X)}$ are the two moment maps for the two punctures.\footnote{There are various interesting dynamical questions one can ask about the gluing. For example, using \eqref{mixtrinion} one can see the the superpotential \eqref{supglues} is relevant. After turning it on and flowing to the IR the gauging becomes again marginal.}

\begin{figure}[t]
	\centering
  	\includegraphics[scale=0.27]{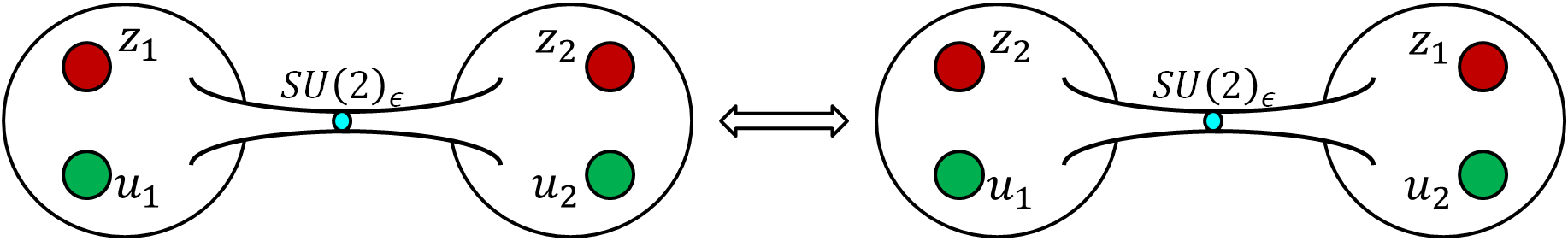}
    \caption{Different duality frames for a four punctured sphere. The fact that the left and right frames are the same implies for example that the index has to be invariant under the exchange of the two $SU(2)$ fugacities. }
    \label{F:EStringDuality}
\end{figure}

A non trivial check of the validity of the conjectured  trinion theory is that the models with more than three punctures should satisfy duality properties. In particular for example the supersymmetric index should be invariant under exchanging any two punctures of the same type as shown in Figure  \ref{F:EStringDuality}. This fact follows from Seiberg and ${\cal N}=2$ S-duality and we prove this invariance under the exchange of two $\epsilon$ type punctures in Appendix \ref{A:MinPuncDuality}.

\subsection{Anomalies}

We can check that the trinion has the predicted anomalies we have derived from six dimensions in the previous section. For example, the following 't Hooft anomalies
\be
&&Tr(U(1)_R) = -10\,,\qquad Tr(U(1)_R^3)=2\,, \nonumber\\
&&Tr(U(1)_R\ SU(2)_{\epsilon}^2) = Tr(U(1)_R\ SU(2)_{z}^2) = Tr(U(1)_R\ SU(2)_{u}^2) = -1\,,
\ee agree perfectly with the ones in \eqref{E:4dAPwPunc} setting $N=1$,  $g=0$ and $s_{tot}=3$, $s_{USp(2N)} +s_{SU(2)^N}=3$.\footnote{Note that for $N=1$ $USp(2N)$ and $SU(2)^N$ punctures are the same.} 

Next we can deduce the flux associated with the suggested trinion. We find this by $\Phi$-gluing two trinions together  to form a genus two Riemann surface. The resulting 't Hooft anomalies are
\be
Tr(U(1)_R) = -11\,,\qquad Tr(U(1)_R^3)=13\,,\qquad Tr(U(1)_R U(1)_{F_i}^2)=-4\,,
\ee
while the rest of the anomalies vanish. $U(1)_{F_i}$ represents all the internal symmetries $-U(1)_t /2$, $U(1)_a /2$ and the Cartans of both $SU(4)$'s. The expected anomalies from six dimensions can be found  by setting $N=1$ in \eqref{E:4dAP}. The anomalies for a rank 1 E-string compactified on a genus $g$ Riemann surface are given by,
\be
Tr(U(1)_R)=-11(g-1) \,,\ \  Tr(U(1)_R^3)=13(g-1) \,,\ \  Tr(U(1)_R U(1)_{F_i}^2)=-4(g-1)\,,\nonumber\\
\ee
where the rest of the anomalies are proportional to the flux. We can see that all anomalies match for genus two if we associate a vanishing flux for all the symmetries of the above trinion.

\begin{figure}[t]
	\centering
  	\includegraphics[scale=0.27]{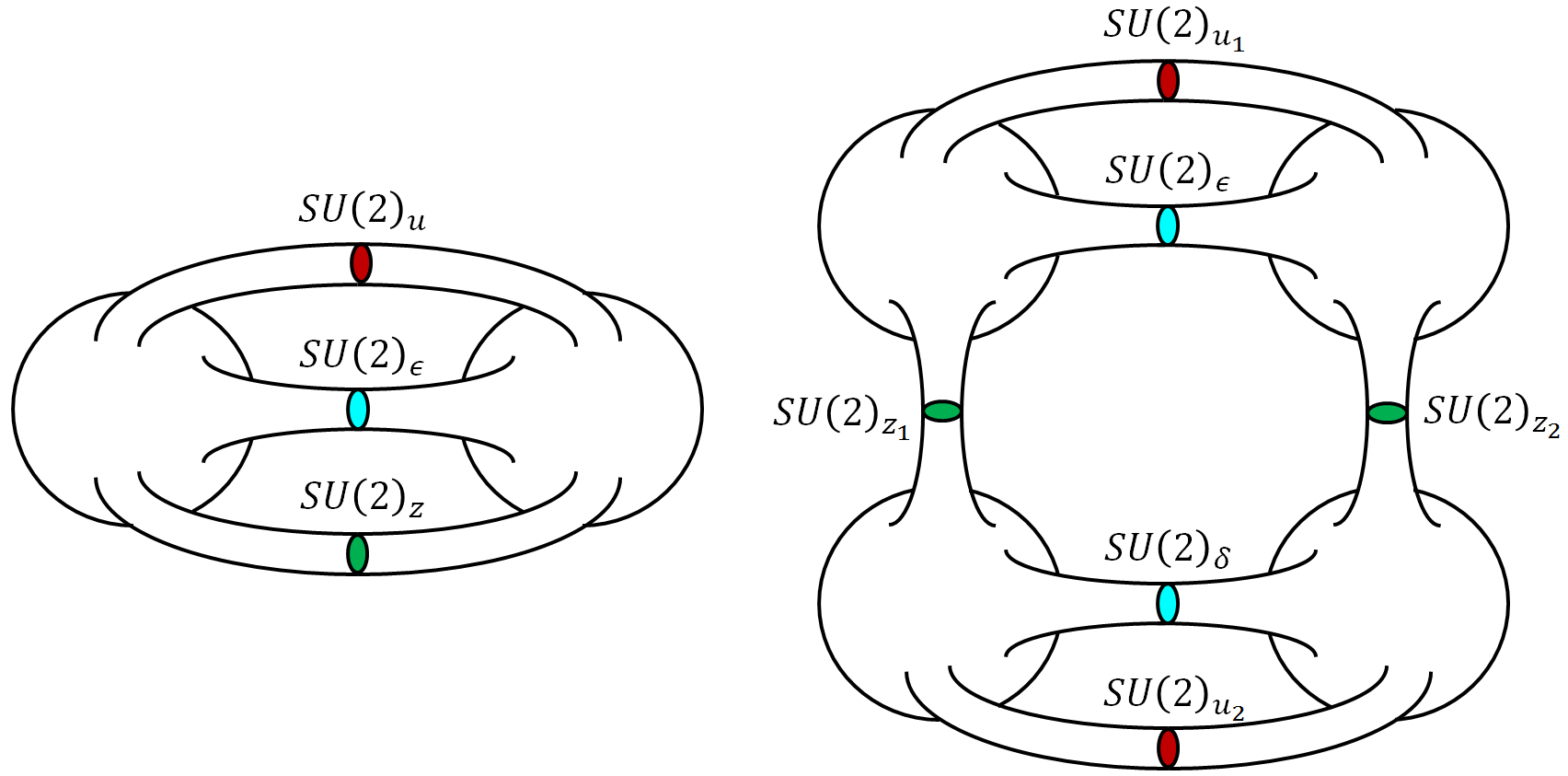}
    \caption{An illustration of a genus $g$ surface in a pair of pants decomposition. On the left is a genus two surface and on the right a genus three surface. The surfaces are constructed by gluing together the trinion with $\Phi$-gluing along matching types of punctures.}
    \label{F:EStringGenusGTrinion}
\end{figure}

\subsection{Higher genus surfaces}

Since we found out that the trinion has zero flux, we expect that gluing it to generate a closed Riemann surface will produce theories with $E_8$ symmetry.
We can test this by computing the supeconformal index.  We expect to be able to express  the index in terms of irreducible representations of $E_8$. In the case of a genus two surface (see Figure  \ref{F:EStringGenusGTrinion} left) built as before, we find the index takes the form
\be
\mathcal{I} = 1 + (\textbf{248}+4)pq+\cdots\,.
\ee
Meaning we get that the index has the expected $E_8$ symmetry. The Adjoint representation of $E_8$ is the \textbf{248} appearing in the index, and it follows the branching rules of the following decompositions
 
\noindent $E_8 \to U(1)_t\times E_7$:
\be
\textbf{248} & \to &  (t^2+t^{-2})\textbf{56} + \textbf{133} + (1+t^4+t^{-4})\,.
\ee
$E_7 \to U(1)_a\times SO(12)$:
\be
\textbf{133} & \to &  (a^2+a^{-2})\textbf{32} + \textbf{66} + (1+a^4+a^{-4})\,,\nonumber\\
\textbf{56} & \to & (a^2+a^{-2})\textbf{12} + \textbf{32}'\,.
\ee
$SO(12) \to SU(4)_c\times SU(4)_{\widetilde c}$:
\be
\label{E:SO12toSU4}
\textbf{66} & \to & (\textbf{15},\textbf{1}) + (\textbf{6},\textbf{6}) + (\textbf{1},\textbf{15})\,,\nonumber\\
\textbf{32} & \to & (\textbf{4},\textbf{4}) + (\overline{\textbf{4}},\overline{\textbf{4}})\,,\nonumber\\
\textbf{32}' & \to & (\textbf{4},\overline{\textbf{4}}) + (\overline{\textbf{4}},\textbf{4})\,,\nonumber\\
\textbf{12} & \to & (\textbf{6},\textbf{1}) + (\textbf{1},\textbf{6})\,.
\ee

\begin{figure}[t]
	\centering
  	\includegraphics[scale=0.2]{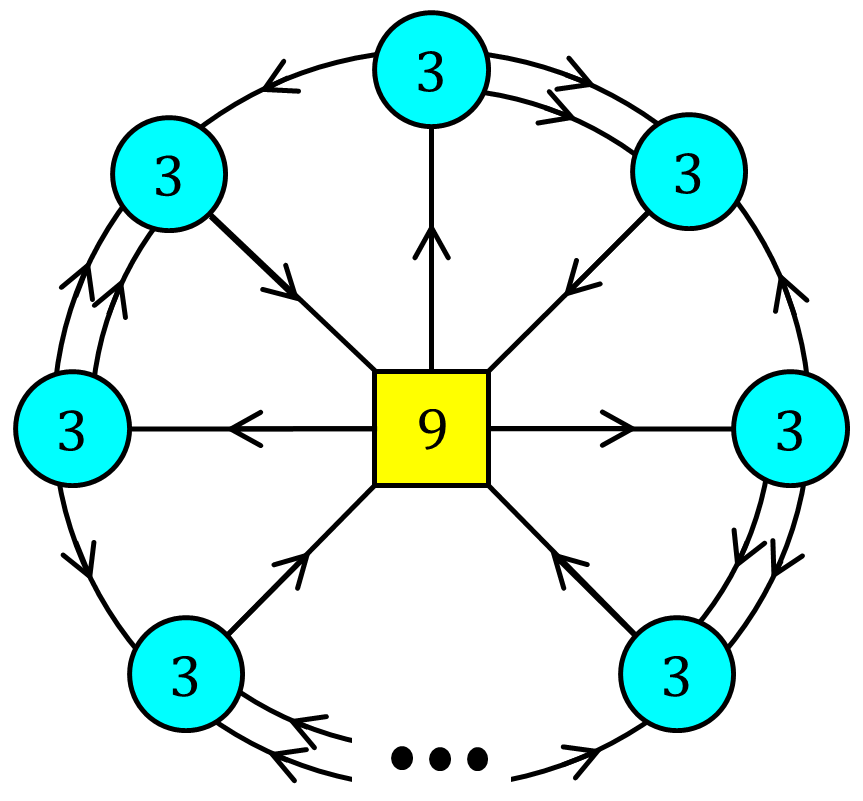}
    \caption{A conformal Lagrangian for E-string compactified on a closed genus $g$ surface obtained in \cite{Razamat:2019vfd}. The quiver has $2g-2$ $SU(3)$ gauge nodes. The $SU(9)$ flavor symmetry group is a maximal subgroup of $E_8$. On a submanifold of the conformal manifold the $SU(9)$ is broken to the Cartan and the conjecture of \cite{Razamat:2019vfd} is that at some locus this becomes the Cartan of the $E_8$ symmetry.}
    \label{F:EstringGenusG}
\end{figure}

We can use the $\Phi$-gluing starting with $2g-2$  trinions to form a genus $g\geq 2$ surface (see Figure  \ref{F:EStringGenusGTrinion}) we find that the index for $g>2$ takes the form,
\be
\mathcal{I} = 1 + (\textbf{248}(g-1)+3(g-1))pq+...
\ee
with the same decomposition and branching rules as before. This matches perfectly the expectation of \eqref{expindex}. For genus two we have an additional marginal deformation the geometric origin of which is not known to us. 
These computations give strong evidence that indeed the above quiver represents the trinion of the rank 1 E-string compactification to $4d$. The general genus results here agree with what was found in \cite{Kim:2017toz, Razamat:2019vfd}. We can check the index unrefined with flavor fugacities to high orders against other conjectured descriptions of genus $g$ compactifications of E-string. For example, both the confromal Lagrangian of \cite{Razamat:2019vfd}, see Figure \ref{F:EstringGenusG}, and the construction using the above trinion give up to order $(pq)^3$ for genus $g=2$ the following index,
\be
&&1+252 pq+251 \left(p+q\right) pq+252 (p^2+q^2)pq+19007 (pq)^2+252(p^3+q^3)pq+\\
&&50130 (p+q)(pq)^2+252 (p^4+q^4)pq+81756 (p^2+q^2)(pq)^2+590764(pq)^3+\cdots\,.\nonumber
\ee  The agreement between the two computations of the index is a rather non trivial consistency check of the conjectures in \cite{Razamat:2019vfd} and in this paper.

\subsection{Higher genus surfaces with flux}

Finally we can construct theories corresponding to compactifications with arbitrary flux for abelian subgroups of the $E_8$ symmetry. To do so we can glue to the aformentioned trinions the theories corresponding to two punctured spheres with flux derived in \cite{Kim:2017toz}. This provides further consistency checks of our procedure. For example the symmetry of the theory corresponding to closed Riemann surface with flux should be the commutant of the flux in $E_8$ and the spectrum of low lying protected operators is expected to be determined for generic values of flux by \eqref{fluxindex}.
\begin{figure}[htbp]
	\centering
  	\includegraphics[scale=0.21]{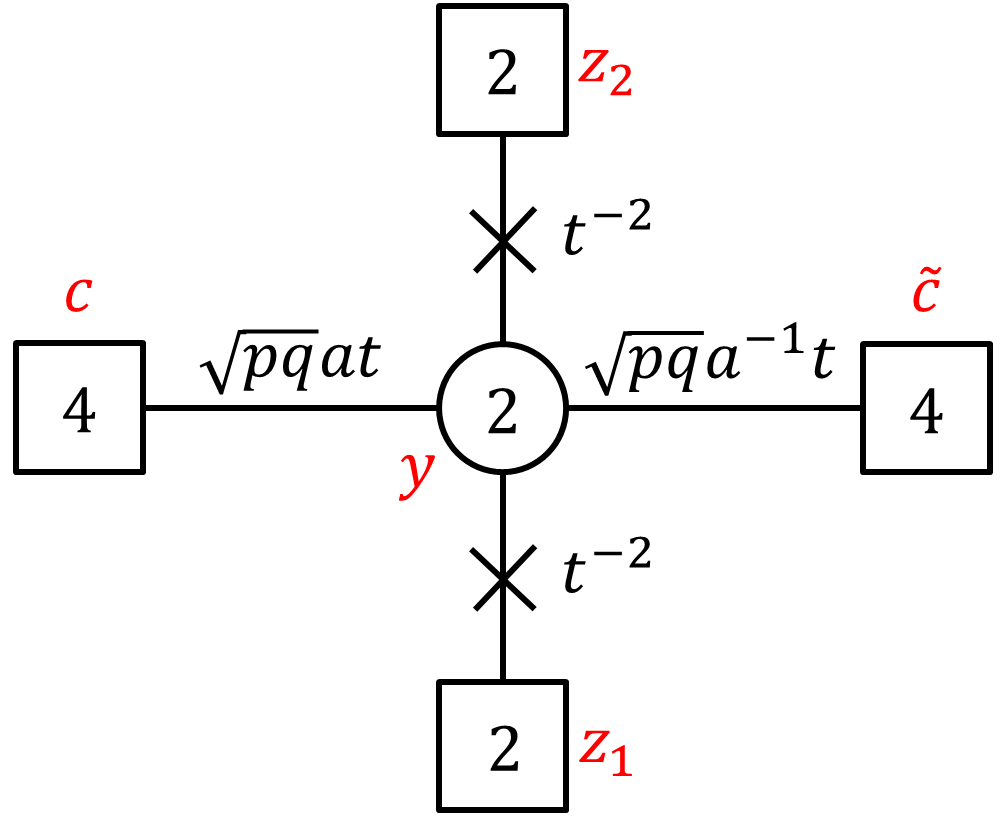}
    \caption{A quiver of a t-flux tube with flux $-1$. This tube can be glued using $\Phi$-gluing to the trinion upper $SU(2)_z$ maximal puncture.}
    \label{F:EStringTFluxTube}
\end{figure}

\begin{figure}[t]
	\centering
  	\includegraphics[scale=0.21]{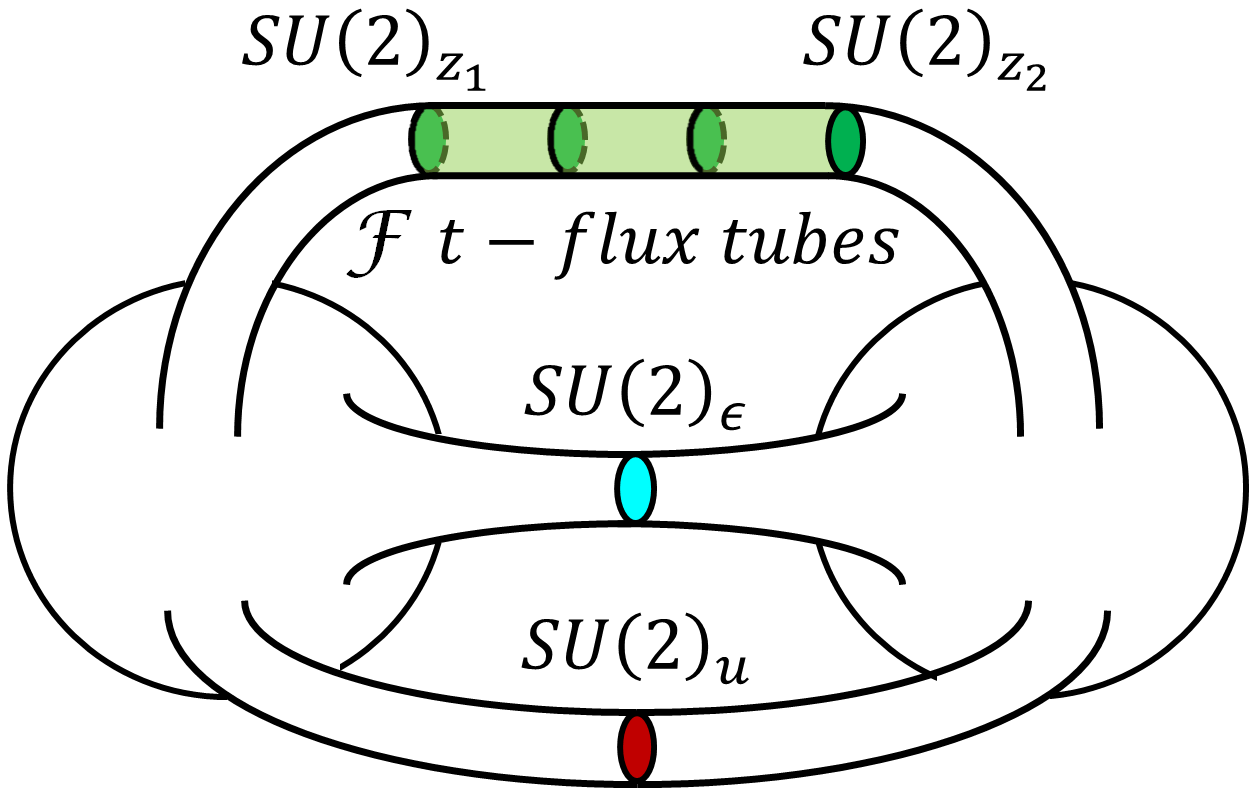}
    \caption{An illustration of a genus $g=2$ surface  with additional $t$-flux. The surfaces is constructed by gluing together two trinions and ${\cal F}$ t-flux tubes.}
    \label{F:EStringGenusGFlux}
\end{figure}

As an example a two punctured sphere theory, flux tube,  with $-1$ unit of flux in a $U(1)$ breaking $E_8$ to $E_7\times U(1)$ is depicted in Figure \ref{F:EStringTFluxTube}. The flux we turn on, by the conventions of \cite{Kim:2017toz} which we follow, is  for $-U(1)_t/2$. We can admix ${\cal F}$ such flux tubes to the construction of a genus $g$ surface to obtain a surface with flux $\mathcal{F}$, see Figure \ref{F:EStringGenusGFlux}.

We find the index takes the form
\be
\mathcal{I} & = & 1 + \left(t^{-4}(g-1+2\mathcal{F}) + t^{-2}(g-1+\mathcal{F})\textbf{56}\right)pq + (g-1)(\textbf{133}+3+1)pq \nonumber\\
 & & + \left(t^4(g-1-2\mathcal{F}) + t^2(g-1-\mathcal{F})\textbf{56}\right)pq+...
\ee
with the expected $E_7$ representations and the structure of \eqref{fluxindex}.
The first and last terms are relevant and (some of the) irrelevant deformations, respectively, for positive t-flux and the other way around for negative t-flux, while the middle term is marginal. This result also perfectly matches the one found in \cite{Kim:2017toz}, and gives further evidence to support the form of the trinion described above.

\section{Compactifications of $D_{N+3}$ with $N>1$}\label{sec:dnmatter}

Let us now generalize the discussion of the previous section to $N>1$. Here we will first define a trinion with two $SU(2)^N$ punctures and one $SU(2)$ puncture. This is the trinion we will derive using flows in the next section. 
We will argue that the $SU(2)$ puncture is a minimal one in the sense that it can be closed by giving a vacuum expectation value to a single operator charged under it. 
We will also show that the $SU(2)$ puncture can be obtained by partially closing a maximal $USp(2N)$ puncture. Finally we will argue that a theory corresponding to two maximal $SU(2)^N$ punctures and $N$ minimal $SU(2)$ punctures has a locus on its conformal manifold where  the minimal punctures collide and combine into a maximal $USp(2N)$ puncture. This will be the trinion from which arbitrary Riemann surfaces can be constructed. Several checks of this proposal will be performed. In particular we will show that the global symmetry of theories constructed from the new trinions and corresponding to compactifications on closed Riemann surfaces enhances to $SO(4N+12)$ as expected.

\subsection{The trinion with one minimal puncture}
\begin{figure}[t]
	\centering
  	\includegraphics[scale=0.31]{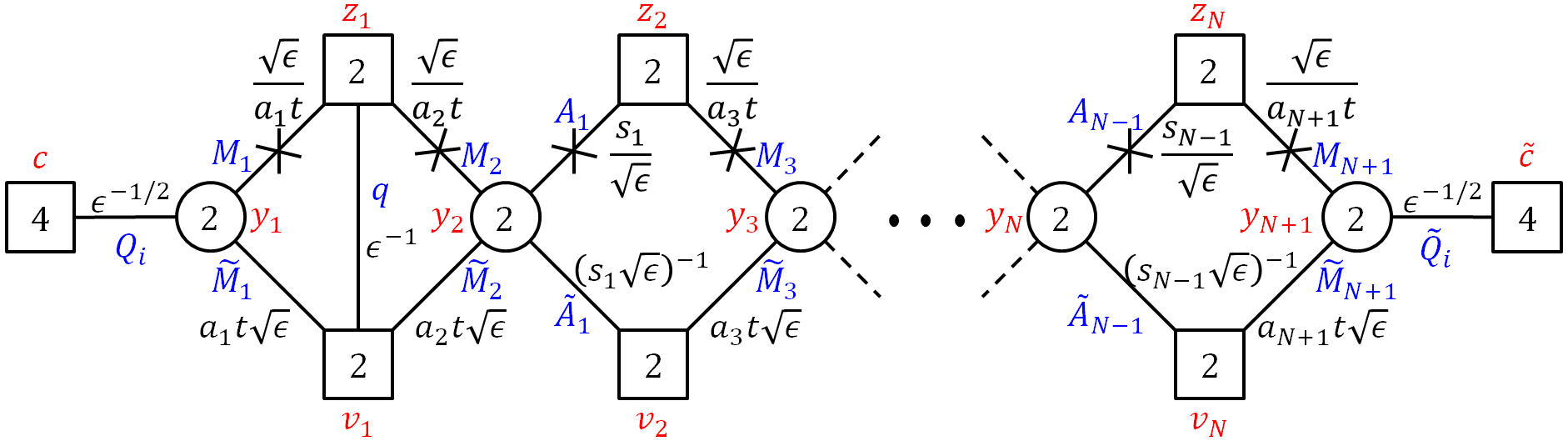}
    \caption{A quiver diagram for a compactification of minimal class $D_{N+3}$ conformal matter on a sphere with two maximal and one minimal punctures. All fields have R-charge $1/2$ except the gauge singlet fields  that have R-charge $1$. $\epsilon$ is the fugacity associated to the additional $SU(2)$ puncture, while $a_i$, $s_j$ and the two $SU(4)$'s are related to the internal symmetries that arise from $6d$. In blue we write the field names, in red we write the fugacities associated to puncture and gauge symmetries, and in black we write the fugacities of each field. The $X$'s on bifundamental fields denote flip fields coupled through the superpotential to  baryonic operators built from the bifundamental fields. Six additional flip fields need to be included as well, three flipping half of the six operators in the symmetric representation of each of the $SU(4)$'s. These break each of the two $SU(4)$ symmetries to $SU(3)\times U(1)$. In addition, each closed loop of fields has a  superpotential term turned on for it.}
    \label{F:MinD6Trinion}
\end{figure}

The $D_{N+3}$ trinion quiver is shown in Figure  \ref{F:MinD6Trinion}. As in the E-string case the two maximal punctures appear explicitly as $SU(2)^N$ global symmetries while the minimal puncture appears as a $U(1)_\epsilon$ symmetry that we expect to be enhanced to $SU(2)_\epsilon$ in the IR. The theory has a superpotential taking the form,
\be
W & = & (M_1 \widetilde{M}_1 + M_2 \widetilde{M}_2)q + \sum_{i=1}^{N-1} M_{i+2} \widetilde{M}_{i+2} A_i \widetilde{A}_i + \sum_{i=1}^{N+1} F_i M_i^2 + \sum_{j=1}^{N-1} \widetilde{F}_j A_j^2\nonumber\\
 & & + F_{12} Q_1 Q_2 + F_{13} Q_1 Q_3 + F_{23} Q_2 Q_3 + \widetilde{F}_{12} \widetilde{Q}_1 \widetilde{Q}_2 + \widetilde{F}_{13} \widetilde{Q}_1 \widetilde{Q}_3 + \widetilde{F}_{23} \widetilde{Q}_2 \widetilde{Q}_3.
\ee
As before we suppress the $SU(2)$ indices, and the $F$ fields are gauge singlet flip fields required for the enhancement of $U(1)_\epsilon$ to $SU(2)_\epsilon$.

The superconformal index for the trinion is
\be
\mathcal{I}_{\boldsymbol{z},\boldsymbol{v},\epsilon}^{trinion} & = & \kappa^{N+1}\prod_{i=1}^{N+1}\oint\frac{dy_{i}}{4\pi iy_{i}}\frac{\prod_{n=1}^{4}\Gamma_{e}\left(\left(pq\right)^{1/4}\epsilon^{-1/2}c_{n}y_{1}^{\pm1}\right)\Gamma_{e}\left(\left(pq\right)^{1/4}\epsilon^{-1/2}y_{N+1}^{\pm1}\widetilde{c}_{n}\right)}{\prod_{i=1}^{N+1}\Gamma_{e}\left(y_{i}^{\pm2}\right)}\nonumber\\
 & & \prod_{n=1}^{3}\underline{\Gamma_{e}\left(\sqrt{pq}\epsilon c_{n}c_{4}\right)\Gamma_{e}\left(\sqrt{pq}\epsilon\widetilde{c}_{n}\widetilde{c}_{4}\right)}\prod_{j=1}^{N-1}\underline{\Gamma_{e}\left(\sqrt{pq}\epsilon s_{j}^{-2}\right)}\prod_{i=1}^{N+1}\underline{\Gamma_{e}\left(\sqrt{pq}\epsilon^{-1}t^{2}a_{i}^{2}\right)}\nonumber\\
 & & \Gamma_{e}\left(\left(pq\right)^{1/4}t^{-1}a_{1}^{-1}\epsilon^{1/2}y_{1}^{\pm1}z_{1}^{\pm1}\right)\Gamma_{e}\left(\left(pq\right)^{1/4}\epsilon^{1/2}ta_{1}y_{1}^{\pm1}v_{1}^{\pm1}\right)\Gamma_{e}\left(\sqrt{pq}\epsilon^{-1}z_{1}^{\pm1}v_{1}^{\pm1}\right)\nonumber\\
 & & \prod_{i=1}^{N}\Gamma_{e}\left(\left(pq\right)^{1/4}\epsilon^{1/2}t^{-1}a_{i+1}^{-1}z_{i}^{\pm1}y_{i+1}^{\pm1}\right)\Gamma_{e}\left(\left(pq\right)^{1/4}\epsilon^{1/2}ta_{i+1}v_{i}^{\pm1}y_{i+1}^{\pm1}\right)\nonumber\\
 & & \prod_{j=1}^{N-1}\Gamma_{e}\left(\left(pq\right)^{1/4}\epsilon^{-1/2}s_{j}z_{j+1}^{\pm1}y_{j+1}^{\pm1}\right)\Gamma_{e}\left(\left(pq\right)^{1/4}\epsilon^{-1/2}s_{j}^{-1}v_{j+1}^{\pm1}y_{j+1}^{\pm1}\right)\,.
\ee
We  underline the flip fields appearing in the superpotential. The fugacities satisfy, $$\prod_{i=1}^4c_i=\prod_{i=1}^4\widetilde c_i=\prod_{i=1}^{N+1} a_i =1\,.$$
In a similar manner to the E-string case, we expect that one should be able to prove using IR dualities that the $U(1)_\epsilon$ symmetry enhances to $SU(2)$. As the details become cumbersome we refrain from doing it here. 

Now we wish  to glue the trinions to one another. Since the specified theory is not a trinion with three maximal punctures, we will not be able to generate higher genus Riemann surfaces yet. Nevertheless, we will note the operators in the fundamental representation of the punctures symmetry. For $SU(2)_\textbf{v}^N$ the operators are $\widetilde{M}_1 Q_n$ and $\widetilde{M}_{N+1} \widetilde{Q}_n$ in the fundamental of $SU(2)_{v_1}$ and $SU(2)_{v_N}$, respectively, and $\widetilde{M}_{j+1} \widetilde{A}_{j}$ in the bifundamental of $SU(2)_{v_j}\times SU(2)_{v_{j+1}}$. For $SU(2)_\textbf{z}^N$ these are $M_1 Q_n$ and $M_{N+1} \widetilde{Q}_n$ in the fundamental of $SU(2)_{z_1}$ and $SU(2)_{z_N}$, respectively, and $M_{j+1} A_{j}$ in the bifundamental of $SU(2)_{z_j}\times SU(2)_{z_{j+1}}$. For $SU(2)_\epsilon$ the operators are all the flip fields, the baryonic operators $\widetilde{M}_i^2,\,\widetilde{A}_j^2$, and the operators $Q_4 Q_n,\,\widetilde{Q}_4 \widetilde{Q}_n$. The ``moment maps'' for the punctures are thus
\be
\label{E:MinDNMomentMaps}
\widehat M^{(\textbf{v})} &:&\quad \{\widehat M^{(v_1)}: \{ t a_1 c_n\}_{n=1}^4,\; \{ \widehat M^{(v_j,v_{j+1})}: \{ t a_{j+1} s_j^{-1}\} \}_{j=1}^{N-1},\; \widehat M^{(v_N)}:\{ t a_{N+1} \widetilde c_n\}_{n=1}^4\}\,,\nonumber\\
\widehat M^{(\textbf{z})} &:&\quad \{\widehat M^{(z_1)}: \{ t^{-1} a_1^{-1} c_n\}_{n=1}^4,\; \{ \widehat M^{(z_j,z_{j+1})}: \{ t^{-1} a_{j+1}^{-1} s_j\} \}_{j=1}^{N-1},\; \widehat M^{(z_N)}:\{ t^{-1} a_{N+1}^{-1} \widetilde c_n\}_{n=1}^4\}\,,\nonumber\\
\widehat M^{(\epsilon)} &:&\quad \{ c_1^{-1} c_2^{-1}\,,\; c_1^{-1} c_3^{-1}\,,\; c_2^{-1}c_3^{-1}\,,\; \widetilde c_1^{-1} \widetilde c_2^{-1}\,,\; \widetilde c_2^{-1} \widetilde c_3^{-1}\,,\;  \widetilde c_3^{-1} \widetilde c_1^{-1}\,,\;\{t^2 a_i^2\}_{i=1}^{N+1}\,,\; \{s_j^{-2}\}_{j=1}^{N-1}\}\,.
\ee
As before, the two maximal punctures differ by charges of the moment map operators, thus are of a different type. In addition, the maximal punctures break the $SO(4N+12)$ symmetry of the $6d$ theory to $SU(4)_c \times U(1)_t U(1)_a^N \times U(1)_s^{N-1} \times SU(4)_{\widetilde c}$. Gluing two trinions using $\Phi$-gluing is done by identifying two maximal punctures of the same type in both trinions, and gauging the diagonal $SU(2)^N$ symmetry of the two punctures. We also add four fundamental fields for each of the $SU(2)$ nodes at the edges of the quiver, and additional $(N-1)$ bifundamental fields, one for each pair of neighboring $SU(2)$ nodes in the quiver. In total we add $N+7$ fields $\Phi_i$, coupled in the superpotential similarly to the E-string case,
\be
\label{E:DPhiSP}
W=\sum_{i=1}^{N+7} \Phi_i \left(\widehat M_i^{(X)} - \widehat N_i^{(X)}\right)\,,
\ee
where $\widehat M_i^{(X)}$ and $\widehat N_i^{(X)}$ are the two moment maps of the two punctures.

As for the E-string case, gluing is a relevant operation. The superconformal R-symmetry of the trinion at the fixed point after turning on the superpotential mixes with the symmetries $U(1)_t$, $U(1)_{c_4}$, $U(1)_{\widetilde c_4}$ and $U(1)_{s_j}$. The superpotential terms are relevant deformations fixing the symmetries of the $\Phi_i$ fields. After this we gauge the puncture symmetries and find that the $\beta$ functions vanishes. In addition, no operator drops bellow the unitarity bound when gluing two trinions. We can also define $S$-gluing as was done for the E-string case. As we will not use this here we will not discuss it and the interested reader can find the details in \cite{ Kim:2018bpg}.

\begin{figure}[t]
	\centering
  	\includegraphics[scale=0.3]{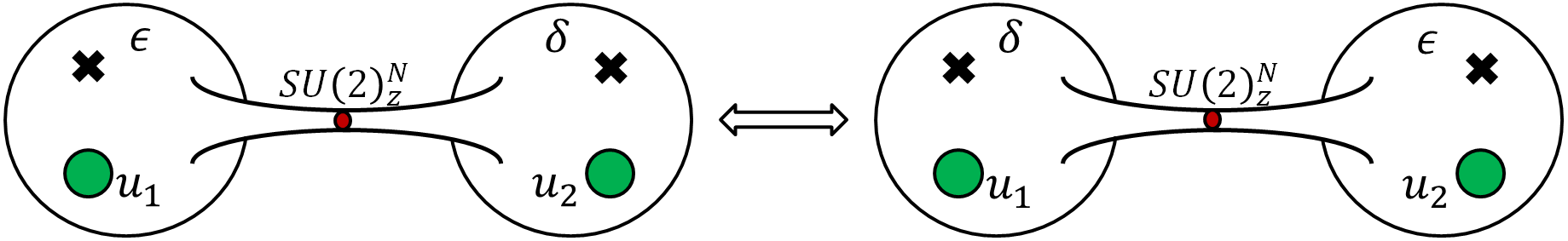}
    \caption{Different duality frames for a four punctured sphere. The fact that the left and right frames are the same implies for example that the index has to be invariant under exchange of the two $SU(2)$ fugacities. }
    \label{F:MinDNDuality}
\end{figure}

For example the index of the four punctured sphere with two maximal punctures and two $SU(2)$ minimal punctures using $\Phi$-gluing the two trinions along a $z$ type of puncture is,
\be
\label{E:MinDNFourPz}
\mathcal{I}_{\textbf{v},\textbf{u},\epsilon,\delta} & = & \kappa^{N} \prod_{i=1}^N\oint\frac{dz_i}{4\pi i z_i}
\mathcal{I}_{\textbf{z},\textbf{v},\epsilon}^{trinion}\mathcal{I}_{\textbf{z},\textbf{u},\delta}^{trinion}
\frac{\prod_{n=1}^{4}\Gamma_{e}\left(\sqrt{pq}t a_1 c_{n}^{-1}z_1^{\pm1}\right)\Gamma_{e}\left(\sqrt{pq}t a_{N+1} \widetilde c_{n}^{-1}z_N^{\pm1}\right)}{\prod_{i=1}^N\Gamma_{e}\left(z_i^{\pm2}\right)}\times\nonumber\\
 & & \prod_{i=1}^{N-1} \Gamma_{e}\left(\sqrt{pq} t a_{i+1} s_i^{-1} z_{i}^{\pm1} z_{i+1}^{\pm1} \right)
\ee
A non-trivial check of the conjectured trinion is that models with more than three punctures satisfy duality properties. One such property is showing that the index is invariant under the exchange of two punctures of the same type, see Figure  \ref{F:MinDNDuality}. We have verified this property in expansion in fugacities.\footnote{As for the $N=1$ case, we expect that the relevant identity satisfied by the index can be deduced from sequences of Seiberg and S-dualities.}

\subsection{Properties of the minimal puncture}
In the case of $N>1$ the new minimal $SU(2)$ puncture has not appeared in the literature so far; therefore, we need to justify the claim that it is indeed a puncture. One check we can perform is the aforementioned duality property; the index needs to be invariant under the exchange of two such punctures.

We also expect to be able to close the new minimal puncture by giving a vacuum expectation value to operators charged under the puncture and the internal symmetries only. These operators are expected to be the  moment maps $\widehat M^{(\epsilon)}$ in analogy to closing punctures in other previously studied setups, say \cite{Gaiotto:2015usa}. Once the puncture is closed by giving a vev to one of the moment map operators, we expect to find a known theory \cite{Kim:2018bpg} of a flux tube after adding possibly some free flip fields. The  flux of the theory obtained in the IR should be related to the operator we gave a vev to.

Considering giving a vev to the moment maps operators, we have $2\times (6+(N+1)+(N-1))=4N+12$  options, all with R-charge $1$, and charges $\{ \epsilon^{\pm1} C_i \}_{i=1}^{2N+6}$, where $C_i$ are the charges appearing in \eqref{E:MinDNMomentMaps} associated with $\widehat M^{(\epsilon)}$.  Closing the puncture shifts the models flux by one quanta matching the charges under internal symmetries of the operator obtaining a vev. For example, giving a vev to an operator with charges $C_i=t^2 a_i^2$ shifts the flux by $+1/(N+1)$ for $U(1)_{t}$ and $U(1)_{a_{j\ne i}}$, in addition to a $-N/(N+1)$ flux shift for $U(1)_{a_{i}}$. To match the 't Hooft anomalies of the IR theory with the expectations from six dimensions we also need to add some flip fields. When giving a vev to an operator of charges $\epsilon^{\pm1} C_i$ one needs to add $2N+6$ flip fields of charges $\{\ \epsilon^{\pm1} C_i^{-1},\,\{ \epsilon^{\mp1} C_j^{-1} \}_{j\neq i} \}$ all with R-charge $1$, and the $\pm$ signs correlated. These flip fields couple through superpotential interactions to corresponding components of the moment map operators.

\begin{figure}[t]
	\centering
  	\includegraphics[scale=0.28]{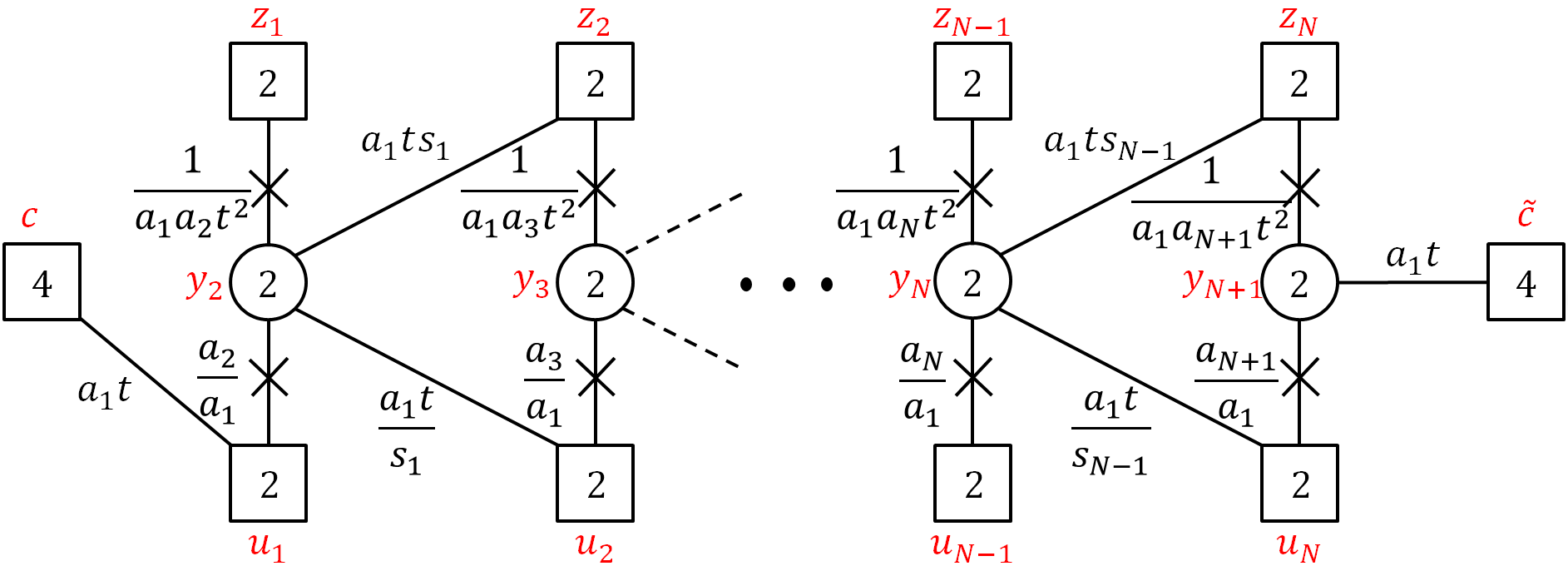}
    \caption{A quiver diagram of the IR theory one finds after closing the minimal puncture by giving a vev to the operator $\widetilde M_1^2$. The squares and circles denote $SU(n)$ global and gauge symmetries, respectively, where $n$ is the number inside. The fields denoted by the vertical lines have R-charge $0$, the flip fields have R-charge $2$ and the rest have R-charge $1$. $a_i$, $s_j$ and the two $SU(4)$'s are related to the internal symmetries that arise from $6d$. In red we write the symmetries associated fugacities, and in black we write the charges of each field. The $X$'s denote as usual flip fields. We emphasize that the six additional flip fields charged under $c$ and $\widetilde c$ were removed. As always, each closed loop of fields has a  superpotential term turned on for it.}
    \label{F:MinDNTube}
\end{figure}

For example, closing the minimal puncture of the trinion by giving a vev to the operator $\widetilde M_1^2$ with charges $\epsilon t^2 a_1^2$ generates a flow to the IR theory described by the quiver in Figure  \ref{F:MinDNTube}.
By construction the remaining theory has two maximal punctures. This flux tube has flux of $1/(N+1)$ for $U(1)_{t}$ and $U(1)_{a_{i\ne 1}}$, and flux $-N/(N+1)$ for $U(1)_{a_{1}}$.

Next, we verify that when we take two such tubes and  $\Phi$-glue them to one another to obtain a torus with flux we get the expected anomalies We find the following anomalies,
\be
&& Tr\left(-U(1)_{t}/2\right)=-4(N+2)\,, \qquad Tr\left((-U(1)_{t}/2)^3\right)=-2(5N+1)\,, \nonumber\\
&& Tr\left(U(1)_{a_1}/2\right)= Tr\left((U(1)_{a_1}/2)^3\right)=4(N+2)\,, \nonumber\\
&& Tr\left((-U(1)_{t}/2)(U(1)_{a_1}/2)^2\right)=-4N\,, \qquad Tr\left((-U(1)_{t}/2)^2(U(1)_{a_1}/2)\right)=2(3N-1)\,, \nonumber\\
&& Tr\left((-U(1)_{t}/2)U(1)_{R}^2\right)=4N\,, \qquad Tr\left((U(1)_{a_1}/2)U(1)_{R}^2\right)= -4N\,, \nonumber\\
&& Tr\left((-U(1)_{t}/2)(U(1)_{a_{i\ne 1}}/2)^2\right)=-4\,, \nonumber\\
&& Tr\left((U(1)_{a_1}/2)(U(1)_{a_{i\ne 1}}/2)^2\right) = Tr\left((U(1)_{a_1}/2)^2(U(1)_{a_{i\ne 1}}/2)\right) =4\,, \nonumber\\
&& Tr\left(U(1)_{R}U(1)_{c(\widetilde c)}^2\right) = Tr\left((-U(1)_{t}/2)U(1)_{c(\widetilde c)}^2\right) = -Tr\left((U(1)_{a_1}/2)U(1)_{c(\widetilde c)}^2\right) = -4\,, \nonumber\\
&& Tr\left(U(1)_{R}U(1)_{s_i}^2\right) = Tr\left((-U(1)_{t}/2)U(1)_{s_i}^2\right) = -Tr\left((U(1)_{a_1}/2)U(1)_{s_i}^2\right) = -8\,,
\ee
where the rest of the anomalies vanish. These anomalies exactly match the expectations given in \eqref{E:4dAP} and \eqref{E:AnomGenFlux}.

\begin{figure}[t]
	\centering
  	\includegraphics[scale=0.18]{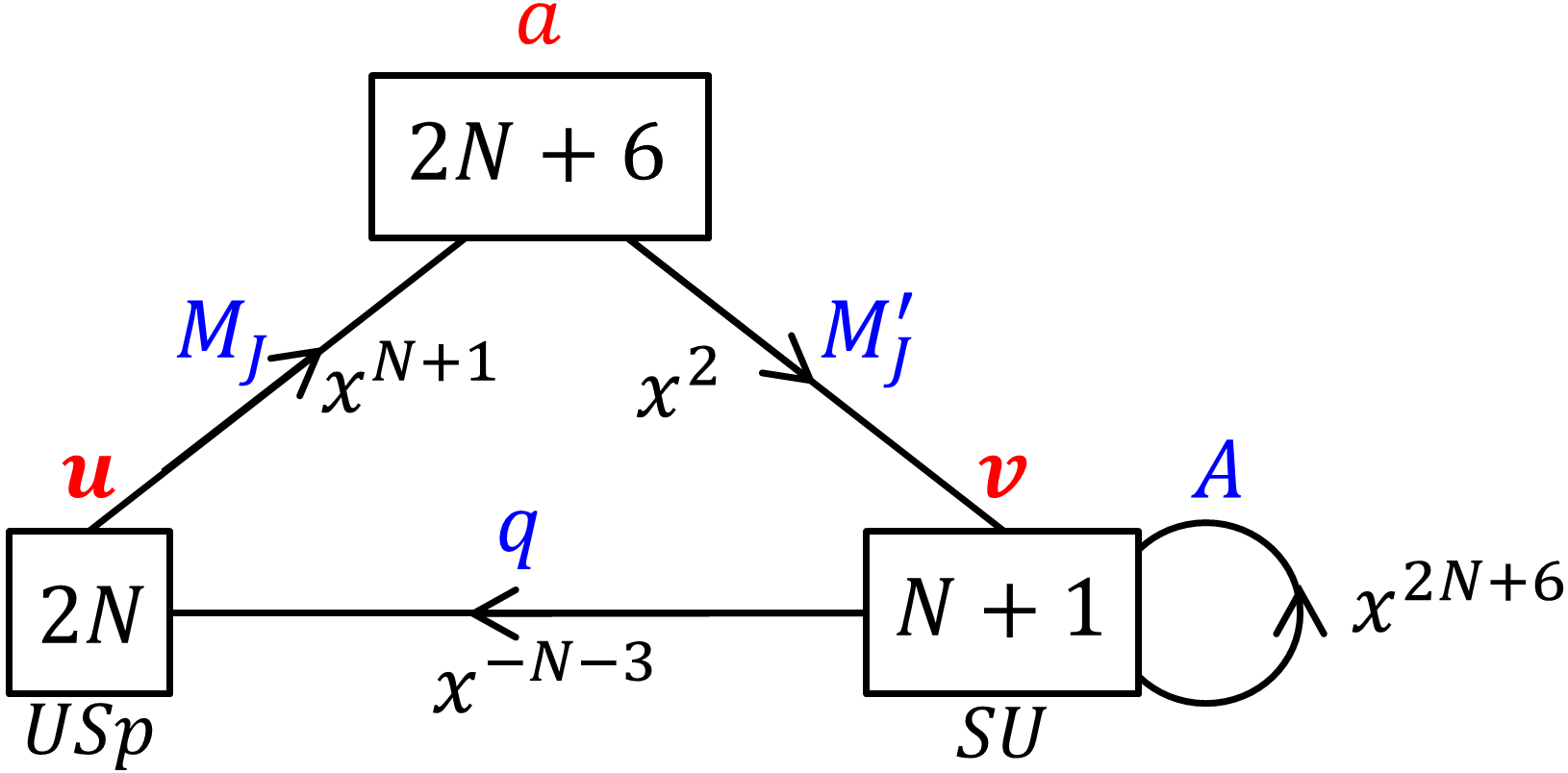}
    \caption{A quiver diagram of the flux tube with one $USp(2N)$ maximal puncture and one $SU(N+1)$ maximal puncture and $1/4$ flux for $U(1)_x$ embedded in $SO(4N+12)$ as $SO(4N+12) \to U(1)_x \times SU(2N+6)$. The $2N$ square denotes the $USp(2N)$ symmetry of one maximal puncture, the $N+1$ square denotes the $SU(N+1)$ symmetry of the second maximal puncture and the $2N+6$ square denotes an $SU(2N+6)$ symmetry. the $SU(2N+6)_a$ symmetry together with $U(1)_x$ are related to the internal symmetries coming from $6d$. The operators $M$ and $M'$ have R-charge $1$, $q$ has R-charge $0$ and $A$ has an R-charge $2$. In blue we write the field names, in red we write the fugacities associated to puncture and gauge symmetries, and in black we write the fugacities of each field. The closed loop of fields has a  superpotential term turned on for it.}
    \label{F:USpSUTube}
\end{figure}

Next, we want to argue that the new $SU(2)$ minimal puncture is related to the known $USp(2N)$ maximal puncture of the minimal $D_{N+3}$ class. To that end we will consider the flux tube described in \cite{Kim:2018bpg} shown as a quiver in Figure \ref{F:USpSUTube}. 
This is a simple example of a theory with $USp(2N)$ puncture from which we can read off the 't Hooft anomalies associated to the puncture.
Let us check that the 't Hooft anomaly $Tr\left(U(1)_R USp(2N)^2\right)$ matches the new minimal puncture anomaly $Tr\left(U(1)_R SU(2)^2\right)$. We find these two are the same and are given by,
\be
Tr\left(U(1)_R USp(2N)^2\right) = Tr\left(U(1)_R SU(2)^2\right) = -\frac{1}{2}(N+1).
\ee

Now we can look at the operators in the fundamental representation of the $USp(2N)$ puncture. These are the $2N+6$ operators $M_i$; thus, the ``moment maps'' for this puncture are
\be
\widehat M^{(u)}_{USp(2N)}\,: \qquad  \{ x^{N+1}\widetilde a_i \}_{i=1}^{2N+6}\,.
\ee 
We can map these moment maps to the ones of $\widehat M^{(\epsilon)}$ as their number is the same. Under such a mapping all anomalies associated to the $USp(2N)$ puncture will be the same as the ones of the $SU(2)$ minimal puncture. By giving a vev to operators charged under the $USp(2N)$ puncture we can close it to the $SU(2)$ minimal puncture. For example, we can give vev to all the operators $M_J^{(i)}$ except one, with $i$ being the $USp(2N)$ index, and this will generate a flow that will leave us with a  puncture of $SU(2)$ symmetry exactly like the new puncture we find in the aforementioned trinion. Because the vacuum expectation values are for components of a bifundamental of $USp(2N)$ and the flavor $SU(2N+6)$ the anomalies of the remaining $SU(2)$ will be the same as the ones of the original $USp(2N)$ which we have argued match our new $SU(2)$ puncture.

\subsection{The trinion with maximal punctures}
\begin{figure}[t]
	\centering
  	\includegraphics[scale=0.31]{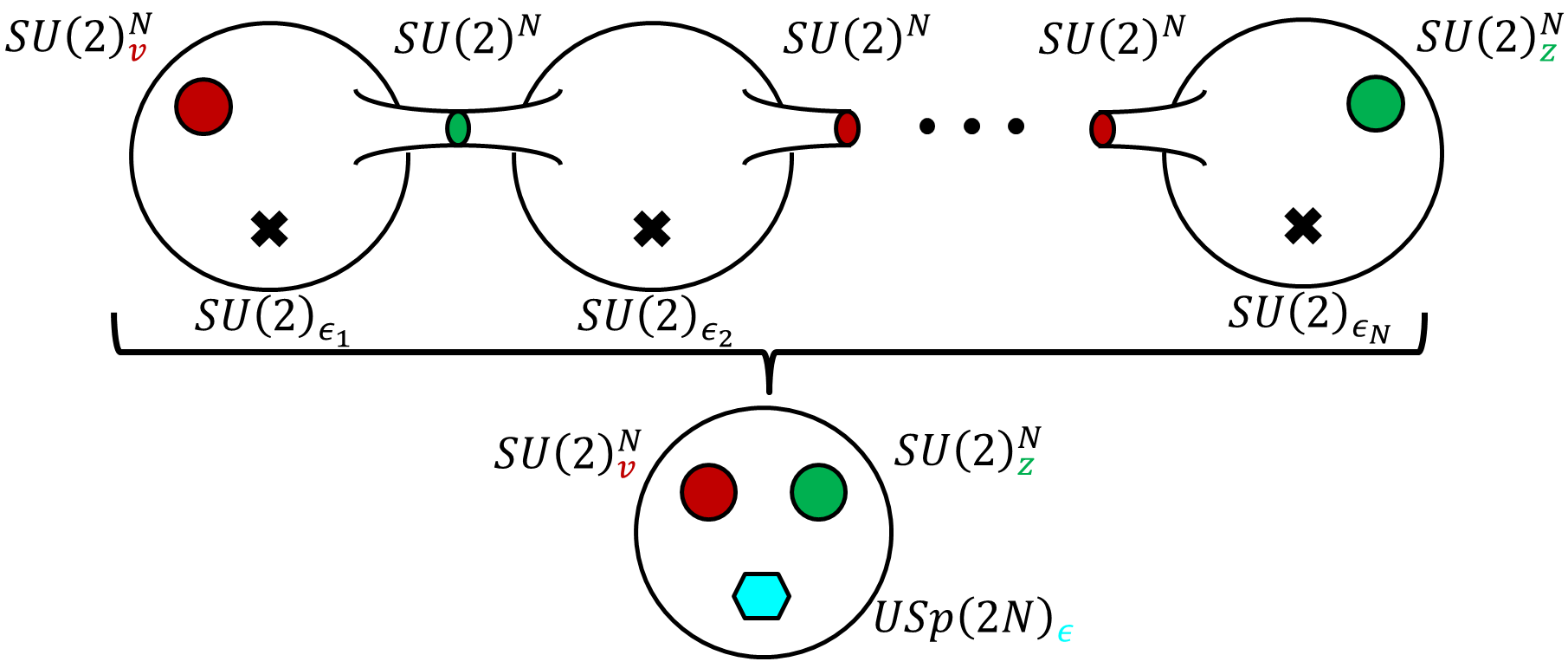}
    \caption{An illustration how $N$ minimal $D_{N+3}$ trinions are glued to form the trinion, with three maximal punctures.}
    \label{F:MinDNTrinion2IT}
\end{figure}

We have established that the conjectured trinion is a three punctured sphere compactification of the minimal $D_{N+3}$ conformal matter, with two maximal punctures with $SU(2)^N$ symmetry each and a minimal puncture with $SU(2)$ symmetry that is related to the $USp(2N)$ maximal puncture by closing it partially. Now we conjecture that by gluing $N$ such trinions we get a theory that is on the same conformal manifold as a  trinion with two maximal punctures with $SU(2)^N$ symmetry and one maximal puncture with $USp(2N)$ symmetry, see Figure  \ref{F:MinDNTrinion2IT}.
This means that the $N$ copies of $SU(2)$ symmetries corresponding to the minimal punctures get enhanced to $USp(2N)$ somewhere on the conformal manifold of the IR SCFT.\footnote{A similarly looking, though different, enhancement of $SU(2)^N$ symmetry to $USp(2N)$ was recently considered in \cite{Pasquetti:2019hxf}.}

Now, with the trinion at hand we wish to combine several of them to form a higher genus Riemann surface. Since we already know how to glue the $SU(2)^N$ maximal punctures , here we will only specify how to glue the new $USp(2N)$ maximal punctures \cite{Kim:2018bpg}. We glue two $USp(2N)$ maximal punctures of the same type by gauging the diagonal $USp(2N)$ symmetry of the two punctures. In addition we add $2N+6$ fields $\Phi_i$ in the fundamental of $USp(2N)$, and couple them in the following manner in the superpotential
\be
W=\sum_{i=1}^{2N+6} \sum_j^{N} \Phi_i^j \left( \widehat M_i^{(\epsilon_j)} - \widehat N_i^{(\epsilon_j)}\right),\,
\ee
where $\widehat M_i^{(\epsilon_j)}$ and $\widehat N_i^{(\epsilon_j)}$ are the moment maps for the pair of $N$ $SU(2)$ minimal punctures composing the two $USp(2N)$ maximal punctures.
We also decomposed $\Phi_i$ into $N$ fundamentals of the $SU(2)^N$ subgroup of $USp(2N)$.

For example the index of the four punctured sphere with four $SU(2)^N$ maximal punctures using $\Phi$-gluing on the two $USp(2N)$ punctures in the aforementioned trinion is given by,
\be
\label{E:MinDNFourP}
\mathcal{I}_{\textbf{v},\textbf{z},\textbf{u},\textbf{w}} & = & \frac{\kappa^{N}}{2^{N}N!}\prod_{a=1}^{N}\oint\frac{d\epsilon_a}{4\pi i \epsilon_a}
\mathcal{I}_{\textbf{v},\textbf{z},\epsilon}^{3MPT}\mathcal{I}_{\textbf{u},\textbf{w},\epsilon}^{3MPT} \prod_{i=1}^{N+1}\Gamma_{e}\left(\sqrt{pq}t^{-2} a_i^{-2} \epsilon_a^{\pm1}\right)\prod_{j=1}^{N-1}\Gamma_{e}\left(\sqrt{pq} s_j^{2} \epsilon_a^{\pm1}\right)\times \nonumber\\
 & & \frac{\prod_{a=1}^{N}\prod_{n=1}^{3}\Gamma_{e}\left(\sqrt{pq}(c_{n}c_4)^{-1}\epsilon_a^{\pm1}\right)\Gamma_{e}\left(\sqrt{pq}(\widetilde c_{i} \widetilde c_4)^{-1}\epsilon_a^{\pm1}\right)}{\prod_{1\le a<b\le N}\Gamma_{e}\left(\epsilon_{a}^{\pm1}\epsilon_{b}^{\pm1}\right)\prod_{a=1}^{N}\Gamma_{e}\left(\epsilon_{a}^{\pm2}\right)}\,,
\ee
where $\mathcal{I}_{\textbf{v},\textbf{z},\epsilon}^{3MPT}$ is the suggested trinion with three maximal punctures.

One of the simplest checks to perform is to verify the anomalies match the expectation from $6d$. The anomalies for the trinion with maximal punctures are given by,
\be
&&Tr(U(1)_R) = -2N(N+4)\,,\qquad Tr(U(1)_R^3)=2N(2N-1)\,, \nonumber\\
&&Tr(U(1)_R\ SU(2)_{z_i}^2) = Tr(U(1)_R\ SU(2)_{u_i}^2) = -1\,,\nonumber\\ 
&& Tr(U(1)_R\ USp(2N)_{\epsilon}^2) = -\frac{1}{2}(N+1).
\ee
These fit perfectly with the predictions of \eqref{E:4dAPwPunc}. In addition, we expect this trinion to be of vanishing flux, as it is built out of vanishing flux trinions. Thus by gluing two such trinions to form a genus two surface we expect the anomalies to match the predictions of \eqref{E:4dAP}. The anomalies of a genus two surface built out of two such trinions $\Phi$-glued to one another are,
\be
& & Tr(U(1)_R) = -N(2N+9)\,,\qquad Tr(U(1)_R^3)=N(10N+3)\,,\nonumber\\
& &  Tr(U(1)_R (-U(1)_{t}/2)^2)=-2N(N+1)\,,\qquad Tr(U(1)_R U(1)_{s_j}^2)=-8N\,,\nonumber\\
& &  Tr(U(1)_R (U(1)_{a_i}/2)^2)= Tr(U(1)_R U(1)_{c_n}^2)=  Tr(U(1)_R U(1)_{\widetilde c_n}^2)= -4N,\, 
\ee
while the rest of the anomalies vanish. These also fit the predictions from $6d$ for a genus two surface with vanishing flux given in \eqref{E:4dAP}. In order to see the match for the $c,\, \widetilde c$ and $s_j$ internal symmetries it is required to know the branching rules of $SO(2N+6)$ given in equation \eqref{E:SO(2N+10)branch}.

Our results are consistent with the symmetry of any number, $\ell$, of minimal punctures   enhancing in the IR somewhere on the conformal manifold  to $USp(2\ell)$. In the case of $\ell=N$ we interpret this enhancement as having a maximal puncture, and in the case of $\ell<N$ we can interpret this puncture as an intermediate puncture. Such an intermediate puncture can be found by giving vevs to a maximal puncture partially closing it via an RG flow in a similar way to how we close a maximal puncture to a minimal one. Finally, for the case of $\ell>N$ we have no puncture interpretation.

\subsection{Higher genus surfaces}
We can perform further checks of the suggested trinion by studying the supersymmetric indices of theories corresponding to closed Riemann surfaces built from it. Since the flux associated to the above trinion is vanishing for all internal symmetries we expect the index to have an apparent $SO(4N+12)$ symmetry coming from the $6d$ model. By $\Phi$-gluing $2(g-1)$ such trinions to form a genus $g$ surface we find the superconformal index takes the form
\be
\mathcal{I} = 1 + (\chi_{\textbf{Adj}}^{SO(4N+12)}(g-1)+3(g-1))pq+...
\ee
Meaning we get that the index has the expected $SO(4N+12)$ symmetry. The $\chi_{\textbf{Adj}}^{SO(4N+12)}$ appearing in the index is the adjoint character of $SO(4N+12)$ and it follows the branching rules of the following decompositions
 
\noindent $SO(4N+12) \to U(1)_t\times SU(N+1) \times SO(2N+10)$:
\be
\chi_{\textbf{Adj}}^{SO(4N+12)} & \to &  \chi_{\textbf{Adj}}^{SO(2N+10)} + \left(\chi_\textbf{F}^{SU(N+1)} t^2 + \chi_{\overline{\textbf{F}}}^{SU(N+1)}t^{-2}\right)\chi_\textbf{V}^{SO(2N+10)}\nonumber \\
 & & + \chi_\textbf{Adj}^{SU(N+1)} + \chi_\textbf{AS}^{SU(N+1)}t^{4} + \chi_{\overline{\textbf{AS}}}^{SU(N+1)} t^{-4} + 1 \,.
\ee
$SO(2N+10) \to SO(2)_{s_i}^{N-1}\times SO(12)$:
\be
\label{E:SO(2N+10)branch}
\chi_{\textbf{Adj}}^{SO(2N+10)} & \to & \textbf{66} +\sum_{i=1}^{N-1}(s_i^2+s_i^{-2})\textbf{12} + \sum_{i>j}^{N-1}(s_i^2+s_i^{-2})(s_j^2+s_j^{-2}) + N-1\,,\nonumber\\
\chi_{\textbf{V}}^{SO(2N+10)} & \to & \textbf{12} + \sum_{i=1}^{N-1}(s_i^2+s_i^{-2})\,.
\ee
Above we denoted the character of the representation $\textbf{R}$ of the symmetry $G$ as $\chi_\textbf{R}^G$.
The $SO(12)$ decomposition to $SU(4)_c \times SU(4)_{\widetilde c}$ follows the branching rules appearing in \eqref{E:SO12toSU4}. The $SU(N+1)$ symmetry is the one parameterized by $a_i$ and $SO(2N+10)$ is the symmetry built out of the two $SU(4)$'s and the $N-1$ $SO(2)$'s represented by $s_i$.

\subsection{Higher genus surfaces with flux}
We can also generate a higher genus surface with flux by also gluing the trinions to flux tubes. We give one such example with flux $\mathcal{F}$ to $U(1)_t$ and find the index takes the form
\be
\mathcal{I} & = & 1 + \left(t^{-4}(g-1+2\mathcal{F})\chi_{\overline{\textbf{AS}}}^{SU(N+1)} + t^{-2}(g-1+\mathcal{F})\chi_{\overline{\textbf{F}}}^{SU(N+1)}\chi_\textbf{V}^{SO(2N+10)}\right)pq \nonumber\\
 & & + (g-1)\left(\chi_\textbf{Adj}^{SO(2N+10)}+3+1\right)pq\nonumber\\
 & & + \left(t^4(g-1-2\mathcal{F})\chi_\textbf{AS}^{SU(N+1)} + t^2(g-1-\mathcal{F})\chi_\textbf{F}^{SU(N+1)}\chi_\textbf{V}^{SO(2N+10)}\right)pq+...
\ee
with the expected $SU(N+1) \times SO(2N+10)$ representations.
The first and last terms are relevant and (some of the) irrelevant operators, respectively, for positive t-flux and the other way around for negative t-flux, while the middle term corresponds to marginal operators.

\section{Derivation of trinions from flows}\label{sec:derivation}
In this section we will derive a trinion with two $SU(2)^N$ maximal punctures and one $SU(2)$ minimal puncture for minimal $D_{N+3}$ conformal matter. We will start by reviewing the results and understandings of \cite{Razamat:2019mdt} as they will be the foundation for this derivation. Afterwards, we will derive the trinion by initiating a flow from a minimal $D_{N+4}$ conformal matter to the $D_{N+3}$ minimal conformal matter. The resulting model will be identified as the trinion glued to several known flux tubes. We will perform the computation in full detail for $N=1$ and outline the, straightforward but technically cumbersome, generalization to higher values of $N$.

\subsection{From $6d$ flows to $4d$ flows: a recap}
\begin{figure}[t]
	\centering
  	\includegraphics[scale=0.31]{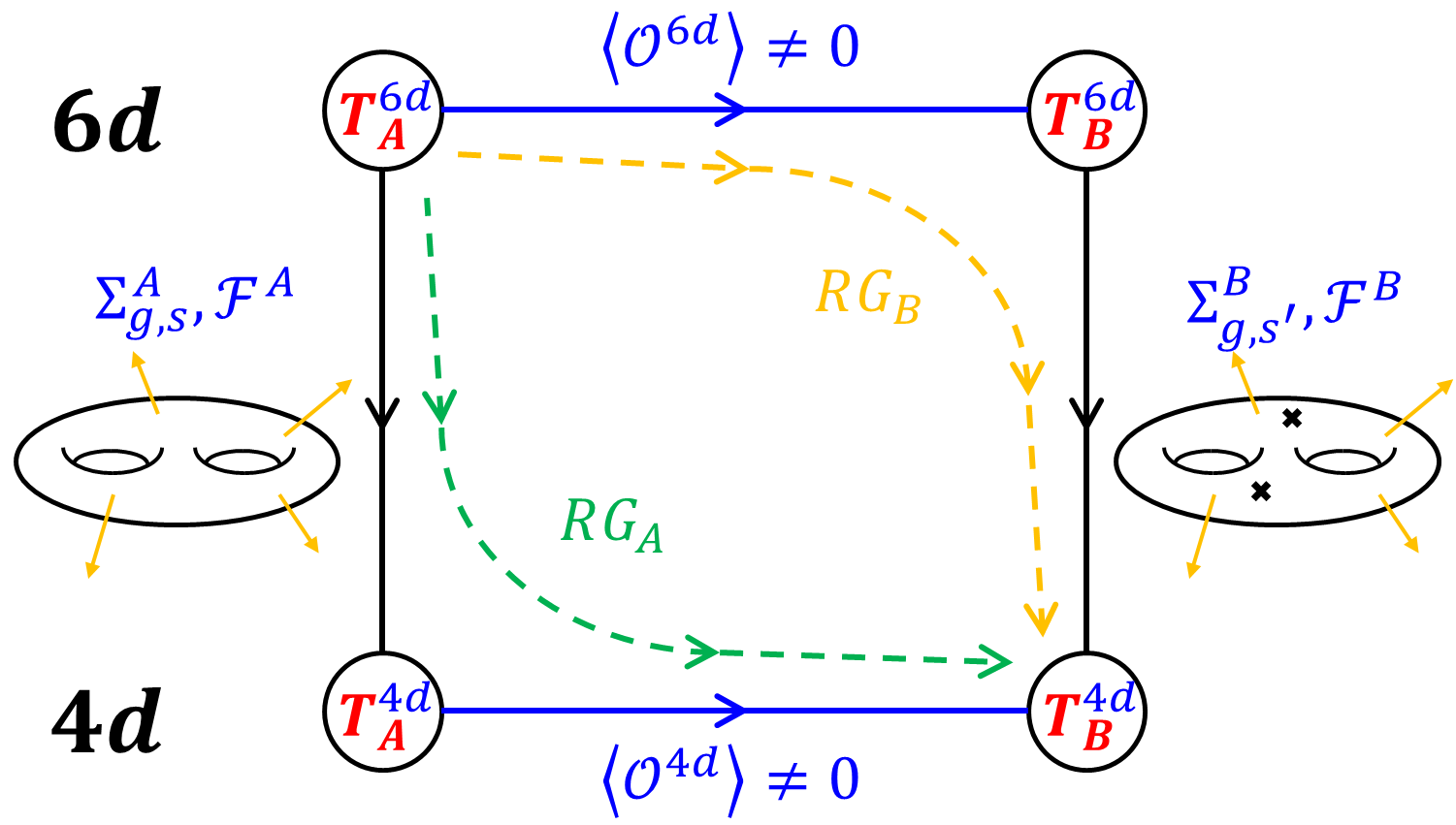}
    \caption{A diagram representing the different RG flows that we can consider. Flow $RG_A$ describes a compactification of a $6d$ model to an effective $4d$ theory followed by a vev to an operator in $4d$. Flow $RG_B$ is defined by first turning the vev to a $6d$ operator and then compactifying the model to $4d$.}
    \label{F:6dRGto4dRG}
\end{figure}

In \cite{Razamat:2019mdt} particular $6d$ $(1,0)$ SCFTs  were considered. These are SCFTs residing on a stack of M5-branes probing a $\mathbb{Z}_k$ singularity, denoted by $\mathcal{T}(SU(k),N)$. When compactified on a Riemann surface to $4d$ via a geometric flow, one finds a class of theories named class $\mathcal{S}_k$ of type $A_{N-1}$ \cite{Gaiotto:2015usa}. One can consider an additional $6d$ flow of such SCFTs triggered by a vev to an operator winding from one end of the $6d$ tensor branch quiver to the other, see Figure \ref{F:End2EndOp}. The operator can be referred to as the ``end to end'' operator. Such a flow reduces $k$ and ends in the $\mathcal{T}(SU(k-1),N)$ SCFT. Next we can consider the two flows, the one triggered by vev in $6d$ and the one triggered by the compact geometry, in two different orders. We denote the first order of flows, where the $6d$ flow is followed by a compactification ending in a $4d$ model, as the $6d\to 6d \to 4d$ flow. The second, is a compactification to $4d$ followed by a flow generated by the matching $4d$ vev ending in the same theory, and is denoted as the $6d\to 4d \to 4d$ flow, see Figure \ref{F:6dRGto4dRG}. An interesting question now is how to map the two flows to one another.

A useful way to think about the mapping between the two flows is that we start from $\mathcal{T}(SU(k),N)$ and turn on both the geometric deformation of compactification and the vev to the $6d$ operator. The order of the two can then be smoothly deformed from $6d\to 6d \to 4d$ to $6d\to 4d \to 4d$ by simply changing the scales of the deformations, meaning the vev scale and the geometry size.
First, one needs to find a $4d$ operator with the same charges as the $6d$ ``end to end'' operator. This operator is a natural candidate for the mapping of the flows, and can be easily found in specific examples. The second part is to match the compactification surface and fluxes of the two flows while flowing from $6d$ to $4d$. It was argued in \cite{Razamat:2019mdt} that the presence of flux forbids turning on a constant vev in $6d$ compactified on the Riemann surface of a finite size, and the vev is needed to be space dependent on the compactified directions.
Next it was argued in \cite{Razamat:2019mdt}  using field theory techniques and brane construction logic that when compactified the spatial profile of the space dependent vev can be thought of as localized at points on the surface and thus interpreted as introducing new punctures. These punctures were found to naturally carry a $U(1)$ symmetry. Looking at the results from the $4d$ point of view: one can take a theory in class $\mathcal{S}_k$ described by a Riemann surface with flux, and give a vev to the $4d$ operator matching the $6d$ ``end to end'' operator. This will result in an RG-flow leading to a theory of class $\mathcal{S}_{k'}$ described by a Riemann surface differing from the original by extra minimal punctures. This theory will  have some value of flux that can be  derived in various ways, for example  by matching anomalies to the ones expected from $6d$.

The process of generating extra punctures by an RG-flow driven by giving vev to a specific operator has only been considered for $\mathcal{T}\left(SU(k),N\right)$ and its compactifications to $4d$. This was based on field theory and brane construction logic which can be extended to other $6d$ $(1,0)$ SCFTs of similar structure. A set of SCFTs with a  similar structure  is the $6d$ $(1,0)$ SCFTs described by a stack of M5-branes probing a $D_{N+3}$ singularity. In what follows we will consider such models with just one M5-brane denoted by $\mathcal{T}\left(SO(2N+6),1\right)$, and their compactifications to $4d$ models.

\subsection{Flows that generate extra punctures in $D_{N+3}$ compactifications}
\begin{figure}[t]
	\centering
  	\includegraphics[scale=0.31]{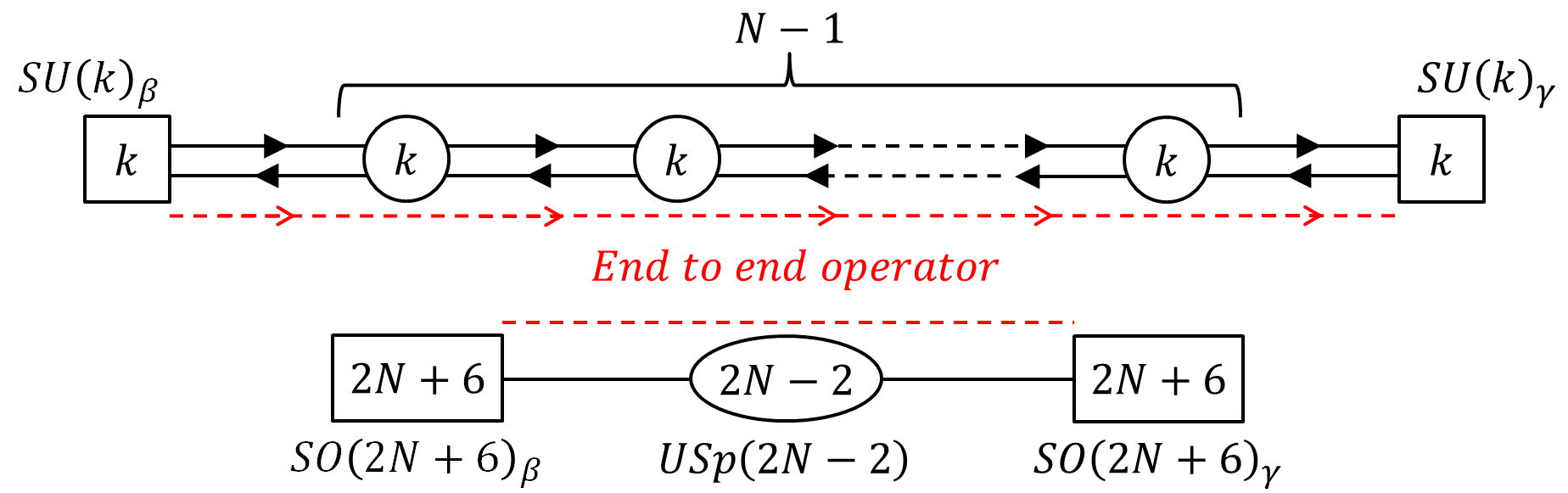}
    \caption{Quiver diagrams of the tensor branch theories of the $6d$ $(1,0)$ $\mathcal{T}\left(SU(k),N\right)$ (above) and $\mathcal{T}\left(SO(2N+6),1\right)$ (below) SCFTs. The arrows represent half hyper-multiplets, while lines represent half hyper-multiplets in the $(\textbf{2N+6},\textbf{2N-2})$ representation of $SO(2N+6)\times USp(2N-2)$. The dashed red line represents the "end to end" operators of each SCFT.}
    \label{F:End2EndOp}
\end{figure}

\begin{figure}[ht]
	\centering
  	\includegraphics[scale=0.31]{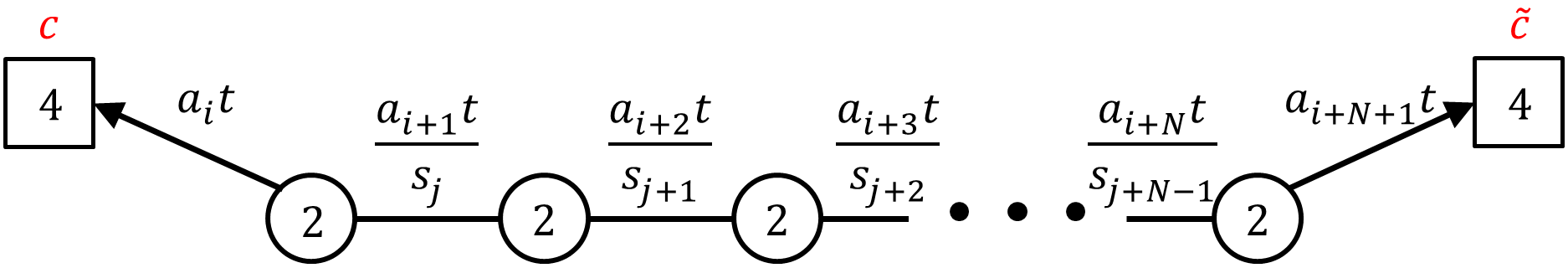}
    \caption{The fields added in $\Phi$-gluing. The baryonic operators $a_i t / s_j$ introduced in the gluing are the ones we give vacuum expectation value to. Not all operators exist in every model, and the spectrum depends on the fluxes and also puncture properties in case there are ones. Thus, it is expected that the flow will depend non trivially on the fluxes. All fields added in the gluing have an R-charge $1$. This R-charge is the one naturally inherited from $6d$ and not necessarily the conformal one. In addition, the representation under the two $SU(4)$'s depends on the type of the punctures glued and can be either fundamental or anti-fundamental.}
    \label{F:DPhiGluingFields}
\end{figure}

Applying the understandings of \cite{Razamat:2019mdt} summarized above to $\mathcal{T}\left(SO(2N+6),1\right)$, we first want to identify the $D_{N+3}$ $6d$ operator analogous to the ``end to end'' operator of $\mathcal{T}\left(SU(k),N\right)$ which should reduce the order of the singularity. The natural candidate is the operator winding from one end of the $\mathcal{T}\left(SO(2N+6),1\right)$ tensor branch quiver to the other as shown in Figure \ref{F:End2EndOp}. These operators  have natural $4d$ operators with same values of charges under the internal symmetries. These operators in known minimal class $\mathcal{S}_{D_{N+3}}$ theories \cite{Kim:2018bpg,Kim:2018lfo}, as in the $A$-type case, are the $\Phi$ fields  added in the process of $\Phi$-gluing (see Figure \ref{F:DPhiGluingFields} for a quiver illustration of the added fields). 

We will divide the derivation into three parts. In the first part we will consider two flux tubes of minimal class $\mathcal{S}_{D_{N+3}}$, with  Lagrangians derived in \cite{Kim:2018lfo}, $\Phi$-glued to one another. As stated before the $\Phi$-gluing adds the fields containing the operator we turn a vev to. We expect such a tube to flow to a similar tube of minimal class $\mathcal{S}_{D_{N+2}}$. This should allow us to identify the  internal symmetries of class $\mathcal{S}_{D_{N+2}}$ from the ones of class $S_{D_{N+3}}$. This expectation comes from an analogous process that can be done in class $\mathcal{S}_k$ flows to class $\mathcal{S}_{k-1}$. 

In the second part we will consider the aforementioned flow applied to a torus built from  the same  flux tubes. In order to preserve all the internal symmetries we will glue $(2-(N\, \mod\, 2))(N+1)$ such minimal tubes \cite{Kim:2018lfo}. We use a torus to avoid complications related to definitions of punctures.
In the third, and last, part we will use Seiberg dualities \cite{Seiberg:1994pq} to bring the resulting IR theory of class $\mathcal{S}_{D_{N+2}}$ to a familiar form \cite{Kim:2018lfo}. In this duality frame we will identify the resulting torus as built out of known flux tubes and an unknown theory with two maximal punctures and an additional symmetry besides the internal symmetries. This unknown theory is the claimed trinion appearing in the beginning of Section \ref{sec:dnmatter}.
In the second and third parts we will only consider explicitly the specific examples of flows starting with  $N=2,3$, as these cases are simple enough to follow, and also easy to generalize to any $N$. 

These steps can be preforemed in a similar fashion for the case of class $\mathcal{S}_k$. There it generates the free trinion, a trinion with two maximal and one minimal puncture from flux tubes. In class $\mathcal{S}_k$ such trinions are already known from \cite{Gaiotto:2015usa} but for class $D$ the trinions derived here are novel.

\subsubsection*{Part I: identifying symmetries}
\begin{figure}[t]
	\centering
  	\includegraphics[scale=0.25]{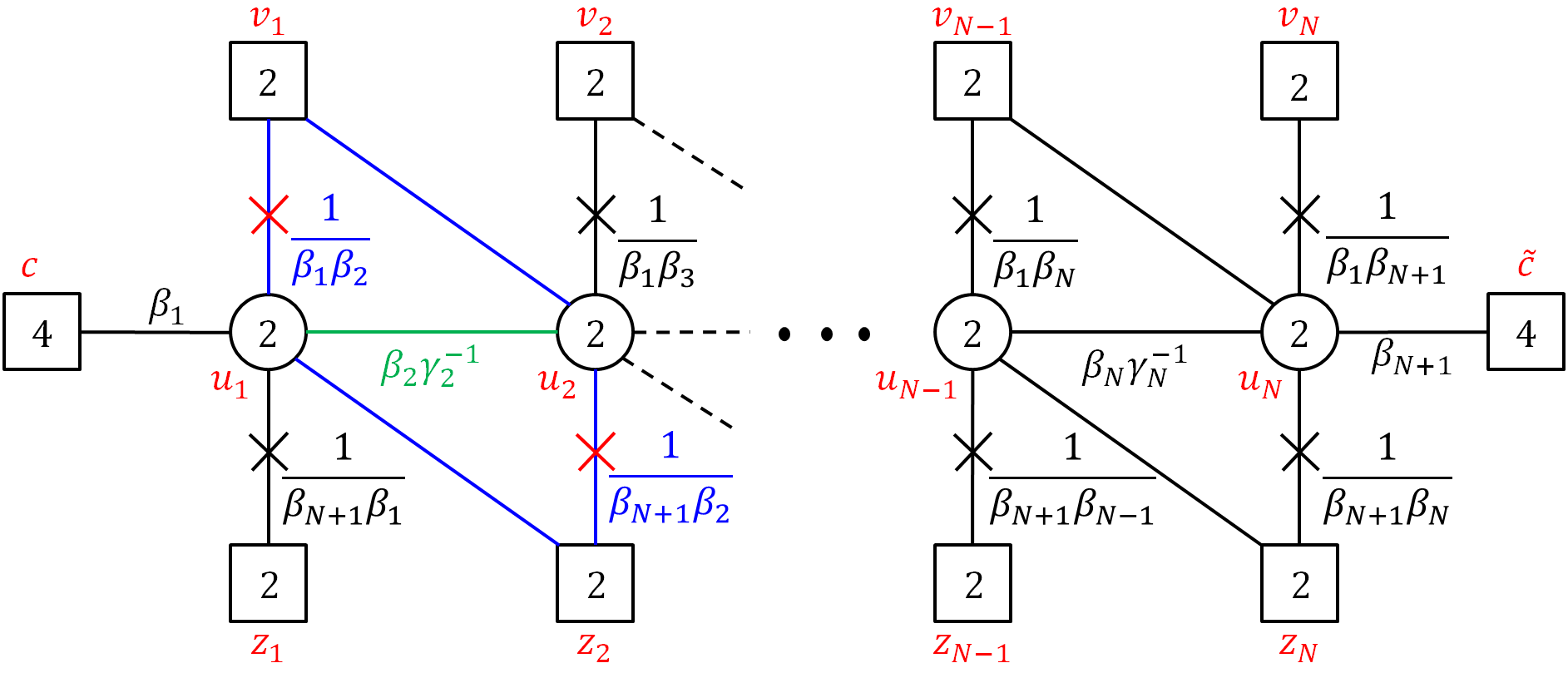}
  	\includegraphics[scale=0.25]{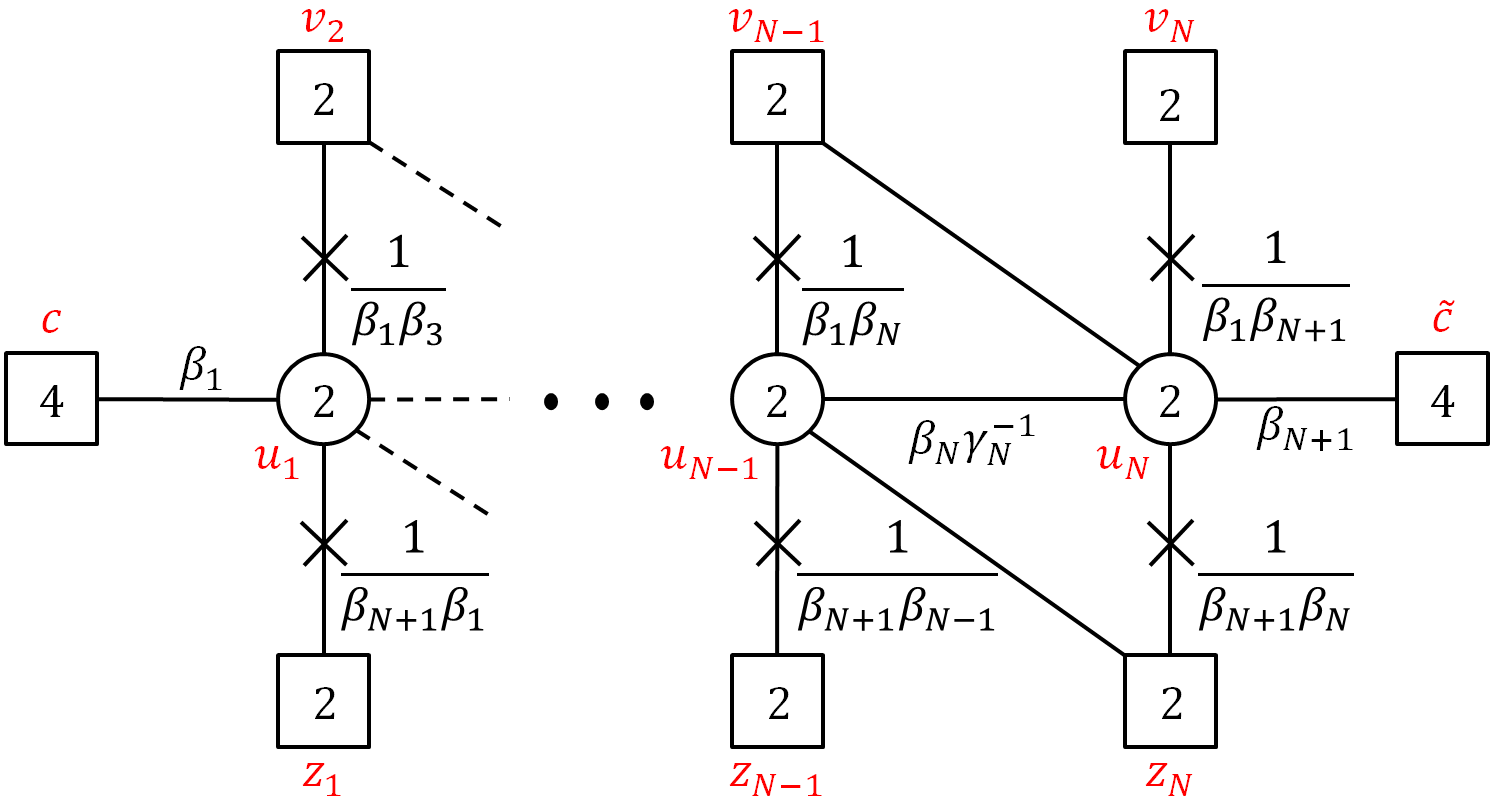}
    \caption{Quiver diagrams of the  flux tube in class $\mathcal{S}_{D_{N+3}}$ (top) which is the UV theory deformed by the vev, and the flux tube in class $\mathcal{S}_{D_{N+2}}$ (bottom), which is the theory we flow to in the IR. The field denoted in green is the one composing the baryonic operator we give a vev to. This vev generates an RG-flow causing the fields marked in blue to become massive. Since these fields become massive their associated flip fields decouple as well in the IR.}
    \label{F:TubeFlow}
\end{figure}

We first start by considering a flow of two flux tubes $\Phi$-glued in minimal class $\mathcal{S}_{D_{N+3}}$ as shown at the upper half of Figure \ref{F:TubeFlow}. The flux of this theory is  \cite{Kim:2018lfo} $\mathcal{F}_{\beta_1}=\mathcal{F}_{\beta_{N+1}}=\frac{N+1}{2N}$ and $\mathcal{F}_{\beta_{i}}=\frac{1}{N}$ for $i=2,...,N$.
As usual it is convenient to organise the matter content, gauge sector, and symmetries of the Lagrangian by writing the superconformal index, which for  this model is given by,
\be
\mathcal{I}_{tube}^{N+3} & = & \kappa^{N}\prod_{i=1}^{N}\oint\frac{du_{i}}{4\pi iu_{i}}\frac{\prod_{n=1}^{4}\Gamma_{e}\left(\sqrt{pq}\beta_{1}c_{n}u_{1}^{\pm1}\right)\Gamma_{e}\left(\sqrt{pq}\beta_{N+1}u_{N}^{\pm1}\widetilde{c}_{n}\right)}{\prod_{i=1}^{N}\Gamma_{e}\left(u_{i}^{\pm2}\right)}\times\nonumber\\
 & & \prod_{i=1}^{N-1}\Gamma_{e}\left(\sqrt{pq}\beta_{i+1}\gamma_{i+1}^{-1}u_{i}^{\pm1}u_{i+1}^{\pm1}\right)\times\nonumber\\
 & & \prod_{i=1}^{N}\Gamma_{e}\left(pq\beta_{N+1}^{2}\beta_{i}^{2}\right)\Gamma_{e}\left(\beta_{N+1}^{-1}\beta_{i}^{-1}z_{i}^{\pm1}u_{i}^{\pm1}\right)\prod_{j=1}^{N-1}\Gamma_{e}\left(\sqrt{pq}\beta_{N+1}\gamma_{j+1}u_{j}^{\pm1}z_{j+1}^{\pm1}\right)\times\nonumber\\
 & & \prod_{i=1}^{N}\Gamma_{e}\left(pq\beta_{1}^{2}\beta_{i+1}^{2}\right)\Gamma_{e}\left(\beta_{1}^{-1}\beta_{i+1}^{-1}u_{i}^{\pm1}v_{i}^{\pm1}\right)\prod_{j=1}^{N-1}\Gamma_{e}\left(\sqrt{pq}\beta_{1}\gamma_{j+1}v_{j}^{\pm1}u_{j+1}^{\pm1}\right)
\ee
We give a vev to the baryonic operator of gauge nodes $u_1$ and $u_2$ matching the $6d$ ``end to end'' operator charges. In the index  we set $\sqrt{pq}\beta_{2}\gamma_{2}^{-1} = 1$, defining $\gamma_{2}=\left(pq\right)^{1/4}\epsilon^{-1/2}$ and $\beta_{2}=\left(pq\right)^{-1/4}\epsilon^{-1/2}$. This initiates an RG-flow Higgsing one gauge node and giving mass to some fields. In addition some of the flip fields decouple as the fields they were coupled to became massive. The index of the IR fixed point theory    is,
\be
\mathcal{I}_{tube}^{N+3,flow} & = & \kappa^{N-1}\prod_{i=1}^{N-1}\oint\frac{du_{i}}{4\pi iu_{i}}\frac{\prod_{n=1}^{4}\Gamma_{e}\left(\sqrt{pq}\beta_{1}c_{n}u_{1}^{\pm1}\right)\Gamma_{e}\left(\sqrt{pq}\beta_{N}u_{N-1}^{\pm1}\widetilde{c}_{n}\right)}{\prod_{i=1}^{N-1}\Gamma_{e}\left(u_{i}^{\pm2}\right)}\times\nonumber\\
 & & \prod_{i=1}^{N-2}\Gamma_{e}\left(\sqrt{pq}\beta_{i+1}\gamma_{i+1}^{-1}u_{i}^{\pm1}u_{i+1}^{\pm1}\right)\times\nonumber\\
 & & \prod_{i=1}^{N-1}\Gamma_{e}\left(pq\beta_{N}^{2}\beta_{i}^{2}\right)\Gamma_{e}\left(\beta_{N}^{-1}\beta_{i}^{-1}z_{i}^{\pm1}u_{i}^{\pm1}\right)\prod_{j=1}^{N-2}\Gamma_{e}\left(\sqrt{pq}\beta_{N}\gamma_{j+1}u_{j}^{\pm1}z_{j+1}^{\pm1}\right)\times\nonumber\\
 & & \prod_{i=1}^{N-1}\Gamma_{e}\left(pq\beta_{1}^{2}\beta_{i+1}^{2}\right)\Gamma_{e}\left(\beta_{1}^{-1}\beta_{i+1}^{-1}u_{i}^{\pm1}v_{i}^{\pm1}\right)\prod_{j=1}^{N-2}\Gamma_{e}\left(\sqrt{pq}\beta_{1}\gamma_{j+1}v_{j}^{\pm1}u_{j+1}^{\pm1}\right)\nonumber\\
 & = & \mathcal{I}_{tube}^{N+2}\,,
\ee
where we identified the internal symmetry of the IR theory with  the UV ones by setting $\beta_{i+1}\to\beta_{i},\gamma_{j+1}\to\gamma_{j}$ with $i=2,...,N$ and $j=3,...,N$. We also shifted the integration variables and puncture fugacities to simplify the writing of the resulting index.

We found that the internal symmetries get mapped in a trivial way in terms of $\beta$ and $\gamma$ fugacities. These are the fugacities used in \cite{Kim:2018lfo} for the non-minimal class $\mathcal{S}_{D_{N+3}}$. Relating them to the fugacities we use for the minimal case is done by setting $\beta_i = t a_i$ and $\gamma_i+1 = s_i$ where $\prod_{i=1}^{N+1} a_i = 1$.

\subsubsection*{Part II: Flow of a torus with $N=2$}
Now we start with the $N=2$ case and consider $\Phi$-gluing six minimal flux tubes to a torus preserving all internal symmetries. The flux of the torus is \cite{Kim:2018lfo} $\mathcal{F}_{\beta_i}=2$ for $i=1,2,3$ and its index is given by,
\be
\mathcal{I}_{g=1}^{N=2} & = & \kappa^{2}\oint\frac{du_{1,1}}{4\pi iu_{1,1}}\oint\frac{du_{1,2}}{4\pi iu_{1,2}}\frac{\prod_{n=1}^{4}\Gamma_{e}\left(\sqrt{pq}\beta_{1}c_{n}^{-1}u_{1,1}^{\pm1}\right)\Gamma_{e}\left(\sqrt{pq}\beta_{3}u_{1,2}^{\pm1}\widetilde{c}_{n}^{-1}\right)}{\Gamma_{e}\left(u_{1,1}^{\pm2}\right)\Gamma_{e}\left(u_{1,2}^{\pm2}\right)}\times\nonumber\\
 & & \kappa^{2}\oint\frac{du_{2,1}}{4\pi iu_{2,1}}\oint\frac{du_{2,2}}{4\pi iu_{2,2}}\frac{\prod_{n=1}^{4}\Gamma_{e}\left(\sqrt{pq}\beta_{2}c_{n}u_{2,1}^{\pm1}\right)\Gamma_{e}\left(\sqrt{pq}\beta_{1}u_{2,2}^{\pm1}\widetilde{c}_{n}\right)}{\Gamma_{e}\left(u_{2,1}^{\pm2}\right)\Gamma_{e}\left(u_{2,2}^{\pm2}\right)}\times\nonumber\\
 & & \kappa^{2}\oint\frac{du_{3,1}}{4\pi iu_{3,1}}\oint\frac{du_{3,2}}{4\pi iu_{3,2}}\frac{\prod_{n=1}^{4}\Gamma_{e}\left(\sqrt{pq}\beta_{3}c_{n}^{-1}u_{3,1}^{\pm1}\right)\Gamma_{e}\left(\sqrt{pq}\beta_{2}u_{3,2}^{\pm1}\widetilde{c}_{n}^{-1}\right)}{\Gamma_{e}\left(u_{3,1}^{\pm2}\right)\Gamma_{e}\left(u_{3,2}^{\pm2}\right)}\times\\
 & & \Gamma_{e}\left(\sqrt{pq}\beta_{2}\gamma_{2}^{-1}u_{1,1}^{\pm1}u_{1,2}^{\pm1}\right)\Gamma_{e}\left(\sqrt{pq}\beta_{3}\gamma_{2}^{-1}u_{2,1}^{\pm1}u_{2,2}^{\pm1}\right)\Gamma_{e}\left(\sqrt{pq}\beta_{1}\gamma_{2}^{-1}u_{3,1}^{\pm1}u_{3,2}^{\pm1}\right)\times\nonumber\\
 & & \Gamma_{e}\left(pq\beta_{1}^{2}\beta_{2}^{2}\right)^{2}\Gamma_{e}\left(pq\beta_{1}^{2}\beta_{3}^{2}\right)^{2}\Gamma_{e}\left(pq\beta_{2}^{2}\beta_{3}^{2}\right)^{2}\times\nonumber\\
 & & \Gamma_{e}\left(\beta_{1}^{-1}\beta_{2}^{-1}u_{1,1}^{\pm1}u_{2,1}^{\pm1}\right)\Gamma_{e}\left(\sqrt{pq}\beta_{1}\gamma_{2}u_{1,2}^{\pm1}u_{2,1}^{\pm1}\right)\Gamma_{e}\left(\beta_{1}^{-1}\beta_{3}^{-1}u_{1,2}^{\pm1}u_{2,2}^{\pm1}\right)\times\nonumber\\
 & & \Gamma_{e}\left(\beta_{2}^{-1}\beta_{3}^{-1}u_{2,1}^{\pm1}u_{3,1}^{\pm1}\right)\Gamma_{e}\left(\sqrt{pq}\beta_{2}\gamma_{2}u_{2,2}^{\pm1}u_{3,1}^{\pm1}\right)\Gamma_{e}\left(\beta_{2}^{-1}\beta_{1}^{-1}u_{2,2}^{\pm1}u_{3,2}^{\pm1}\right)\times\nonumber\\
 & & \Gamma_{e}\left(\beta_{3}^{-1}\beta_{1}^{-1}u_{3,1}^{\pm1}v_{1,1}^{\pm1}\right)\Gamma_{e}\left(\sqrt{pq}\beta_{3}\gamma_{2}u_{3,2}^{\pm1}v_{1,1}^{\pm1}\right)\Gamma_{e}\left(\beta_{3}^{-1}\beta_{2}^{-1}u_{3,2}^{\pm1}v_{1,2}^{\pm1}\right)\times\nonumber\\
 & & (u\leftrightarrow v,c\to c^{-1},\widetilde{c}\to\widetilde{c}^{-1})\,.\nonumber
\ee
The last line implies that  the terms in the lines above need to be multiplied by the same expression with the  denoted transformations.\footnote{This includes the integrations to be put in as well.}

\begin{figure}[htbp]
	\centering
  	\includegraphics[scale=0.31]{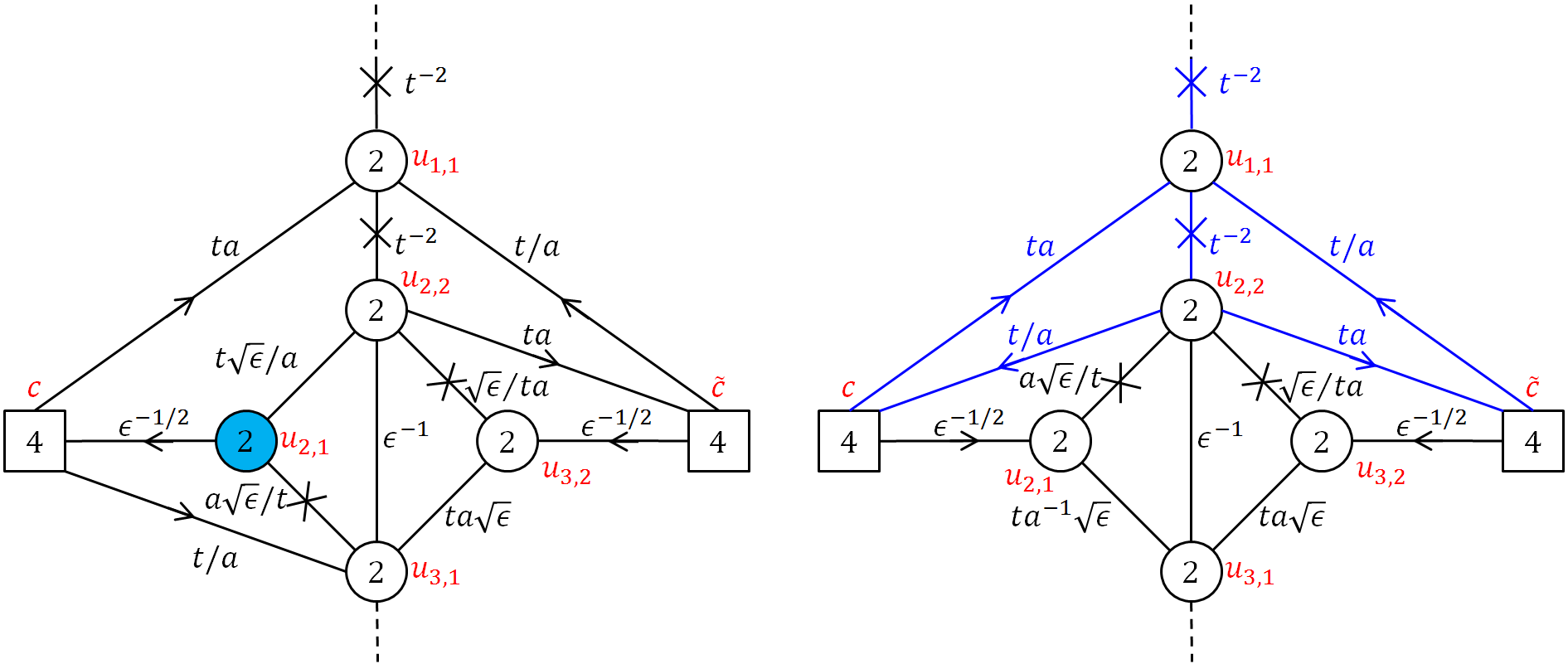}
    \caption{Quiver diagrams of the theory resulting from the RG-flow starting at the class $\mathcal{S}_{D_5}$ torus. The dashed lines indicate that one needs to add the same figure again with $u \to v$ and replacing all the fundamental and anti-fundamental representations of both $SU(4)$ symmetries. \textbf{Left:} The theory resulting from the flow before using Seiberg duality on the $SU(2)$ gauge node denoted by $u_{2,1}$ and marked in blue. \textbf{Right:} The theory after using the aforementioned Seiberg duality. We mark in blue the fields composing the two $-1/2$ t-flux tubes glued together.}
    \label{F:DTorus2Trinion}
\end{figure}

We again initiate the flow with the same vev setting in the index $\sqrt{pq}\beta_{2}\gamma_{2}^{-1} = 1$, defining $\gamma_{2}=\left(pq\right)^{1/4}\epsilon^{-1/2}$ and $\beta_{2}=\left(pq\right)^{-1/4}\epsilon^{-1/2}$. The flow Higgses some of the gauge symmetries and some fields become massive. In addition the flip fields, which couple to composite operators constructed from  fields that acquire mass in the flow, decouple in the IR. Finally, we translate to the fugacity conventions of minimal class $\mathcal{S}_{D_4}$ and the resulting theory quiver appears on the left of Figure \ref{F:DTorus2Trinion}
with the matching index
\be
\mathcal{I}_{g=1}^{N=2,flow} & = & \kappa\oint\frac{du_{1,1}}{4\pi iu_{1,1}}\frac{\prod_{n=1}^{4}\Gamma_{e}\left(\sqrt{pq}tac_{n}^{-1}u_{1,1}^{\pm1}\right)\Gamma_{e}\left(\sqrt{pq}ta^{-1}u_{1,1}^{\pm1}\widetilde{c}_{n}^{-1}\right)}{\Gamma_{e}\left(u_{1,1}^{\pm2}\right)}\times\nonumber\\
 & & \kappa^{2}\oint\frac{du_{2,1}}{4\pi iu_{2,1}}\oint\frac{du_{2,2}}{4\pi iu_{2,2}}\frac{\prod_{n=1}^{4}\Gamma_{e}\left(\left(pq\right)^{1/4}\epsilon^{-1/2}c_{n}u_{2,1}^{\pm1}\right)\Gamma_{e}\left(\sqrt{pq}tau_{2,2}^{\pm1}\widetilde{c}_{n}\right)}{\Gamma_{e}\left(u_{2,1}^{\pm2}\right)\Gamma_{e}\left(u_{2,2}^{\pm2}\right)}\times\nonumber\\
 & & \kappa^{2}\oint\frac{du_{3,1}}{4\pi iu_{3,1}}\oint\frac{du_{3,2}}{4\pi iu_{3,2}}\frac{\prod_{n=1}^{4}\Gamma_{e}\left(\sqrt{pq}ta^{-1}c_{n}^{-1}u_{3,1}^{\pm1}\right)\Gamma_{e}\left(\left(pq\right)^{1/4}\epsilon^{-1/2}u_{3,2}^{\pm1}\widetilde{c}_{n}^{-1}\right)}{\Gamma_{e}\left(u_{3,1}^{\pm2}\right)\Gamma_{e}\left(u_{3,2}^{\pm2}\right)}\times\nonumber\\
 & & \Gamma_{e}\left(\left(pq\right)^{1/4}ta^{-1}\epsilon^{1/2}u_{2,1}^{\pm1}u_{2,2}^{\pm1}\right)\Gamma_{e}\left(\left(pq\right)^{1/4}ta\epsilon^{1/2}u_{3,1}^{\pm1}u_{3,2}^{\pm1}\right)\times\nonumber\\
 & & \Gamma_{e}\left(pqt^{4}\right)^{2}\Gamma_{e}\left(\sqrt{pq}t^{2}a^{\pm2}\epsilon^{-1}\right)\Gamma_{e}\left(t^{-2}u_{1,1}^{\pm1}u_{2,2}^{\pm1}\right)\Gamma_{e}\left(t^{-2}u_{3,1}^{\pm1}v_{1,1}^{\pm1}\right)\times\nonumber\\
 & & \Gamma_{e}\left(\left(pq\right)^{1/4}t^{-1}a\epsilon^{1/2}u_{2,1}^{\pm1}u_{3,1}^{\pm1}\right)\Gamma_{e}\left(\sqrt{pq}\epsilon^{-1}u_{2,2}^{\pm1}u_{3,1}^{\pm1}\right)\Gamma_{e}\left(\left(pq\right)^{1/4}\left(ta\right)^{-1}\epsilon^{1/2}u_{2,2}^{\pm1}u_{3,2}^{\pm1}\right)\nonumber\\
 & & \times(u\leftrightarrow v,c\to c^{-1},\widetilde{c}\to\widetilde{c}^{-1})\,.
\ee
Note that the theory has ten $SU(2)$ gaugings which is the same as the number of integrals. We will interpret this to be given by combining together two three punctured spheres with zero flux and four flux tubes. Each trinion will be a quiver gauge theory with two $SU(2)$ groups. To identify this decomposition we need to perform a sequence of Seiberg dualities, and we do this next. 

\subsubsection*{Part III: The trinion with $N=1$}
\begin{figure}[t]
	\centering
  	\includegraphics[scale=0.21]{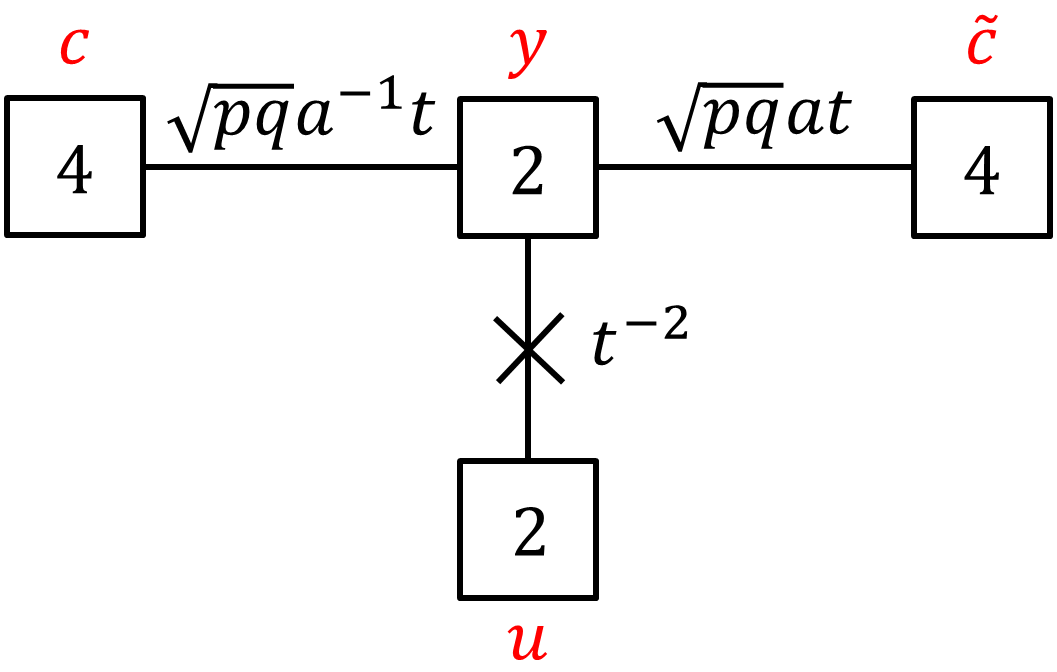}
    \caption{A quiver of a t-flux tube with flux $-1/2$.}
    \label{F:EStringHalfTFluxTube}
\end{figure}

We now wish to isolate the trinion. To that end we will use Seiberg duality on the $SU(2)$ gauge symmetries associated to $u_{2,1}$ and $v_{2,1}$. Since this is an $SU(2)$ gauge symmetry we can choose which fields we will denote as fundamentals and which as anti-fundamentals. We choose the four fields charged under $c_n$ as the fundamental and the rest as anti-fundamental. The resulting theory quiver after the duality appears on the right of Figure \ref{F:DTorus2Trinion}, and its index is,
\be
\label{E:TorusFlow}
\mathcal{I}_{g=1}^{N=2,flow} & = & \kappa\oint\frac{du_{1,1}}{4\pi iu_{1,1}}\frac{\prod_{n=1}^{4}\Gamma_{e}\left(\sqrt{pq}tac_{n}^{-1}u_{1,1}^{\pm1}\right)\Gamma_{e}\left(\sqrt{pq}ta^{-1}u_{1,1}^{\pm1}\widetilde{c}_{n}^{-1}\right)}{\Gamma_{e}\left(u_{1,1}^{\pm2}\right)}\times\qquad\qquad\qquad\qquad\nonumber\\
 & & \kappa\oint\frac{du_{2,2}}{4\pi iu_{2,2}}\frac{\prod_{n=1}^{4}\Gamma_{e}\left(\sqrt{pq}ta^{-1}c_{n}u_{2,2}^{\pm1}\right)\Gamma_{e}\left(\sqrt{pq}tau_{2,2}^{\pm1}\widetilde{c}_{n}\right)}{\Gamma_{e}\left(u_{2,2}^{\pm2}\right)}\times\nonumber\\
 & & \kappa^{3}\oint\frac{du_{2,1}}{4\pi iu_{2,1}}\oint\frac{du_{3,1}}{4\pi iu_{3,1}}\oint\frac{du_{3,2}}{4\pi iu_{3,2}}\frac{1}{\Gamma_{e}\left(u_{2,1}^{\pm2}\right)\Gamma_{e}\left(u_{3,1}^{\pm2}\right)\Gamma_{e}\left(u_{3,2}^{\pm2}\right)}\times\nonumber\\
 & & \Gamma_{e}\left(pqt^{4}\right)^{2}\Gamma_{e}\left(t^{-2}u_{1,1}^{\pm1}u_{2,2}^{\pm1}\right)\Gamma_{e}\left(t^{-2}u_{3,1}^{\pm1}v_{1,1}^{\pm1}\right)\times\nonumber\\
 & & \prod_{n=1}^{4}\Gamma_{e}\left(\left(pq\right)^{1/4}\epsilon^{-1/2}c_{n}^{-1}u_{2,1}^{\pm1}\right)\Gamma_{e}\left(\left(pq\right)^{1/4}\epsilon^{-1/2}u_{3,2}^{\pm1}\widetilde{c}_{n}^{-1}\right)\times\nonumber\\
 & & \Gamma_{e}\left(\sqrt{pq}t^{2}a^{\pm2}\epsilon^{-1}\right)\Gamma_{e}\left(\left(pq\right)^{1/4}t^{-1}a\epsilon^{1/2}u_{2,1}^{\pm1}u_{2,2}^{\pm1}\right)\Gamma_{e}\left(\left(pq\right)^{1/4}t^{-1}a^{-1}\epsilon^{1/2}u_{2,2}^{\pm1}u_{3,2}^{\pm1}\right)\times\nonumber\\
 & & \Gamma_{e}\left(\sqrt{pq}\epsilon^{-1}u_{2,2}^{\pm1}u_{3,1}^{\pm1}\right)\Gamma_{e}\left(\left(pq\right)^{1/4}ta^{-1}\epsilon^{1/2}u_{2,1}^{\pm1}u_{3,1}^{\pm1}\right)\Gamma_{e}\left(\left(pq\right)^{1/4}ta\epsilon^{1/2}u_{3,1}^{\pm1}u_{3,2}^{\pm1}\right)\times\nonumber\\
 & & (u\leftrightarrow v,c\to c^{-1},\widetilde{c}\to\widetilde{c}^{-1})\,.
\ee
Before we interpret the different contributions to this result let us recall the quiver of the $-1/2$ $t$-flux tube of the E-string theory \cite{Kim:2017toz} appearing in Figure \ref{F:EStringHalfTFluxTube}. 
For the readers convenience we also write the superconformal index associated to this flux tube theory
\be
\mathcal{I}_{-1/2\,t\, flux\, tube}^{N=1} & = & \Gamma_{e}\left(pqt^{4}\right)\Gamma_{e}\left(t^{-2}u^{\pm1}y^{\pm1}\right)\prod_{n=1}^{4}\Gamma_{e}\left(\sqrt{pq}ta^{-1}c_{n}y^{\pm1}\right)\Gamma_{e}\left(\sqrt{pq}tay^{\pm1}\widetilde{c}_{n}\right).\,
\ee 
Looking at \eqref{E:TorusFlow} we can identify two such $-1/2$ $t$-flux tubes glued to a third unknown theory in the first part and also in the second part denoted by the last line (see the right side of Figure \ref{F:DTorus2Trinion}). The additional unknown pieces we interpret, up to certain gauge singlet fields, as the new trinions. In this case the resulting theory is an E-string compactification and we identify the E-string trinions index as,
\be
\mathcal{I}_{v,z,\epsilon}^{trinion} & = & \kappa^{2}\oint\frac{dy_{2}}{4\pi iy_{1}}\oint\frac{dy_{2}}{4\pi iy_{2}}\frac{\prod_{i=1}^{4}\Gamma_{e}\left(\left(pq\right)^{\frac{1}{4}}\epsilon^{-\half}c_{i}y_{1}^{\pm1}\right)\Gamma_{e}\left(\left(pq\right)^{\frac{1}{4}}\epsilon^{-\half}y_{2}^{\pm1}\widetilde{c}_{i}\right)}{\Gamma_{e}\left(y_{1}^{\pm2}\right)\Gamma_{e}\left(y_{2}^{\pm2}\right)}\nonumber\\
 & & \prod_{i=1}^{3}\underline{\Gamma_{e}\left(\sqrt{pq}\epsilon c_{i}c_{4}\right)\Gamma_{e}\left(\sqrt{pq}\epsilon \widetilde{c}_{i}\widetilde{c}_{4}\right)}\Gamma_{e}\left(\sqrt{pq}t^{2}a^{\pm2}\epsilon^{-1}\right)\nonumber\\
 & & \Gamma_{e}\left(\left(pq\right)^{1/4}t^{-1}a\epsilon^{1/2}y_{1}^{\pm1}z^{\pm1}\right)\Gamma_{e}\left(\left(pq\right)^{1/4}t^{-1}a^{-1}\epsilon^{1/2}z^{\pm1}y_{2}^{\pm1}\right)\Gamma_{e}\left(\sqrt{pq}\epsilon^{-1}z^{\pm1}v^{\pm1}\right)\nonumber\\
 & & \Gamma_{e}\left(\left(pq\right)^{1/4}ta^{-1}\epsilon^{1/2}y_{1}^{\pm1}v^{\pm1}\right)\Gamma_{e}\left(\left(pq\right)^{1/4}ta\epsilon^{1/2}v^{\pm1}y_{2}^{\pm1}\right)\,.
\ee
For this  result we added additional flip fields underlined in the index. These are added such that the $U(1)_\epsilon$ symmetry will enhance to $SU(2)$ in the IR and will be on the same footing as other puncture symmetries.

\subsubsection*{Part II+III: $N=3$ torus flow and $N=2$ trinion}

As another example let us more succinctly discuss an additional example of flow between $N=3$ to $N=2$.
 The number of  flux tubes we glue to form a flux torus preserving all internal symmetries is four. The flux of the torus is $\mathcal{F}_{\beta_i}=1$ for $i=1,...,4$ and its superconformal index is given by
\be
\mathcal{I}_{g=1}^{N=3}&=&\kappa^{12}\prod_{i=1}^{4}\prod_{j=1}^{3}\oint\frac{du_{i,j}}{4\pi iu_{i,j}}\frac{\prod_{i,n=1}^{4}\Gamma_{e}\left(\sqrt{pq}\beta_{i}c_{n}^{1-2(i\,mod\,2)}u_{i,1}^{\pm1}\right)\Gamma_{e}\left(\sqrt{pq}\beta_{i+3}u_{i,4}^{\pm1}\widetilde{c}_{n}^{1-2(i\,mod\,2)}\right)}{\prod_{i=1}^{4}\prod_{j=1}^{3}\Gamma_{e}\left(u_{i,j}^{\pm2}\right)}\nonumber\\
 & & \times\prod_{i=1}^{4}\Gamma_{e}\left(\sqrt{pq}\beta_{i+1}\gamma_{2}^{-1}u_{i,1}^{\pm1}u_{i,2}^{\pm1}\right)\Gamma_{e}\left(\sqrt{pq}\beta_{i+2}\gamma_{3}^{-1}u_{i,2}^{\pm1}u_{i,3}^{\pm1}\right)\nonumber\\
 & & \times\prod_{i=1}^{4}\prod_{j=1}^{3}\Gamma_{e}\left(pq\beta_{i}^{2}\beta_{i+j}^{2}\right)\Gamma_{e}\left(\beta_{i}^{-1}\beta_{i+j}^{-1}u_{i,j}^{\pm1}u_{i+1,j}^{\pm1}\right)\nonumber\\
 & & \times\prod_{i=1}^{4}\Gamma_{e}\left(\sqrt{pq}\beta_{i}\gamma_{2}u_{i,2}^{\pm1}u_{i+1,1}^{\pm1}\right)\Gamma_{e}\left(\sqrt{pq}\beta_{i}\gamma_{3}u_{i,3}^{\pm1}u_{i+1,2}^{\pm1}\right)\,.
\ee
In the above index the indices are defined such that $u_{i+4,j+3}=u_{i,j+3}=u_{i+4,j}=u_{i,j}$ and $\beta_{i+4}=\beta_i$.

We initiate the flow in the same manner as before, giving a vev that sets $\sqrt{pq}\beta_{2}\gamma_{2}^{-1} = 1$. We can parameterize then the fugacities as $\gamma_{2}=\left(pq\right)^{1/4}\epsilon^{-1/2}$ and $\beta_{2}=\left(pq\right)^{-1/4}\epsilon^{-1/2}$. The flow is similar to the case of $N=2$ and we find a flux torus compactification of class  $\mathcal{S}_{D_5}$ with an additional $U(1)_\epsilon$ symmetry. This symmetry is the symmetry of the extra puncture.

Next, we want to isolate the trinion. For that end, we again use  a sequence of Seiberg dualties. The  superconformal index of the trinion extracted from the IR theory is,
\be
\mathcal{I}_{\boldsymbol{z},\boldsymbol{u},\epsilon}^{trinion} & = & \kappa^{3}\prod_{i=1}^{3}\oint\frac{dy_{i}}{4\pi iy_{i}}\frac{\prod_{n=1}^{4}\Gamma_{e}\left(\left(pq\right)^{1/4}\epsilon^{-1/2}c_{n}y_{1}^{\pm1}\right)\Gamma_{e}\left(\left(pq\right)^{1/4}\epsilon^{-1/2}y_{3}^{\pm1}\widetilde{c}_{n}\right)}{\Gamma_{e}\left(y_{1}^{\pm2}\right)\Gamma_{e}\left(y_{2}^{\pm2}\right)\Gamma_{e}\left(y_{3}^{\pm2}\right)}\times\nonumber\\
 & & \underline{\Gamma_{e}\left(\sqrt{pq}\epsilon s_{1}^{-2}\right)}\prod_{n=1}^{3}\Gamma_{e}\left(\sqrt{pq}\epsilon^{-1}t^{2}a_{n}^{2}\right)\underline{\Gamma_{e}\left(\sqrt{pq}\epsilon c_{n}c_{4}\right)\Gamma_{e}\left(\sqrt{pq}\epsilon\widetilde{c}_{n}\widetilde{c}_{4}\right)}\times\nonumber\\
 & & \Gamma_{e}\left(\left(pq\right)^{1/4}t^{-1}a_{2}^{-1}\epsilon^{1/2}y_{1}^{\pm1}z_{1}^{\pm1}\right)\Gamma_{e}\left(\left(pq\right)^{1/4}\epsilon^{1/2}ta_{2}y_{1}^{\pm1}u_{1}^{\pm1}\right)\Gamma_{e}\left(\sqrt{pq}\epsilon^{-1}z_{1}^{\pm1}u_{1}^{\pm1}\right)\times\nonumber\\
 & & \Gamma_{e}\left(\left(pq\right)^{1/4}\epsilon^{1/2}t^{-1}a_{3}^{-1}z_{1}^{\pm1}y_{2}^{\pm1}\right)\Gamma_{e}\left(\left(pq\right)^{1/4}ta_{3}\epsilon^{1/2}u_{1}^{\pm1}y_{2}^{\pm1}\right)\times\nonumber\\
 & & \Gamma_{e}\left(\left(pq\right)^{1/4}\epsilon^{-1/2}s_{1}z_{2}^{\pm1}y_{2}^{\pm1}\right)\Gamma_{e}\left(\left(pq\right)^{1/4}\epsilon^{-1/2}s_{1}^{-1}y_{2}^{\pm1}u_{2}^{\pm1}\right)\times\nonumber\\
 & & \Gamma_{e}\left(\left(pq\right)^{1/4}\epsilon^{1/2}t^{-1}a_{1}^{-1}z_{2}^{\pm1}y_{3}^{\pm1}\right)\Gamma_{e}\left(\left(pq\right)^{1/4}t\epsilon^{1/2}a_{1}y_{3}^{\pm1}u_{2}^{\pm1}\right)\,,
\ee
where we marked the additional flip fields added in order to get the enhancement of $U(1)_\epsilon$ to $SU(2)_\epsilon$ in the IR.

This procedure can be generalized to any $N$ by repeating the same steps. The only difference is that the number of Seiberg dualities required to isolate the trinion grows, as one needs $N$ Seiberg dualities in the general case.

\section{Two $M5$ branes on $\Z_k$ singularity on arbitrary surface}\label{sec:atrinions}

\begin{figure}[htbp]
	\centering
  	\includegraphics[scale=0.26]{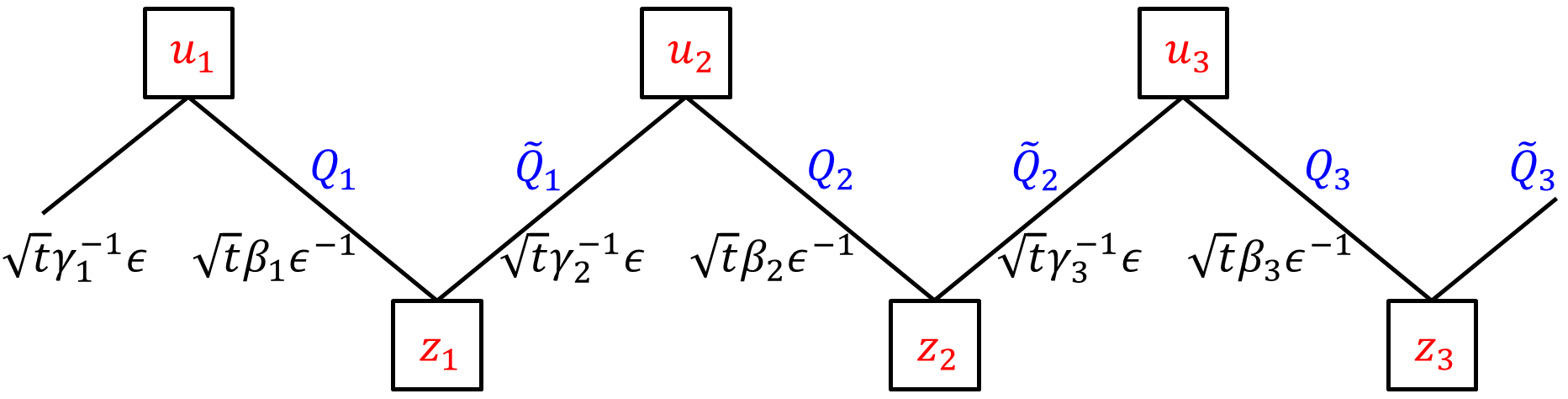}
    \caption{The free trinion in the case of $k=3$. All the groups are of $SU(2)$ symmetry, the $\beta$ and $\gamma$ charges are under the Cartan of $SU(k)_\beta\times SU(k)_\gamma$ internal symmetry, and the $t$ charge is under the $U(1)_t$ internal symmetry. The fugacity $\epsilon$ corresponds to the minimal puncture $U(1)$ symmetry. The quiver is a circular one as the left and right edge are implicitly connected. The moment map operators for the maximal punctures are $M_i = Q_i\widetilde Q_i$ for the $u$ puncture and $M'_i=\widetilde Q_i Q_{i+1}$ for the $z$ puncture. The moment maps for the minimal punctures are the $2k$ baryons $Q_i^2$ and $\widetilde Q_i^2$.}
    \label{F:FreeTrinion}
\end{figure}

Two of the general insights one can deduce from the previous sections are that a collection of simpler punctures can be equivalent on the conformal manifold to a maximal puncture, and that this maximal puncture can be of different type than one would naively expect. Let us discuss an example of compactifications in which these insights can be used to deduce new results. The models we will discuss are of compactifications of two $M5$ branes probing a $\Z_k$ singularity. There are two reasons to pick these particular models. First, trinions with two maximal punctures and one minimal puncture are explicitly known in this case \cite{Gaiotto:2015usa}, see Figure \ref{F:FreeTrinion}. Moreover, two punctured spheres with various values of flux have been also derived \cite{Razamat:2016dpl,Kim:2018lfo,Bah:2017gph}. The symmetry of the maximal puncture is $SU(2)^k$ and of the minimal puncture is $U(1)$. Second,  gluing these trinions together is done by gauging $SU(2)^k$ symmetry. The moment map operators charged under the maximal puncture symmetries are mesons and under the minimal puncture symmetry are baryons which, as in the cases we discussed till now, are on the same footing for $SU(2)$ groups. Thus we can ask whether taking some number of such free trinions together we obtain some type of a maximal puncture. We will answer this question positively and will derive a trinion with three maximal punctures from which arbitrary surfaces can be constructed.

Before carrying on with the derivation let us mention some facts about the $6d$ theories at hand and what is known about their reductions to four dimensions.  We refer the reader to the references above for detailed derivations and here we give the bare essentials. First, the symmetry in six dimensions is $SU(2k)$ for $k\neq 2$ and is $SO(7)$ for $k=2$.
We remind the reader that we refer to these symmetries as internal symmetries as opposed to puncture symmetries. For $k=1$ this is just the interacting $(2,0)$ SCFT residing on two $M5$ branes. It is convenient to think about the global symmetry in its $SU(k)_\beta\times SU(k)_\gamma\times U(1)_t$ decomposition as this is the symmetry of $N$ $M5$ branes on $\Z_k$ singularity for general values of $N$. The free trinion of Figure \ref{F:FreeTrinion} has only the Cartan of the global symmetry manifest and it corresponds to compactification on sphere with two maximal and one minimal punctures with flux in $U(1)_t$ which in our normalization is $\frac{1}{2k}$ (we use same normalization as in \cite{Bah:2017gph,Razamat:2018zus}). There are two ways to glue two surfaces by gauging $SU(2)^k$ symmetry of the two maximal punctures. The maximal punctures come with a set of $k$ moment map operators $M_i$ which are in bifundamental representation of the $i$-th and the $(i+1)$-th $SU(2)$ group where we think of the groups as circularly ordered.\footnote{ Note that for $k=1$ the bifundamental becomes an adjoint and a singlet of the single $SU(2)$  group. This is the case with no orbifold singularity and corresponds to the familiar case of class ${\cal S}$ \cite{Gaiotto:2009we,Gaiotto:2009hg} where usually the singlet is dropped.}
 Then we can perform $\Phi$-gluing \cite{Gaiotto:2015usa} in which a $k$-tuple of bifundamental fields $\Phi_i$ is introduced and the gauging is accompanied by a superpotential,
\be
W=\sum_{i=1}^k \Phi_i\,\left(M_i-M'_i\right)\,,
\ee 
with $M$ and $M'$ being the moment maps of the two punctures. Another way to glue two theories together is $S$ gluing \cite{Gaiotto:2015usa,Franco:2015jna,Hanany:2015pfa,Razamat:2016dpl} where no additional fields are introduced but the superpotential,
\be 
W=\sum_{i=1}^k M_i M'_i\,,
\ee 
is turned on. Note that in the former gluing the charges  of $M$ and $M'$ are identified while in the latter they are conjugated. In particular this means that in the first gluing the flux corresponding to the glued theory is the sum of the fluxes of the ingredients while in the latter it is the difference. 
Finally let us also mention that it was noticed in \cite{Razamat:2016dpl} that $S$-gluing two free trinions in the case of $k=2$ causes the two $U(1)$ symmetries corresponding to minimal punctures to enhance to $SU(2)^2$ symmetry of a maximal puncture. This is achieved when turning on marginal deformations breaking some of the internal symmetries, and results in a trinion from which arbitrary surfaces can be constructed.
Our results refine and generalize this observation for arbitrary $\Z_k$ orbifold. 

\subsection{A heuristic argument}

Let us start with a heuristic argument.
We assume that gluing $k$ free trinions together we obtain an SCFT corresponding to a trinion with two maximal punctures, of $SU(2)^k$ symmetry each visible in the Lagrangian, and one additional puncture with symmetry $G$, which is to be determined. We also assume that the superconformal R-symmetry of all the chiral fields is $\half$.  We are combining the trinions with $S$-gluing and ignoring all the internal symmetries making the argument heuristic but simple. We will soon reintroduce all the moving parts and make the statement precise.
We can compute the $a$ and $c$ anomalies of the theory corresponding to gluing such trinions to closed Riemann surfaces by gauging the puncture symmetries. Again, we assume that we glue trinions as in $S$-gluing just by gauging the puncture symmetries and not introducing any additional fields. Each trinion has $k(k-1)$ $SU(2)$ gauge groups and $8k^2$ chiral fields with R-charge $\frac12$. We have $2g-2$ such trinions, $2g-2$ $SU(2)^k$ gaugings, and $g-1$ $G$ gaugings. 
The result is,
\be
a(g)=\frac1{16}(g-1)(3\, \text{dim}\, G+21k^2)\,,\qquad c(g)=\frac1{16}(g-1)(2\,\text{dim}\, G+23k^2)\,.
\ee 
On the other hand computing these anomalies from six dimensions one then obtains assuming the compactification surface has zero flux,\footnote{We detail the relevant anomalies in Appendix \ref{A:Atypeanoms}.}
\be
a^{6d}(g)=\frac1{16}(g-1)(6+24k^2)\,,\qquad c^{6d}(g)=\frac1{16}(g-1)(4+25k^2)\,.
\ee 
The two results agree perfectly if $\text{dim}\, G=k^2+2$. The natural candidate for the group $G$ is then $SU(k)\times SU(2)$. This is consistent with the trivial $k=1$ case where the free trinion itself has three maximal punctures with symmetry $SU(2)$ each, and with the $k=2$ case  \cite{Razamat:2016dpl} where the two minimal puncture symmetries enhance to $SU(2)^2$ when some of the internal symmetries are broken.

We thus will conjecture that the theory corresponding to gluing together $k$ free trinions,  moving on its conformal manifold, also can be interpreted as a sphere with three maximal punctures, two $SU(2)^k$ and one $SU(k)\times SU(2)$. In fact for reasons we next discuss we will refer to the new puncture as $SU(k)\times U(1)$ one.

Let us further check this conjecture by looking at the simplest operators, the {\it ``moment maps''},  charged under the minimal puncture $U(1)$ symmetries. These operators are the baryons which have R-charge one and contribute to the index as,
\be
(pq)^{\frac12}\left(k\sum_{i=1}^k\left(\e_i^2+\frac1{\e_i^2}\right)\right)\,.
\ee 
Here $\e_i$ are the minimal puncture $U(1)$ symmetries. We want to relate these symmetries to the Cartan of $SU(k)\times SU(2)$. Clearly the above character cannot be interpreted in a natural way as a character of some representation of $SU(k)\times SU(2)$ for any simple redefinition of fugacities for general $k$. However, by reparametrizing $\e_i= c^{\frac1k} a_i^{\frac12}$ with $\prod_{i=1}^k a_i=1$, we can write the above as,
\be
\label{momentmaps}
(pq)^{\frac12}\left(k\,c^{\frac2k} \chi_\textbf{F}^{\ }({\bf a}) +k\,c^{-\frac2k} \chi_{\overline{\textbf{F}}}({\bf a})\right)\,,
\ee 
with $ \chi_\textbf{F}({\bf a})$ being the character of the fundamental of $SU(k)$. So our first claim will be that somewhere on the conformal manifold the $k$ punctures with $U(1)$ symmetries enhance to $SU(k)_a\times U(1)_{c}$.  When we glue two surfaces together we then gauge the $SU(k)$ symmetry so we can check whether this symmetry is anomalous and whether the $R$ symmetry is anomalous. First, as the representation of the matter fields is real there is no cubic anomaly. Second, let us compute $Tr U(1)_R SU(k)^2$. As we only see the Cartan explicitly in the Lagrangian let us evaluate the contribution of the matter to $Tr U(1)_R U(1)_{a_i}^2$. Each trinion has $k$ $SU(2)$ bifundamental fields with charge $+1/2$ and $k$ fields with charge $-1/2$ all with R charge $1/2$. Thus, the matter contribution of the two trinions is 
\be
2\times \left(-\frac12\right) \times 4\times \left(k\times \left(+\frac12\right)^2+k\times \left(-\frac12\right)^2\right)\times2=-4 k\,.\nonumber
\ee 
the above contribution is the same as the contribution of $2k$ fundamentals and antifundamentals of $SU(k)$ with R-charge $1/2$. Thus the R-symmetry is not anomalous assuming the enhancement to $SU(k)$ happens somewhere on the conformal manifold. 
Next we assume that the symmetry $U(1)_c$ enhances after gauging the $SU(k)$ symmetry to $SU(2)_c$. Evidence for this assumption will be shown using the superconformal index in the next subsection.
Following this assumption we can also gauge the $SU(2)$ symmetry. Again, let us compute the contribution of the matter to the $Tr U(1)_R U(1)_c^2$ anomaly. This is given by,
\be
2\times \left(-\frac12\right)\times 4\times k^2 \times \left( \left(+\frac1k\right)^2+\left(-\frac1k\right)^2\right)=-8\,,\nonumber
\ee 
which again agrees with $8$ fundamentals of $SU(2)$ with R-charge $1/2$, and thus the R-symmetry is not anomalous.

\begin{figure}[t]
	\centering
  	\includegraphics[scale=0.32]{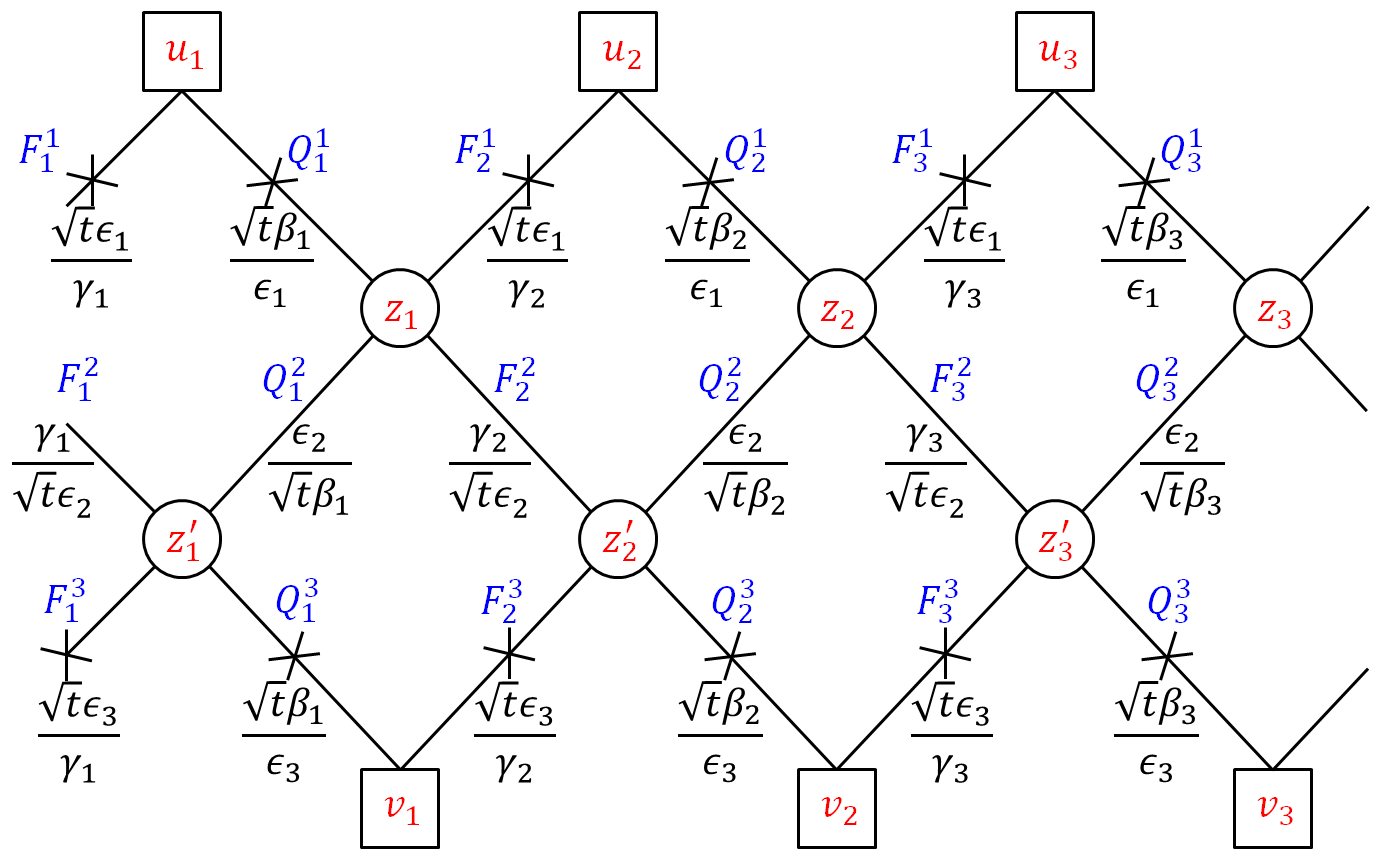}
    \caption{The trinion with three maximal punctures. Here depicted in the case of $k=3$ where the two edges of the quiver are identified as this is a circular quiver. The $k$ $U(1)_{\e_i}$ symmetries are conjectured to enhance somewhere on the conformal manifold to $SU(k)\times U(1)$. Gluing trinions together by gauging the $SU(k)$ symmetry the $U(1)$ is further conjectured to enhance to $SU(2)$ which is also gauged. The flip fields are added to every other glued free trinion.}
    \label{F:Atypek3Trinion}
\end{figure}

\subsection{The trinion with three maximal punctures}

Let us now make our statement more precise motivated by the heuristic argument above. 
We conjecture that gluing together $k$ free trinions as depicted in Figure \ref{F:Atypek3Trinion} we obtain a trinion with two maximal punctures with symmetry $SU(2)^k$ and a new puncture which is glued by gauging $SU(k)\times SU(2)$, first the $SU(k)$ factor and then the $SU(2)$ factor. To be precise we will also need to add a collection of gauge singlet fields, as depicted in Figure \ref{F:Atypek3Trinion}.\footnote{Note that the trinion for the $k=2$ case suggested in  \cite{Razamat:2016dpl}  does not have these flip fields as these become 
massive once one breaks some of the symmetries by going on the conformal manifold as is done there.}
 This is done so that the structure of the moment map operators \eqref{momentmaps} still remains after we turn on the set of symmetries $U(1)_t\times \prod_{i=1}^k U(1)_{\beta_i}\times U(1)_{\gamma_i}$. These are expected to be the Cartan generators of the internal $SU(2k)$ symmetry. More precisely we decompose $SU(2k)$ to $U(1)_t\times SU(k)_\beta\times SU(k)_\gamma$ such that the characters of the fundamental representations are,
\be
\label{charactersu2k}
&&\chi^{SU(2k)}_\textbf{F} =t\, \chi_\textbf{F}^{SU(k)_\beta}(\beta_i)+t^{-1} \chi_\textbf{F}^{SU(k)_\gamma}(\gamma_j)\,,\nonumber\\
&&\chi_\textbf{F}^{SU(k)_\beta}(\beta_i) =\sum_{i=1}^k \beta_i^2\,,\qquad \chi_\textbf{F}^{SU(k)_\gamma}(\gamma_j)= \sum_{j=1}^k \gamma_j^{2}\,,
\ee
with $\prod_{i=1}^k \beta_i=1$ and $\prod_{j=1}^k \gamma_j=1$. In particular the moment maps have the following charges,
\be
\label{momentmapsfugacities}
(pq)^{\frac12}t^{-1}\left(\chi_{\overline{\textbf{F}}}^{SU(k)_\beta}(\beta_i)c^{\frac2k} \chi_{\textbf{F}}^{\ }({\bf a}) +\chi_{\textbf{F}}^{SU(k)_\gamma}(\gamma_j)\,c^{-\frac2k} \chi_{\overline{\textbf{F}}}({\bf a})\right)\,.
\ee 
We denote the moment map operators as $M_\beta$ and $M_\gamma$ with the former being in the bi-fundamental representation of the puncture $SU(k)$ symmetry and the internal symmetry $SU(k)_\beta$ and the latter in the bi-fundamental representation of the puncture $SU(k)$ symmetry and the internal symmetry $SU(k)_\gamma$. We $\Phi$-glue the new puncture by gauging $SU(k)$ and then $SU(2)$ and introducing operators $\Phi_\beta$ and $\Phi_\gamma$ coupling them to moment map operators as,
\be
W= \sum_{i,J=1}^k \left(\Phi_\beta\right)^i_J\left(\left(M_\beta\right)^J_i-\left(M'_\beta\right)^J_i\right) +  \sum_{i,J=1}^k \left(\Phi_\gamma\right)^i_J\left(\left(M_\gamma\right)^J_i-\left(M'_\gamma\right)^J_i\right)\,.
\ee 
Here $M$ and $M'$ are the moment maps of the two glued punctures and we denote by $i$ the $SU(k)$ puncture index and by $J$ the $SU(k)_{\beta}$ or $SU(k)_\gamma$ index.
We $S$-glue the puncture by turning on the superpotential,
\be
W=\sum_{i,J=1}^k \left(\left(M_\beta\right)^J_i\left(M'_\beta\right)^J_i+\left(M_\gamma\right)^J_i\left(M'_\gamma\right)^J_i\right)\,,
\ee and gauging first the $SU(k)$ symmetry and then the conjecturally emergent $SU(2)$ symmetry.
Using these gluing rules we can now construct theories corresponding to arbitrary surfaces.  

\begin{figure}[t]
	\centering
  	\includegraphics[scale=0.3]{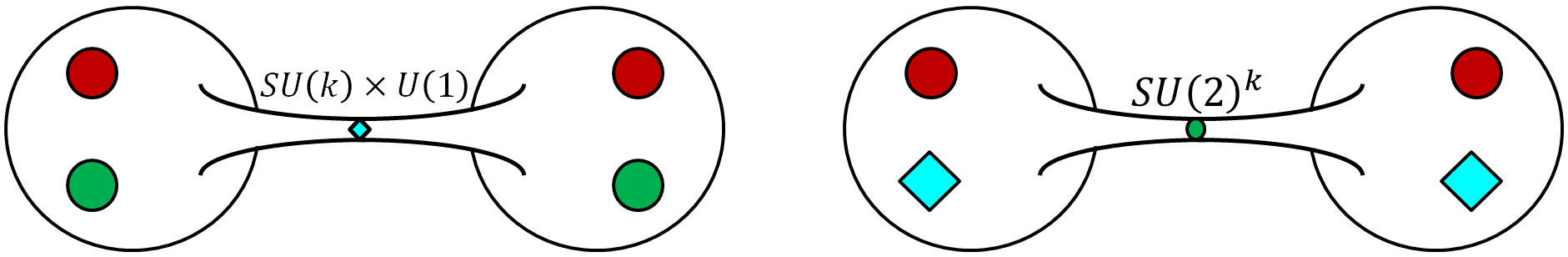}
    \caption{Two types of four punctured spheres. If the suggestion is correct the corresponding models should enjoy dualities exchanging same types of punctures.}
    \label{F:AtypeDualities}
\end{figure}

For example the basic four punctured spheres then has either two $SU(2)^k$ punctures and two $SU(k)\times U(1)$ punctures or four $SU(2)^k$ punctures depending which puncture is used to glue two trinions together. 
Consistency of our proposal implies that the four punctured spheres should enjoy duality exchanging punctures of the same type, See Figure \ref{F:AtypeDualities}. This can be verified computing the supersymmetric index for each type of the four punctured sphere at least in expansion in fugacities.

Typically one can understand the allowed punctures in compactifications of $6d$ SCFTs on Riemann surfaces by studying first reductions on a circle, as we already mentioned. One can thus wonder what our novel puncture corresponds to. In particular we are after a five dimensional theory obtained by circle compactification of the next to minimal $A$ type conformal matter which has $SU(k)$ and $SU(2)$ gauge groups. There is a natural candidate for this appearing in \cite{Ohmori:2015pia}. There it's claimed that by reducing the $6d$ SCFT residing on $N$ $M5$ branes probing a $\Z_k$ singularity on a circle with no holonomies for the global symmetry, one obtains a five dimensional theory built by gauging an $SU(N)$ flavor symmetry group of certain $5d$ SCFT. This $5d$ SCFT has a relevant deformation which takes it to a gauge theory with $SU(k)^{N-1}$ gauge group. This deformation breaks the $SU(N)$ flavor symmetry to the Cartans and thus to gauge it we need to switch it off by going to the SCFT point. This echoes the structure we see in four dimension for $N=2$. In particular it suggests that if one is to understand the generalization of our statements for non minimal conformal matter one could look for $SU(k)^{N-1}\times U(1)^{N-1}$ punctures such that gluing is performed by first gauging the $SU(k)^{N-1}$ symmetry and then the $U(1)^{N-1}$ enhances to $SU(N)$ and subsequently gauged.\footnote{We are grateful to Gabi Zafrir for suggesting this connection.}

In this section we have argued that combining $k$ minimal punctures leads to a new maximal puncture of symmetry $SU(k)\times U(1)$ such that we find a trinion with three maximal punctures with minimal flux. We discussed a concrete procedure of $S$-gluing together $k$ free trinions. As the minimal punctures of the $S$-glued trinions are of different type\footnote{Minimal punctures of a different sign in notations of \cite{Razamat:2016dpl}.} we needed to add appropriate flip fields. As the procedure of combining punctures to a maximal one is local on the Riemann surfaces we could define a trinion with the novel maximal puncture by say $\Phi$-gluing $k$ free trinion, albeit this trinion would have flux $1/2$. The computation of the index is consistent with the symmetry of $\ell$ minimal punctures of the same type  enhancing in the IR somewhere on the conformal manifold to $SU(\ell)\times U(1)$. When $\ell = k$ we find the above described maximal puncture, and when $\ell<k$ we can interpret this as an intermediate puncture that one can find by partially closing an $SU(k)\times U(1)$ maximal puncture. In the case of $\ell>k$ we have no interpretation as a puncture.

\subsection{Constructing Riemann surfaces with the new trinion}

\begin{figure}[t]
	\centering
  	\includegraphics[scale=0.2]{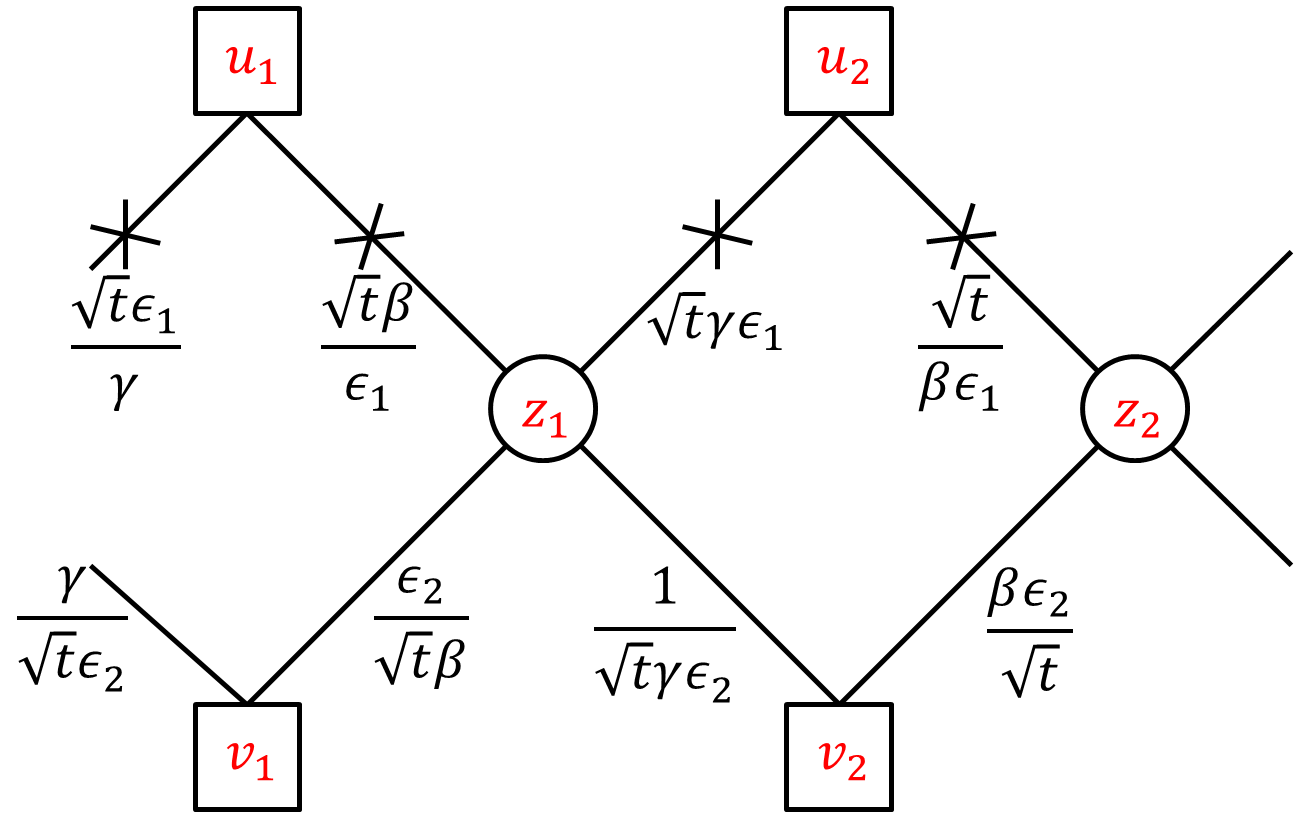}
    \caption{The trinion with three maximal punctures. Here depicted is the case of $k=2$.}
    \label{F:Atypek2Trinion}
\end{figure}

Let us now discuss some properties of theories corresponding to closed Riemann surfaces constructed from the conjectured trinions. One issue we still need to settle is what is the flux corresponding to the trinion and we also need to verify that the symmetry properties of the theories are consistent with this flux. To do this we need to distinguish the case of even $k$ as opposed to odd $k$.
From the definition of the trinion one can see that the two cases are qualitatively different. If $k$ is even then the number of free trinions with flipped baryons and with unflipped baryons is the same (see Figure \ref{F:Atypek2Trinion}), while if $k$ is odd we have one more free trinion with flipped baryons (see Figure \ref{F:Atypek3Trinion}). The free trinions correspond to compactifications on punctured spheres with flux $\pm \frac1{2k}$ for $U(1)_t$ symmetry (we use the normalization of \cite{Razamat:2018zus}). Thus as these are $S$-glued to each other the flux in the even case is zero while in the odd case it is $\pm\frac1{2k}$ and we will use the negative sign for concreteness.
These are the expected fluxes of the trinions with the three maximal punctures we conjecture. 

The simplest check to verify this is to compute the anomalies of theories corresponding to closed Riemann surfaces obtained by $\Phi$-gluing the trinions so that the flux will be $(2g-2) \mathcal{F}_{trinion}$. Here $\mathcal{F}_{trinion}$ is the flux associated with the trinion. It is straightforward to compute the anomalies for the field theoretic construction. It is convenient to package them into trial $a$ and $c$ anomalies \cite{Intriligator:2003jj} by defining a trial $R$ symmetry. Here for the sake of brevity we will  define the trial $R$ symmetry by admixing only the $U(1)_t$ symmetry as it is sufficient to fix the value of the flux. We thus take $R$ to   $R+e\, q_t$ with $q_t$ being the charge under $U(1)_t$ and $e$ an arbitrary number. Then the trial $a$ and $c$ anomalies for the even $k$ case are,
\be
&&a(e) =\frac{3}{16} (g-1) \left(2-\left(9e^2-8\right) k^2\right)\,,\nonumber\\
&&c(e) = \frac{1}{16} (g-1) \left(4-\left(27 e^2-25\right) k^2\right)\,,
\ee 
matching perfectly with the anomalies predicted from six dimensions for $U(1)_t$ flux $\mathcal{F}_{trinion}=0$ genus $g$ Riemann surface. These 't Hooft anomalies were obtained by integrating the $6d$ anomaly polynomial on a genus $g$ Riemann surface. The relevant anomalies are detailed in Appendix \ref{A:Atypeanoms}. For the odd $k$ case the anomalies computed from field theory are,
\be
&&a(e) =\frac{3}{16} (g-1) \left(12 e^3 k-9 e^2 k^2-10 e k+8 k^2+2\right)\,,\nonumber\\
&&c(e) = \frac{1}{16} (g-1) \left(36 e^3 k-27 e^2 k^2-32 e k+25 k^2+4\right)\,.
\ee 
These match the six dimensional computations when setting the $U(1)_t$ flux to $F_{trinion}=-\frac1{2k}$. These fluxes are precisely as expected above.

A further check of the flux assignment is to compute the supersymmetric index. 
Computing the index by gluing the contributions of the trinions together one can explicitly observe that gauging the $SU(k)$ symmetry of the novel puncture causes the fugacities of the $U(1)_c$ symmetry to form representations of $SU(2)_c$. This is consistent with the assumption that the symmetry enhances. In particular, considering the expansion of the index in fugacities one can observe it is invariant under $c\to 1/c$. This is a necessary property in order to get the $U(1)$ enhancement to $SU(2)$ following the Weyl symmetry of the latter, but it is not a sufficient property. Thus, this is a check of the enhancement but not a proof. 
In the case of even $k$ as the flux is zero the $\Phi$- and the $S$- gluings should lead to equivalent theories and in particular the index should be the same. Indeed using either gluings we find that to leading order the index is given by,
\be\label{fluxzeroindex}
{\cal I}=1+\left(\chi_{\textbf{Adj}}^{SU(2k)}(\beta_i,\gamma_j,t)(g-1)+3g-3\right)pq+\cdots\,, 
\ee
where,
\be
\chi_{\textbf{Adj}}^{SU(2k)}(\beta_i,\gamma_j,t)=\chi_{\textbf{F}}^{SU(2k)}(\beta_i,\gamma_j,t)\;\chi_{\overline{\textbf{F}}}^{SU(2k)}(\beta_i,\gamma_j,t)-1\,.
\ee $\chi_{\textbf{F}}^{SU(2k)}$ is defined in \eqref{charactersu2k}.
First, this is exactly the index we expect to obtain using the general expectations \eqref{expindex} when the flux is zero. Second, the fugacities for all the $U(1)$ internal symmetries arrange themselves into characters of $SU(2k)$, the expected symmetry of the theory. As the flux is zero we expect that the four dimensional models will have a locus on their conformal manifold with the full internal symmetry apparent. Moreover, for the case of $k=2$, where the $SU(4)$ symmetry is expected to enhance to $SO(7)$ \cite{DelZotto:2015rca,Ohmori:2015pia}, one finds that the index is for generic $g$,\footnote{For genus two there are additional $4$ marginal deformations. These can be also seen using the description of \cite{Razamat:2016dpl} , however they have been regretfully missed in the said reference.}
\be
{\cal I}=1+\left(\chi_{\textbf{Adj}}^{SO(7)}(\beta,\gamma,t)(g-1)+3g-3\right)pq+\cdots\,, 
\ee
where,
\be
\chi_{\textbf{Adj}}^{SO(7)}(\beta,\gamma,t)=\chi_{\textbf{Adj}}^{SU(4)}(\beta,\gamma,t)+\chi_{\textbf{AS}}^{SU(4)}(\beta,\gamma,t)\,.
\ee Here $\textbf{AS}$ stands for two index six dimensional antisymmetric representation of $SU(4)$. This is yet another non trivial check of our procedure.

Next for odd $k$ we find that $S$ gluing together $2g-2$ trinions the index is as in \eqref{fluxzeroindex}, this is expected as the total flux of the theory is zero: we have $g-1$ trinions with flux $+\frac1{2k}$ and $g-1$ with flux $-\frac1{2k}$. However when we $\Phi$-glue $2g-2$ trinions to form closed genus $g$ surface the index takes the form,
\be
{\cal I} & = & 1+ pq\left(3g-3+(g-1)\left(1+\chi_{\textbf{Adj}}^{SU(k)}(\beta_i)+\chi_{\textbf{Adj}}^{SU(k)}(\gamma_j)\right)\right.\\
&&\left.+\left(-2\frac{2g-2}{2k}+g-1\right)t^{-2}\chi_{\overline{\textbf{F}}}^{SU(k)}(\beta_i)\chi_{\overline{\textbf{F}}}^{SU(k)}(\gamma_j)\right.\nonumber\\
&&\left. +\left(2\frac{2g-2}{2k}+g-1\right)t^{2}\chi_{\textbf{F}}^{SU(k)}(\beta_i)\chi_{\textbf{F}}^{SU(k)}(\gamma_j)\right)+\cdots, \nonumber
\ee 
again precisely as expected from \eqref{fluxindex} assuming the flux of the trinion is $-\frac1{2k}$. Note that if $g-1$ is not an integer multiple of $k$ gluing the trinions together will break 
some of the internal symmetries due to anomalies. This can be understood as in these cases there are no punctures but the flux is not integer. The proper way to think about these models is that one turns on a non trivial $w_2$ Stiefel-Whitney class supported on the surface. See \cite{Bah:2017gph} for a discussion of this effect. In particular if $g-1$ is not multiple of $k$ in the above expression for the index the fugacities for broken symmetries have to be set to one.  One can further check the proposal by changing the value of flux by also admixing two punctured spheres with flux derived in \cite{Bah:2017gph,Kim:2018lfo}.

\section{Discussion}\label{sec:discussion}

Let us briefly summarize our results and make several comments. We have derived four dimensional theories corresponding to compactifications of  $6d$ minimal $(D_{N+3},D_{N+3})$ conformal matter SCFTs on a three punctured sphere, with two maximal and one minimal puncture. This was done by using the RG arguments relating compactifications of {\it different} SCFTs on {\it different } Riemann surfaces.  The maximal punctures in the derived theory had an $SU(2)^N$ symmetry while the minimal punctures had an $SU(2)$ symmetry.
We then observed that $N$ minimal punctures can be combined into another type of maximal puncture with symmetry $USp(2N)$. Thus, a trinion with three maximal punctures can be constructed allowing to study compactifications on any surface. Finally we argued that a similar procedure works also for $A_{k-1}$ type next to minimal conformal matter SCFTs. In compactifications of these SCFTs to $4d$, a collection of $k$ minimal $U(1)$ punctures can be combined into a new type of maximal puncture with $SU(k)\times U(1)$ symmetry. We have discussed how to glue such punctures to obtain theories corresponding to general Riemann surfaces. 

The constructions in this paper have several possible generalizations. First, the same type of flows can be used to study surfaces with minimal punctures for not necessarily minimal $D$-type conformal matter $6d$ SCFTs. The theories corresponding to spheres with two maximal punctures and flux are known \cite{Kim:2018lfo} and thus, using the flows, theories with minimal punctures can be constructed. Here the gauge groups appearing in the four dimensional constructions are not just $SU(2)$ and thus the analysis is more involved but we expect it to be doable. In our analysis the minimal punctures could be related to maximal ones because the baryons and mesons are on the same footing for $SU(2)$. This will not be the case for higher rank unitary groups and thus we expect such a relation, if any, to be more involved.

In principle we have at the moment an algorithm to understand compactifications of a generic $6d$ SCFT (see \cite{Heckman:2018jxk} for a recent beautiful review of $6d$ SCFTs) down to four dimensions. First, one should understand the compactification of the $6d$ SCFTs on a circle to five dimensions. A lot of progress have been made on this front in recent years \cite{Zafrir:2015rga,Bhardwaj:2018yhy,Hayashi:2015vhy,Bhardwaj:2018vuu,Apruzzi:2019vpe,Apruzzi:2019opn,Apruzzi:2019enx,Ohmori:2015pia,Ohmori:2015tka,Hayashi:2015zka,Hayashi:2015fsa,Hayashi:2016abm,Jefferson:2017ahm,Jefferson:2018irk,Ohmori:2018ona,Bhardwaj:2019fzv,Razamat:2018gbu,Yonekura:2015ksa}. In case the five dimensional theory has a gauge theory effective description it teaches us about the punctures we can have and how to glue four dimensional theories together. The gauge symmetry in five dimensions becomes the puncture symmetry in four dimensions while gluing punctures is done by gauging this symmetry. Different five dimensional theories related by UV dualities can be useful to define different types of punctures. Thus understanding five dimensional dualities should be of great interest, see for example references above and \cite{Bergman:2013aca,Bergman:2014kza,Bhardwaj:2019ngx}.
 Then one can study domain walls in these five dimensional models which can lead to understanding of compactifications of six dimensional SCFTs on two punctured spheres and tori with flux \cite{Kim:2017toz,Kim:2018bpg,Kim:2018lfo}. Once we understand such compactifications we can use the RG flows connecting different SCFTs in six dimensions to introduce more punctures. In favorable situations we can construct from these simple punctures maximal ones. This will produce surfaces with more than two maximal punctures from which general compactifications can be constructed.

For the above general program to succeed a lot of different parts need to be better understood. For example in addition to understanding the reductions of $6d$ SCFTs mentioned above, we need to better understand the constructions of punctures (see {\it e.g.} \cite{Heckman:2016xdl,Hassler:2017arf}), of six dimensional RG flows (see {\it e.g.} \cite{Heckman:2018pqx}), and of five dimensional  domain walls \cite{Gaiotto:2015una} (see {\it e.g.} for lower dimensional discussions \cite{Dimofte:2017tpi}). We leave these exciting problems for future research.

Yet another interesting venue for the use of our results is to derive integrable systems associated to the minimal $D$ conformal matter.  Many relations between supersymmetric gauge theories and quantum mechanical models are known, most notably in the case of extended supersymmetry, {\it e.g.} \cite{Nekrasov:2009rc}. One way to obtain
such integrable systems  is by studying surface defects in the four dimensional theories \cite{Gaiotto:2012xa}. For example, in case of $(2,0)$ theories one obtains  the Ruijsenaars-Schneider model \cite{Gaiotto:2012xa} and in case of E-string the van-Diejen model \cite{Nazzal:2018brc}. In case of $A$-type conformal matter the models were discussed in \cite{Gaiotto:2015usa,Maruyoshi:2016caf,Yagi:2017hmj,Ito:2016fpl}.  It would be very interesting to further understand this map between six dimensional SCFTs and quantum mechanical integrable models.

\section*{Acknowledgments}

We are grateful to Gabi Zafrir for useful discussions and comments.
The research of SSR and ES is supported by Israel Science Foundation under grant no. 2289/18 and by I-CORE  Program of the Planning and Budgeting Committee.
SSR is grateful to the Simons Center for Geometry and Physics  for hospitality during initial stages of this project.

\vspace{10pt}
\begin{appendix}
\vspace{10pt}

\section{$\mathcal{N}=1$ superconformal index}\label{A:indexdefinitions}
In this appendix we give a short introduction of the $\mathcal{N}=1$ superconformal index \cite{Kinney:2005ej,Romelsberger:2005eg,Dolan:2008qi}, some related notations, and usful results. For more comprehensive explanations and definitions see \cite{Rastelli:2016tbz}.
An SCFT index is defined as the Witten index of the theory in radial quantization. In four dimensions it is defined as a trace over the Hilbert space of the theory quantized on $\mathbb{S}^3$,
\be
\mathcal{I}\left(\mu_i\right)=Tr(-1)^F e^{-\beta \delta} e^{-\mu_i \mathcal{M}_i},
\ee
where $\delta\triangleq \half \left\{\mathcal{Q},\mathcal{Q}^{\dagger}\right\}$, with $\mathcal{Q}$ one of the Poincar\'e supercharges, and $\mathcal{Q}^{\dagger}=\mathcal{S}$ its conjugate conformal supercharge, $\mathcal{M}_i$ are $\mathcal{Q}$-closed conserved charges and $\mu_i$ their associated chemical potentials. The non-vanishing contributions came from states with $\delta=0$ making the index independent on $\beta$, since states with $\delta>0$ come in boson/fermion pairs.

For $\mathcal{N}=1$, the supercharges are $\left\{\mathcal{Q}_{\alpha},\,\mathcal{S}^{\alpha} \triangleq \mathcal{Q}^{\dagger\alpha},\,\widetilde{\mathcal{Q}}_{\dot{\alpha}},\,\widetilde{\mathcal{S}}^{\dot{\alpha}} \triangleq \widetilde{\mathcal{Q}}^{\dagger\dot{\alpha}}\right\}$, with $\alpha=\pm$ and $\dot{\alpha}=\dot{\pm}$ the respective $SU(2)_1$ and $SU(2)_2$ indices of the isometry group of $\mathbb{S}^3$ ($Spin(4)=SU(2)_1 \times SU(2)_2$).
Since different choices of $\mathcal{Q}$ in the definition of the index lead to physically equivalent indices, we choose $\mathcal{Q}=\widetilde{\mathcal{Q}}_{\dot{-}}$. Under this choice the index formula takes the form,
\be
\mathcal{I}\left(p,q\right)=Tr(-1)^F p^{j_1 + j_2 +\half r} q^{j_2 - j_1 +\half r}.
\ee
where $j_1$ and $j_2$ are the Cartan generators of $SU(2)_1$ and $SU(2)_2$, and $r$ is the generator of the $U(1)_r$ R-symmetry.

The index is computed by listing all gauge singlet operators one can construct from modes of the fields. The modes and operators are conventionally called "letters" and "words", respectively. The single-letter index for a vector multiplet and a chiral multiplet transforming in the $\mathcal{R}$ representation of the gauge$\times$flavor group is,
\be
i_V \left(p,q,U\right) & = & \frac{2pq-p-q}{(1-p)(1-q)} \chi_{adj}\left(U\right), \nonumber\\
i_{\chi(r)}\left(p,q,U,V\right) & = & 
\frac{(pq)^{\half r} \chi_{\mathcal{R}} \left(U,V\right) - (pq)^{\frac{2-r}{2}} \chi_{\overline{\mathcal{R}}} \left(U,V\right)}{(1-p)(1-q)},
\ee
where $\chi_{\mathcal{R}} \left(U,V\right)$ and $\chi_{\overline{\mathcal{R}}} \left(U,V\right)$ denote the characters of $\mathcal{R}$ and the conjugate representation $\overline{\mathcal{R}}$, with $U$ and $V$ gauge and flavor group matrices, respectively.

With the single letter indices at hand, we can write the full index by listing all the words and projecting them to gauge singlets by integrating over the Haar measure of the gauge group. This takes the general form
\be
\mathcal{I} \left(p,q,V\right)=\int \left[dU\right] \prod_{k} PE\left[i_k\left(p,q,U,V\right)\right],
\ee
where $k$ labels the different multiplets in the theory, and $PE[i_k]$ is the plethystic exponent of the single-letter index of the $k$-th multiplet, responsible for listing all the words. The plethystic exponent is defined by
\be
PE\left[i_k\left(p,q,U,V\right)\right] \triangleq \exp \left\{ \sum_{n=1}^{\infty} \frac{1}{n} i_k\left(p^n,q^n,U^n,V^n\right) \right\}.
\ee
First focusing on the case of $SU(N_c)$ gauge group. The full contribution for a chiral superfield in the fundamental representation of $SU(N_c)$ with R-charge $r$ can be written in terms of elliptic gamma functions, as follows
\be
PE\left[i_k\left(p,q,U\right)\right] & \equiv & \prod_{i=1}^{N_c} \Gamma_e \left((pq)^{\half r} z_i \right), \nonumber \\
\Gamma_e(z)\triangleq\Gamma\left(z;p,q\right) & \equiv & \prod_{n,m=0}^{\infty} \frac{1-p^{n+1} q^{m+1}/z}{1-p^n q^m z},
\ee
where $\{z_i\}$ with $i=1,...,N_c$ are the fugacities parameterizing the Cartan subalgebra of $SU(N_c)$, with $\prod_{i=1}^{N_c} z_i = 1$. In addition, in many occasions we will use the shorten notation
\be
\Gamma_e \left(u z^{\pm n} \right)=\Gamma_e \left(u z^{n} \right)\Gamma_e \left(u z^{-n} \right).
\ee

In a similar manner we can write the full contribution of the vector multiplet in the adjoint of $SU(N_c)$, together with the matching Haar measure and projection to gauge singlets as
\be
\frac{\kappa^{N_c-1}}{N_c !} \oint_{\mathbb{T}^{N_c-1}} \prod_{i=1}^{N_c-1} \frac{dz_i}{2\pi i z_i} \prod_{k\ne \ell} \frac1{\Gamma_e(z_k/z_\ell)}\cdots,
\ee
where the dots denote that it will be used in addition to the full matter multiplets transforming in representations of the gauge group. The integration is a contour integration over the maximal torus of the gauge group. $\kappa$ is the index of $U(1)$ free vector multiplet defined as
\be
\kappa \triangleq (p;p)(q;q),
\ee
where
\be
(a;b) \triangleq \prod_{n=0}^\infty \left( 1-ab^n \right)
\ee
is the q-Pochhammer symbol.

As we also use $USp(2N)$ groups in this paper we will write the contributions to the index for such groups as well. The contribution for a chiral superfield in the fundamental representation of $USp(2N)$ with R-charge $r$ is given by
\be
\prod_{a=1}^N \Gamma_e\left(\left(pq\right)^{\half r}z_a^{\pm1}\right),
\ee
where ${z_a}$ with $a=1,...,N$ being the fugacities parameterizing the Cartan subalgebra of $USp(2N)$. The full contribution of the vector multiplet in the adjoint of $USp(2N)$, with the matching Haar measure and projection to gauge singlets is
\be
\frac{\kappa^{N}}{2^N N!} \oint_{\mathbb{T}^{N}} \prod_{a=1}^{N} \frac{dz_a}{2\pi i z_a} \frac1{\Gamma_e\left(z_a^{\pm 2}\right)} \prod_{1<a<b<N} \frac1{\Gamma_e\left(z_a^{\pm 1} z_b^{\pm 1}\right)}\cdots,
\ee
where again the dots are to be replaced with the full matter multiplets transforming in representations of the gauge group, and the integration is over the maximal torus of the gauge group. 

\section{Duality proof of symmetry enhancement}\label{sec:duality} 

Let us first discuss a proof of the fact that the $U(1)_\epsilon$ symmetry of the $N=1$ $D$-type trinion enhances to $SU(2)$ in the IR using Seiberg dualities. The proof for $N>1$ should follow similar arguments and we refrain from detailing it here, leaving the proof to an enthusiastic reader.

\begin{figure}[htbp]
	\centering
  	\includegraphics[scale=0.25]{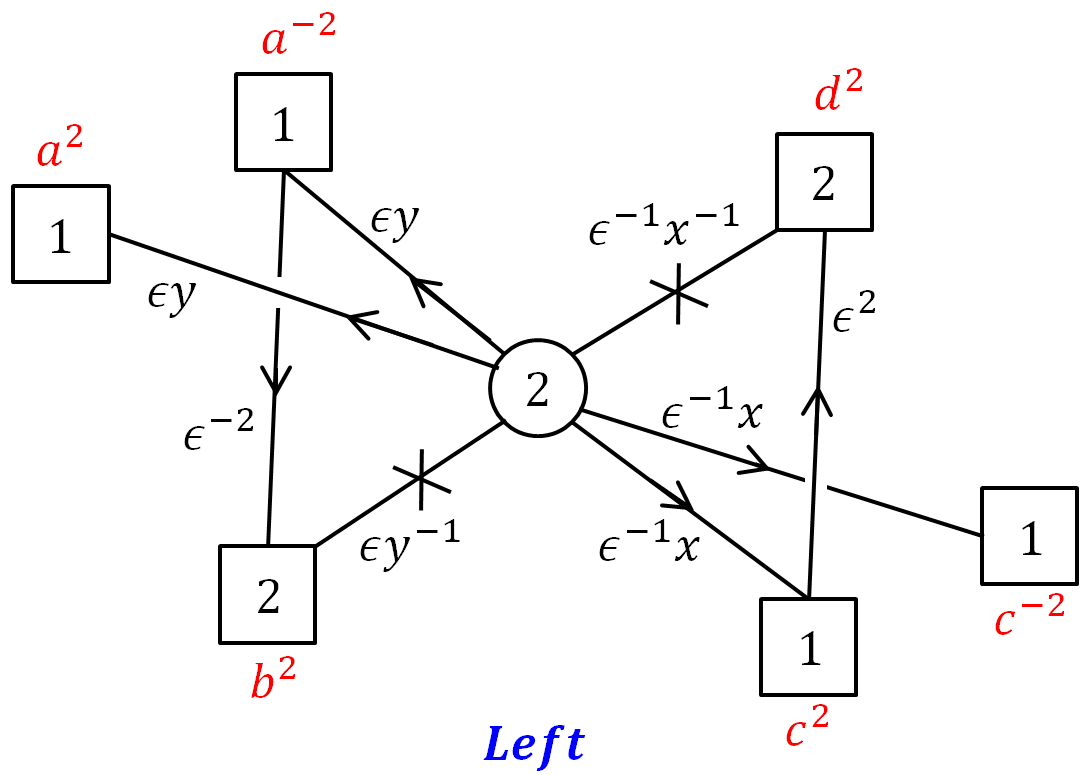}\;\;\;\;\; \includegraphics[scale=0.25]{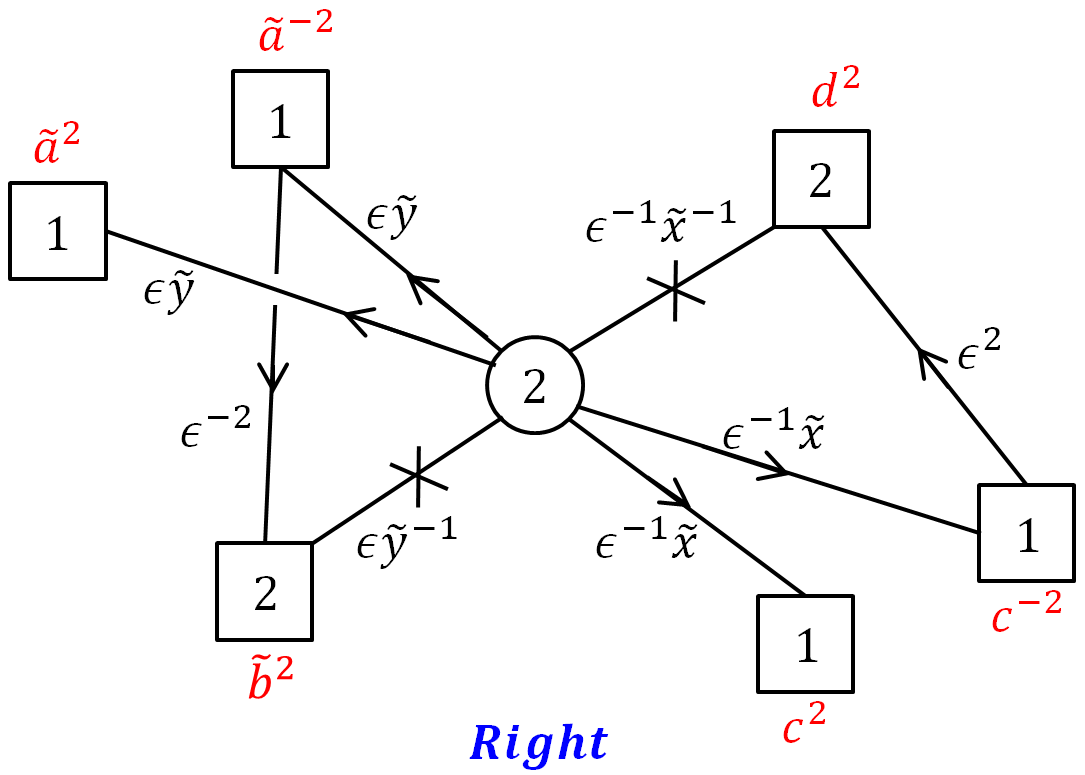}
    \caption{Decomposition of the trinion into left and right part. The two parts are glued with superpotential which identifies the symmetries as indicated. We will study Seiberg duality sequences of each half. Note the two halves are almost identical modulo different gauge singlet fields charged under $c$ and $d$ symmetries. R-charges of gauge singlet fields are $1$ and fields transforming under the gauge $SU(2)$ have R charge $\frac12$.}
    \label{F:original}
\end{figure}

 First, the trinion consists of two halves, two $SU(2)$ gauge sectors which are glued together using a superpotential. We will apply same sequence of Seiberg dualities on each one of the halves. For the sake of future convenience we decompose the trinion into two halves as depicted in Figure \ref{F:original}.\footnote{The map of the charges between Figure \ref{F:original} to Figure \ref{F:EStringTrinion} is: $(c^2,d^2,\epsilon^2|x,a^2y,a^{-2}y, b^2 y^{-1}, b^{-2} y^{-1}|\widetilde x,\widetilde a^2 \widetilde y ,\widetilde a^{-2} \widetilde y,\widetilde b^2\widetilde y^{-1},\widetilde b^2\widetilde y^{-1})\to
 (v,z,\epsilon^{-1}| t a^{-1}, c_1,c_4,c_2,c_3| t a, \widetilde c_1,\widetilde c_4,\widetilde c_2, \widetilde c_3)$. The halves are glued by identifying the $d$ and the $c$ symmetries. }

\begin{figure}[htbp]
	\centering
  	\includegraphics[scale=0.27]{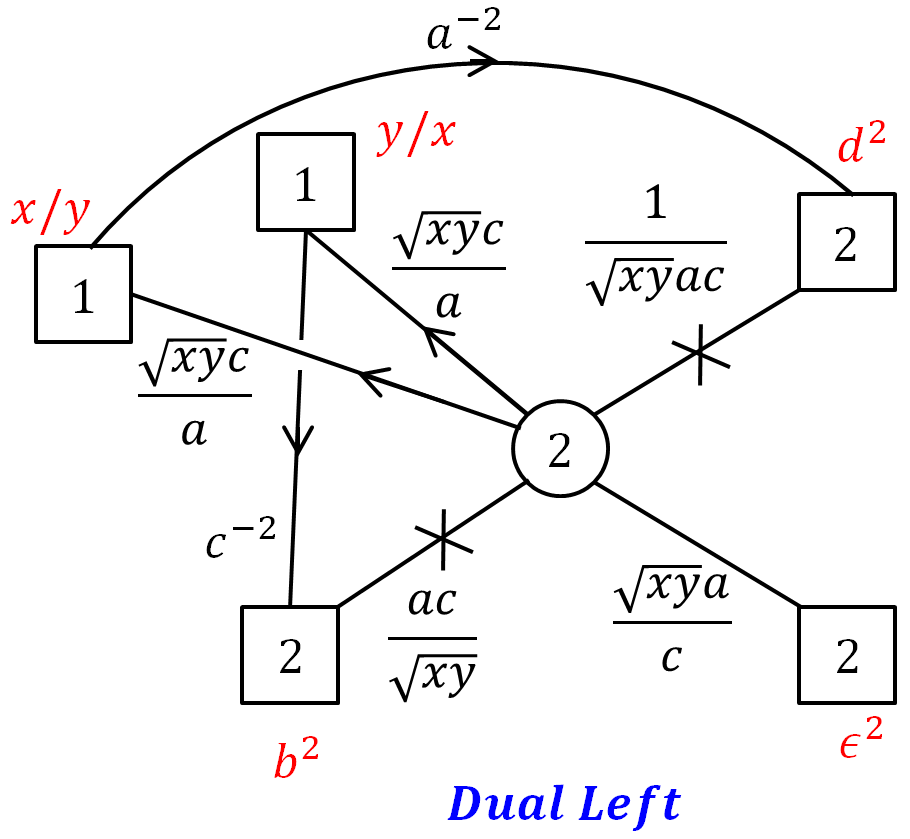}\qquad \includegraphics[scale=0.27]{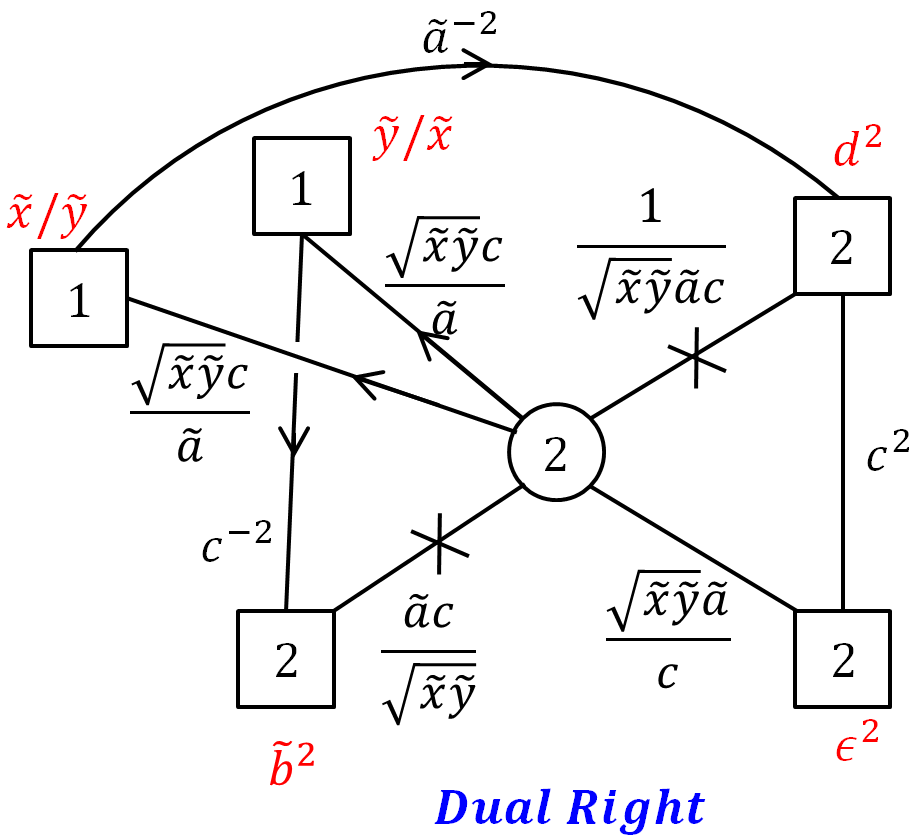}
    \caption{The duals after the sequence of Seiberg dualities of the left and the right part. The two parts are glued identifying $\epsilon$ and $d$ symmetries. Note that now $\epsilon$ symmetry is manifestly $SU(2)$ while $c$ is $U(1)$. }
    \label{F:originaldual}
\end{figure}

Let us focus on the left half and apply certain sequence of dualities. Precisely same sequence will be applied to the right half. The gauge node has an octet of $SU(2)$ fundamentals which have charges,
\be
\left\{ \frac{c^2 x}{\epsilon },\frac{x}{c^2 \epsilon },\frac{d^2}{x \epsilon },\frac{1}{d^2 x \epsilon },a^2 y \epsilon ,\frac{y \epsilon }{a^2},\frac{b^2 \epsilon }{y},\frac{\epsilon }{b^2 y}\right\}\,.
\ee 
To perform Seiberg duality we need to divide this octet into quartet of quarks $Q$ and quartet of antiquarks $\widetilde Q$. We do so first in the following manner,
\be
\{Q\; |\;\widetilde Q\}_1 = \left\{\frac{c^2 x}{\epsilon },\frac{d^2}{x \epsilon },a^2 y \epsilon ,\frac{b^2 \epsilon }{y}\;|\;\frac{x}{c^2 \epsilon },\frac{y \epsilon }{a^2},\frac{1}{d^2 x \epsilon },\frac{\epsilon }{b^2 y}\right\}\,.
\ee 
We will have two Seiberg dualities and the notation $\{Q\; |\;\widetilde Q\}_k$ indicates the division into quarks and antiquarks for $k$-th duality. We apply Seiberg duality which has two effects. First it changes charges of the fundamental fields and second it produces a collection of gauge singlet fields with the same charges as the mesons. We will not detail the mesons here for the sake of brevity and only quote their charges in the end of the sequence of the dualities. The charges of the dual quarks are,
\be
\{q\; |\;\widetilde q\}_1 = \left\{\frac{a b d \epsilon }{c x},\frac{a b c x \epsilon }{d},\frac{b c d}{a y \epsilon },\frac{a c d y}{b \epsilon }\;|\;\frac{c \epsilon }{a b d x},\frac{a}{b c d y \epsilon },\frac{d x \epsilon }{a b c},\frac{b y}{a c d \epsilon }\right\}\,.
\ee 
We rearrange the quarks again as,
\be
\{Q\; |\;\widetilde Q\}_2 = \left\{\frac{d x \epsilon }{a b c},\frac{a b c x \epsilon }{d},\frac{b c d}{a y \epsilon },\frac{a b d \epsilon }{c x}\;|\;\frac{a}{b c d y \epsilon },\frac{a c d y}{b \epsilon },\frac{c \epsilon }{a b d x},\frac{b y}{a c d \epsilon }\right\}\,,
\ee 
and apply Seiberg duality to get,
\be
\{q\; |\;\widetilde q\}_2= \left\{\frac{a b^2 c}{\sqrt{x y}},\frac{d^2}{a c \sqrt{x y}},\frac{a \epsilon ^2 \sqrt{x y}}{c},\frac{c x \sqrt{x y}}{a y}\;|\;\frac{c y \sqrt{x y}}{a x},\frac{1}{a c d^2 \sqrt{x y}},\frac{a \sqrt{x y}}{c \epsilon ^2},\frac{a c}{b^2 \sqrt{x y}}\right\}\,.
\ee 
The result is depicted in Figure \ref{F:originaldual} where we also included all the gauge singlet fields. 

We perform exactly the same duality sequence on the right half. As the two halves differ slightly by gauge singlet fields the final result differs also by the same. We can now combine the two halves together and observe that the Lagrangian has $\epsilon$ symmetry manifestly as $SU(2)$. This implies that the conformal manifold has loci on which any two of the three puncture symmetries are $SU(2)$  with the only assumption being validity of Seiberg duality.  It is then plausible that these loci intersect and at the intersection locus all three symmetries are $SU(2)$. The index identity following from Seiberg duality was proven mathematically in \cite{rains}.
This proof is similar in spirit to say arguing that $SU(3)$ SQCD with five flavors and mesons flipped has $SU(10)$ symmetry, though only $SU(5)\times SU(5)\times  U(1)$ is visible in the Lagrangian, as it is dual to $SU(2)$ SQCD with five flavors. 

To get further insights into the enhancement we can compute the index with superconformal R-symmetry.  We can think of the model as first taking two copies of $SU(2)$ SQCD with four flavors, and then the superconformal R-symmetry of all the chiral fields is $\frac12$, and next gluing them together through superpotentials. As superpotentials are cubic and involve free flip fields they are relevant. Performing a-maximization we obtain that the only symmetries which mix with the R-symmetry are the $t$, $c_4$, and $\widetilde c_4$. In particular the superconformal R-symmetry is,

\be\label{mixtrinion}
R+\frac{1}{6} \left(3-\sqrt{10}\right) q_t+ \left(\frac12-\frac1{ \sqrt{3}}\right) \left(q_{c_4}+q_{\widetilde c_4}\right)\,.
\ee All the gauge invariant operators are above the unitarity bound. The index at order $pq$ captures the marginal operators minus the conserved currents. Computing it we find,

\be
{\bf 3}_\epsilon \times {\bf 3}_v\times {\bf 3}_z -{\bf 3}_\epsilon-{\bf 3}_v-{\bf 3}_z -{\bf 8}_c -{\bf 8}_{\widetilde c}-1-1-1-1\,.
\ee 
Here the interpretation is that all the terms with negative signs are conserved currents. We have explicitly the three puncture $SU(2)$s, two $SU(3)$s parametrized by $c_{1,2,3}$ and by $\widetilde c_{1,2,3}$, and four $U(1)$'s, the $U(1)_t$, $U(1)_a$, $U(1)_{c_4}$, and  $U(1)_{\widetilde c_4}$. Note that this index is consistent with a theory that has a point on the conformal manifold where the three puncture symmetries are $SU(2)$ and the marginal operators are in ${\bf 3}_\epsilon \times {\bf 3}_v\times {\bf 3}_z$. It is also consistent with all three $SU(2)$'s broken to the Cartan, but we cannot make sense of it if only two puncture symmetries are $SU(2)$ and the third one is $U(1)$. Thus we conclude that, with the caveat that there are no accidental symmetries and we flow to an SCFT, indeed there is a locus on the conformal manifold with three puncture $SU(2)$ symmetries.

\section{Duality proof of exchanging minimal punctures}\label{A:MinPuncDuality}
In this appendix we will prove, at least at the level of supersymmetric indices, using Seiberg duality and S-duality that two minimal punctures with $SU(2)$ symmetry can be exchanged when gluing two $N=1$ D-type trinions. We do not show this is true for $N>1$ as such a proof becomes increasingly more complicated as $N$ increases, but we expect the property of exchanging two minimal punctures is true for these cases as well by indirect ways.  One can show all anomalies related to the  punctures match and that the index is symmetric under the exchange of punctures. For this proof we  use $S$-gluing. In addition, we will ignore all the flip fields of the two glued trinions as these are symmetric under the exchange of the two minimal puncture symmetries.

\begin{figure}[t]
	\centering
  	\includegraphics[scale=0.4]{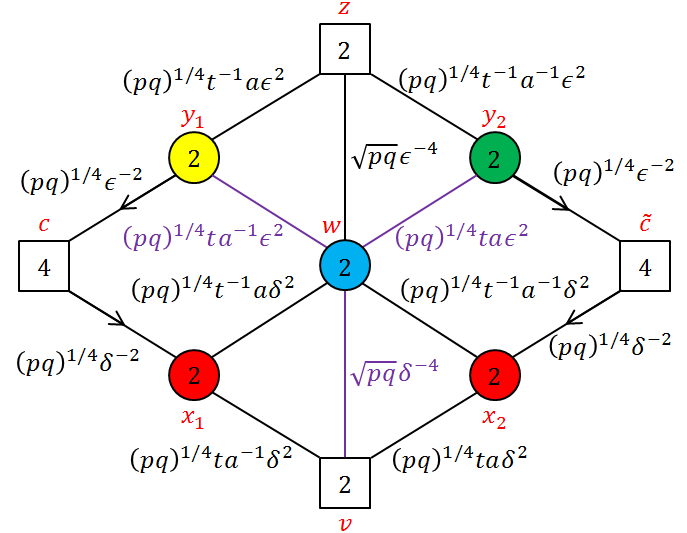}\ \  \includegraphics[scale=0.4]{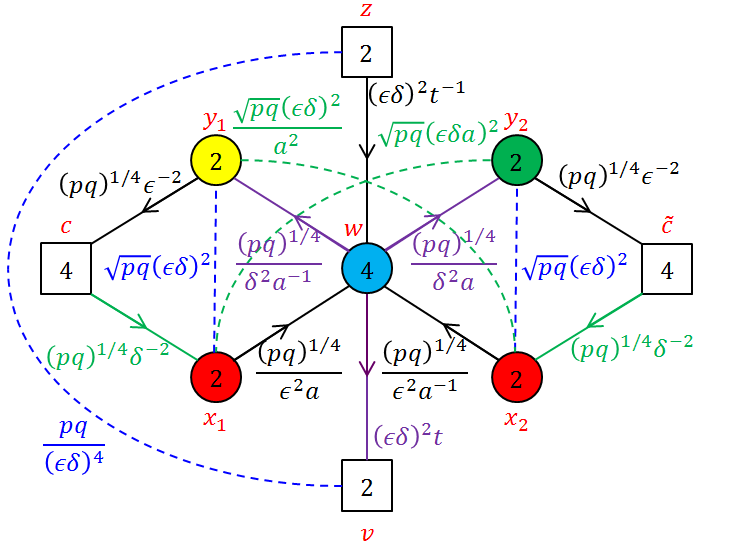}
    \caption{Quivers of two E-string trinions $S$-glued together. On the right we have the initial two glued trinions. Using Seiberg duality on an $SU(2)$ gauge symmetry we need to decide how to divide the fields to fundamental and anti-fundamental. We mark in black the fields we consider as fundamentals and in purple the ones considered anti-fundamentals of the $SU(2)_w$ gauge symmetry (Blue) on which we perform the duality. On the left, the resulting quiver after the duality. On this quiver we perform Seiberg duality on the two $SU(2)_{x_i}$ symmetries (Red). We mark in green the fields considered anti-fundamentals and the rest are considered fundamentals.}
    \label{F:2MinExchange1}
\end{figure}

We  start by considering two trinions $S$-glued to one another as appearing on the left of Figure  \ref{F:2MinExchange1}. For convenience we redefined the minimal punctures symmetry fugacities as $\epsilon \to \epsilon^4$ and $\delta \to \delta^4$. The first Seiberg duality is performed on the middle $SU(2)_w$ gauge node which has six flavors. We identify the fields going to nodes $x_1,x_2$ and $z$ as the fundamentals and the res as anti-fundamentals for the duality. The resulting quiver appears on the right of Figure \ref{F:2MinExchange1}, where the $SU(2)_w$ gauge node gets replaced with an $SU(4)$ gauge node.

\begin{figure}[t]
	\centering
  	\includegraphics[scale=0.4]{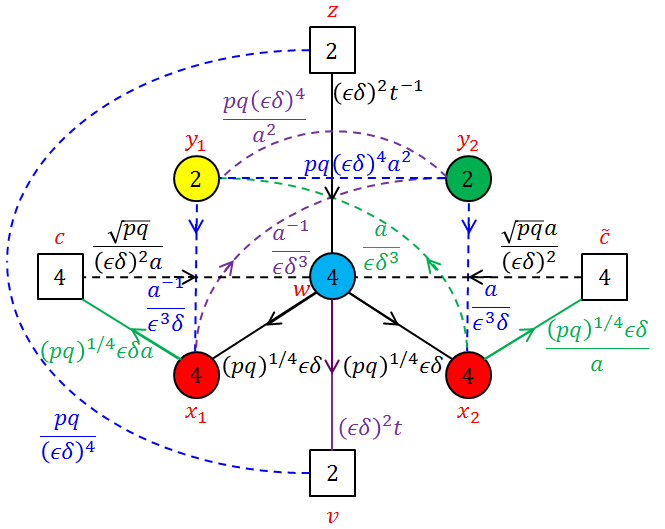}\ \  \includegraphics[scale=0.4]{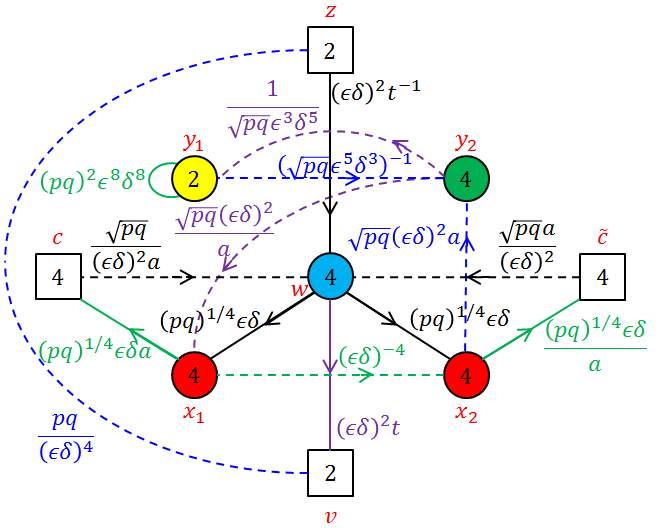}
    \caption{Quivers of two E-string trinions glued together after several Seiberg dualities. On the right we have the resulting quiver of two $S$-glued trinions after three Seiberg dualities.  We mark in blue the fields we consider as fundamentals and the rest are considered anti-fundamentals of the $SU(2)_{y_2}$ gauge symmetry (Green) on which we perform the duality. On the left, the resulting quiver after the duality. On this quiver we perform S-duality on the $SU(2)_{y_2}$ gauge symmetry (Yellow). We pick a duality frame where the field in the fundamental representation of $y_2$ becomes anti-fundamental, but with the same charges and vice-versa.}
    \label{F:2MinExchange2}
\end{figure}

The next step is to perform two more Seiberg dualities on the $SU(2)_{x_1}$ and $SU(2)_{x_2}$ gauge nodes both with six flavors. For $x_1$ node we consider the fields coming out of  $w$ and $y_1$ as the fundamentals and the rest as anti-fundamentals for the Seiberg duality. Similarly for $x_2$ we consider the fields coming out of $w$ and $y_2$ as fundamentals. In the resulting quiver both $SU(2)$ gauge nodes get replaced with $SU(4)$ gauge nodes, see Figure \ref{F:2MinExchange2} on the left.

The fourth Seiberg duality we employ is on the $SU(2)_{y_2}$ gauge node with six flavors. The fields coming out of $x_2$ and $y_1$ nodes are considered to be fundamental for the duality. The resulting quiver appears on the right of Figure \ref{F:2MinExchange2} with the $SU(2)_{y_2}$ gauge exchanged with an $SU(4)$ gauge node.

After these four Seiberg duality we can see that the quiver is symmetric under the exchange of $\delta$ and $\epsilon$ except for the two fundamental fields coming out of the $y_1$ node. This $SU(2)_{y_1}$ gauge node has  one adjoint and four fundamental fields; thus, we can perform S-duality on it. One of the S-dual frames is the same as the original only exchanging the fundamental field transforming under $SU(4)_{y_2}$ to an anti-fundamental field with all the other charges unchanged, and vice-versa for the anti-fundamental. The resulting picture is the same as if we exchanged $\delta$ and $\epsilon$. Finally one can use the same Seiberg dualities we described above in reverse and end up with the same theory we started with only with $\delta$ and $\epsilon$ exchanged. This proves that the new $SU(2)$ minimal punctures obey the expected S-duality exchanging them at the level of indices. This is a strong indication that the duality indeed holds.

\section{Anomalies of $A$-type compactifications}\label{A:Atypeanoms}

Let us here detail the 't Hooft anomalies for compactifications of $N$ $M5$ branes probing a $\Z_k$ singularity on a surface of genus $g$ with fluxes in Abelian subgroups of the internal symmetry.
The expressions are for general $N$ but in this paper we will only use the case of $N=2$. The anomalies were computed in \cite{Bah:2017gph,Razamat:2018zus} and here we just quote the results,

\be
\label{RS $6d$ anomalies}
 & & TrR'=-\frac{1}{2}(k^{2}-2)(N-1)(2g-2)\,,\qquad Trt=-k^{2}NN_{e}\nonumber\\
 & & TrR'^{3}=\frac{1}{2}(N-1)(k^{2}(N^{2}+N-1)+2)(2g-2)\,,\qquad TrR'^{2}t=\frac{1}{3}k^{2}N(N^{2}-1)N_{e}\nonumber\\
 & & TrR't^{2}=-\frac{1}{6}k^{2}N(N^{2}-1)(2g-2)\,,\qquad Trt^{3}=-k^{2}N^{3}N_{e}\nonumber\\
 & & TrR'^{2}\beta_{i}/\gamma_{i}=-kN(N-1)\left(N_{b_{i}/c_{i}}-N_{b_{k}/c_{k}}\right)\,,\qquad Trt^{2}\beta_{i}/\gamma_{i}=kN^{2}\left(N_{b_{i}/c_{i}}-N_{b_{k}/c_{k}}\right)\nonumber\\
 & & TrR'\beta_{i}^{2}/\gamma_{i}^{2}=-kN^{2}(N-1)(2g-2)\,,\qquad TrR'\left(\beta_{i}\beta_{j}/\gamma_{i}\gamma_{j}\right)=-\frac{1}{2}kN^{2}(N-1)(2g-2)\nonumber\\
 & & Trt\beta_{i}^{2}=-kN^{2}\left(2NN_{e}-\left(N_{b_{i}}+N_{b_{k}}\right)\right)\,,\qquad Trt\gamma_{i}^{2}=-kN^{2}\left(2NN_{e}+\left(N_{c_{i}}+N_{c_{k}}\right)\right)\nonumber\\
 & & Trt\beta_{i}\beta_{j}=-kN^{2}\left(NN_{e}-N_{b_{k}}\right)\,,\qquad Trt\gamma_{i}\gamma_{j}=-kN^{2}\left(NN_{e}+N_{c_{k}}\right)\nonumber\\
 & & Tr\beta_{i}/\gamma_{i}=kN\left(N_{b_{i}/c_{i}}-N_{b_{k}/c_{k}}\right)\,,\qquad Tr\beta_{i}^{3}/\gamma_{i}^{3}=N^{2}\left(kN+6(N-1)\right)\left(N_{b_{i}/c_{i}}-N_{b_{k}/c_{k}}\right)\nonumber\\
 & & Tr\beta_{i}^{2}\beta_{j}=2N^{2}(N-1)\left(N_{b_{i}}+N_{b_{j}}-2N_{b_{k}}\right)+kN^{3}\left(N_{e}-N_{b_{k}}\right)\nonumber\\
 & & Tr\gamma_{i}^{2}\gamma_{j}=2N^{2}(N-1)\left(N_{c_{i}}+N_{c_{j}}-2N_{c_{k}}\right)-kN^{3}\left(N_{e}+N_{c_{k}}\right)\nonumber\\
 & & Tr\beta_{i}\beta_{j}\beta_{\ell}=N^{2}(N-1)\left(N_{b_{i}}+N_{b_{j}}+N_{b_{\ell}}-3N_{b_{k}}\right)+kN^{3}\left(N_{e}-N_{b_{k}}\right)\nonumber\\
 & & Tr\gamma_{i}\gamma_{j}\gamma_{\ell}=N^{2}(N-1)\left(N_{c_{i}}+N_{c_{j}}+N_{c_{\ell}}-3N_{c_{k}}\right)-kN^{3}\left(N_{e}+N_{c_{k}}\right)\nonumber\\
 & & Tr\left(\beta_{i}^{2}\gamma_{j}/\gamma_{i}^{2}\beta_{j}\right)=2N^{2}\left(N_{c_{j}/b_{j}}-N_{c_{k}/b_{k}}\right)\nonumber\\
 & & Tr\left(\beta_{i}\beta_{j}\gamma_{\ell}/\gamma_{i}\gamma_{j}\beta_{\ell}\right)=N^{2}\left(N_{c_{\ell}/b_{\ell}}-N_{c_{k}/b_{k}}\right)\,.
\ee
Here $N_{b_i}$, $N_{c_j}$ and $N_e$ are fluxes for the $U(1)_{\beta_i}$, $U(1)_{\gamma_j}$ and $U(1)_t$ symmetries respectively. $R'$ denotes the natural $U(1)_{R'}$ symmetry coming from $6d$ which is not necessarily the conformal one. The slashes appearing in some of the formulas are correlated, and the anomalies not written vanish for all parameters.

\end{appendix}


\bibliographystyle{ytphys}
\bibliography{refs}

\end{document}